\documentclass[manuscript,screen]{acmart}
\AtBeginDocument{%
  \providecommand\BibTeX{{%
    \normalfont B\kern-0.5em{\scshape i\kern-0.25em b}\kern-0.8em\TeX}}}

\setcopyright{acmcopyright}
\copyrightyear{2023}
\acmYear{2023}
\acmDOI{XXXXXXX.XXXXXXX}

\acmPrice{15.00}
\acmISBN{978-1-4503-XXXX-X/18/06}

\usepackage{makecell}
\usepackage{tabularx}
\usepackage{bbm}
\usepackage{subfig}
\usepackage{graphicx}
\begin{document}

\title{Self-supervised Learning for Electroencephalogram: A Systematic Survey}

\author{Weining Weng}
\email{wengweining21b@ict.ac.com}
\affiliation{%
  \institution{Institute of Computing Technology, Chinese Academy of Sciences}
  \city{Haidian}
  \state{Beijing}
  \country{China}
  \postcode{100083}
}

\author{Yang Gu}
\email{guyang@ict.ac.cn}
\affiliation{%
  \institution{Institute of Computing Technology, Chinese Academy of Sciences}
  \city{Haidian}
  \state{Beijing}
  \country{China}
  \postcode{100083}
}

\author{Shuai Guo}
\email{guoshuai20g@ict.ac.cn}
\affiliation{%
  \institution{Institute of Computing Technology, Chinese Academy of Sciences}
  \city{Haidian}
  \state{Beijing}
  \country{China}
  \postcode{100083}
}
\author{Yuan Ma}
\email{mayuan20z@ict.ac.cn}
\affiliation{%
  \institution{Institute of Computing Technology, Chinese Academy of Sciences}
  \city{Haidian}
  \state{Beijing}
  \country{China}
  \postcode{100083}
}
\author{Zhaohua Yang}
\email{yangzhaohua21s@ict.ac.cn}
\affiliation{%
  \institution{Institute of Computing Technology, Chinese Academy of Sciences}
  \city{Haidian}
  \state{Beijing}
  \country{China}
  \postcode{100083}
}
\author{Yuchen Liu}
\email{liuyuchen232@mails.ucas.ac.cn}
\affiliation{%
  \institution{Institute of Computing Technology, Chinese Academy of Sciences}
  \city{Haidian}
  \state{Beijing}
  \country{China}
  \postcode{100083}
}
\author{Yiqiang Chen}
\authornote{Corresponding author.}
\email{yqchen@ict.ac.com}
\affiliation{%
  \institution{Institute of Computing Technology, Chinese Academy of Sciences}
  \city{Haidian}
  \state{Beijing}
  \country{China}
  \postcode{100083}
}

\begin{abstract}
Electroencephalogram (EEG) is a non-invasive technique to record bioelectrical signals. Integrating supervised deep learning techniques with EEG signals has recently facilitated automatic analysis across diverse EEG-based tasks. However, the label issues of EEG signals have constrained the development of EEG-based deep models. Obtaining EEG annotations is difficult that requires domain experts to guide collection and labeling, and the variability of EEG signals among different subjects causes significant label shifts. 
To solve the above challenges, self-supervised learning (SSL) has been proposed to extract representations from unlabeled samples through well-designed pretext tasks. This paper concentrates on integrating SSL frameworks with temporal EEG signals to achieve efficient representation and proposes a systematic review of the SSL for EEG signals. In this paper, 1) we introduce the concept and theory of self-supervised learning and typical SSL frameworks. 2) We provide a comprehensive review of SSL for EEG analysis, including taxonomy, methodology, and technique details of the existing EEG-based SSL frameworks, and discuss the difference between these methods. 3) We investigate the adaptation of the SSL approach to various downstream tasks, including the task description and related benchmark datasets. 4) Finally, we discuss the potential directions for future SSL-EEG research.
\end{abstract}




\maketitle

\section{Introduction}
Electroencephalography (EEG) is a neurophysiological technique that records and measures the brain's electrical activity. The EEG signals are collected in a non-invasive way that involves placing electrodes on the scalp to measure and record the electrical impulses generated by the brain \cite{re1}. Due to the characteristic that the EEG signals are the external representation of the inner brain neural activity, which contains abundant neural information related to various brain stimuli, EEG signals have been widely studied to deal with different real-world tasks: for example, epilepsy recognition \cite{re13}, emotion recognition \cite{re15}, sleep research \cite{re17}, and the brain-computer interface application \cite{re20}. Therefore, the EEG signal is an incredible tool in neuroscience and possesses an exceptionally high clinical utility, which generally became the research focus of physiological signals. 
 
Recently, with the fast development of deep learning and artificial intelligence, machine learning and deep learning models are integrated with labeled samples to complete different classification \cite{re21}, regression \cite{re22}, and generation \cite{re23} tasks. The combination of intelligent algorithms and labeled EEG datasets with supervised learning modes has emerged as a powerful tool to enhance the analysis and interpretation of EEG data. Traditional machine learning methods such as Support Vector Machine (SVM), RandomForeast, and Multi-Layer Perceptron (MLP) demonstrated their efficiency in detecting significant patterns from different hand-crafted EEG features \cite{re27}. Some simple EEG-based tasks, such as EEG-based event classification, emotion recognition, epilepsy detection, and motor imagery classification, can be automatically performed by machine learning models \cite{re28,re29,re30}. The end-to-end deep learning frameworks composed of the Convolutional Neural Network (CNN), Long-Short Term Memory Network (LSTM), Transformer\cite{re34}, and other networks are implemented to model the spatial correlation between electrodes and the temporal variation of the EEG signals. Deep learning methods contain more parameters and complex network structures, with stronger learning and expression abilities to extract physiological information and recognize complex patterns. Adequate labeled EEG data and powerful deep learning models are critical elements for intelligent EEG analysis. Moreover, relying on the large amount of high-quality labeled EEG data, deep models trained with supervised modes can accomplish complex EEG tasks.

The most critical challenge intelligent EEG analysis faces is the scarcity of labeled samples. While training deep models demands extensive labeled data. However, obtaining a large-scale labeled EEG signal for model training is impractical. The annotation of the EEG signal necessitates manual intervention from experts well-versed in neurophysiology, possessing a profound familiarity with the distinctive features of interest embedded within the EEG data. The high costs and the need for expert knowledge in the annotation process make constructing EEG datasets extremely challenging. In addition, the scarcity of specific brain states significantly affects the acquisition of EEG signals \cite{re40}. For example, abnormal emotion states and seizure states are relatively rare among the subjects, making it more difficult for sample collection. Therefore, building annotated EEG datasets for training deep models is constrained by various factors. It necessitates the involvement of domain experts \cite{re41}, demanding substantial time and cost \cite{re42}, which poses a significant challenge for the application of supervised learning in EEG analysis.
\begin{figure}[t]
  \centering
  \includegraphics[width=0.5\linewidth]{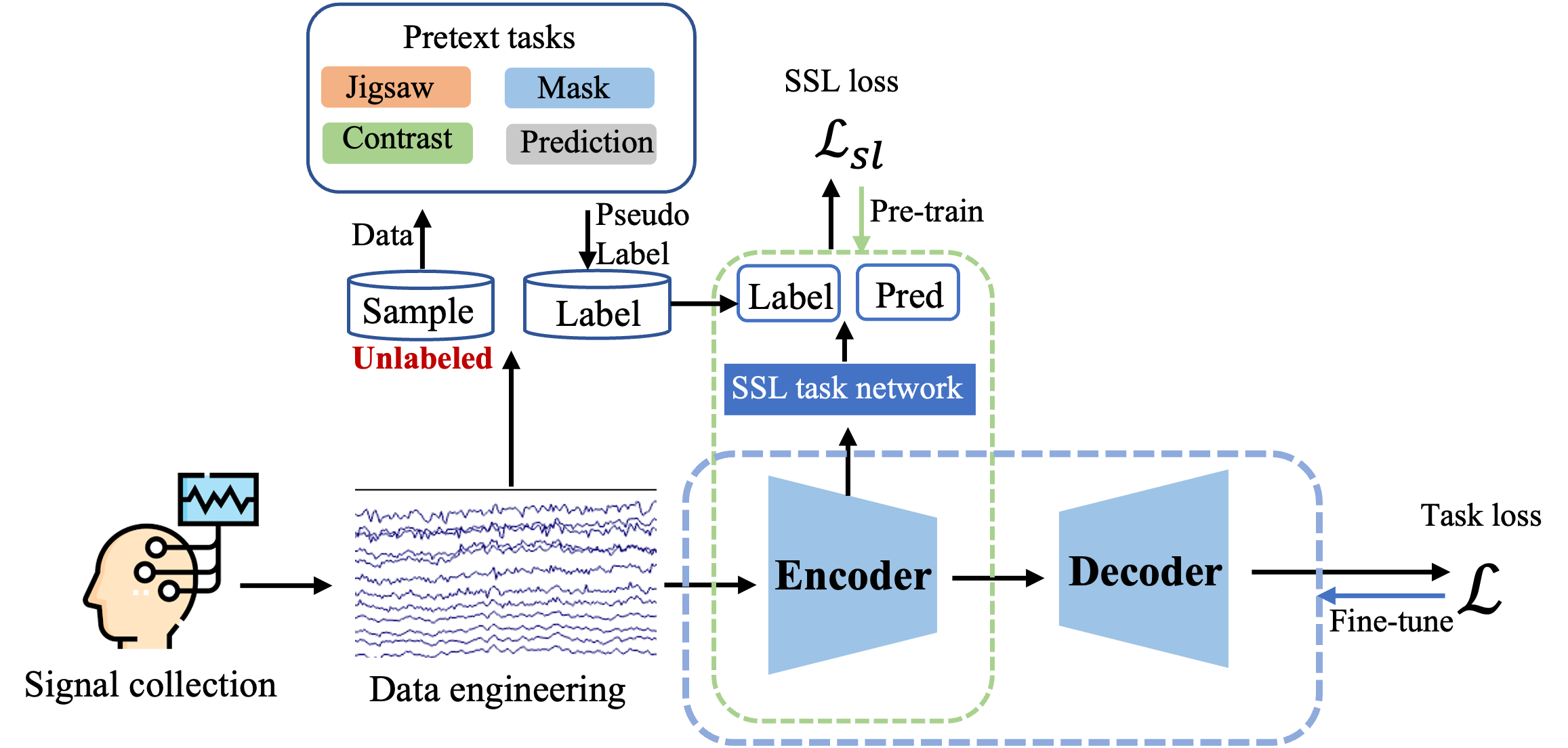}`
  \caption{The general process of SSL-integrated EEG analysis. The black arrows represent forward propagation, the green and blue arrows denote backpropagation based on pretext task loss and downstream task loss, respectively. }
  \label{fig_1}
\vspace{-\baselineskip}
\end{figure}

Besides, supervised EEG analysis faces the inconsistency problem, which severely impacts the effectiveness of supervised learning. Interpreting EEG signals often involves subjectivity and variability among the subjects and evaluators \cite{re43}. First, some tasks generally collected signals annotated by the participants, which have a vital subjective component and do not necessarily represent the actual states generated by their brains, leading to inconsistencies in the labels. Besides, owing to the distinctive differences in each individual's brain, substantial variations exist in brain signals among different subjects. This diversity may result in evaluative discrepancies in labels annotated by domain experts, as different experts might assign different labels to the same EEG segment \cite{re46}. Such variations introduce an inconsistency in the labeled samples. Therefore, mitigating the significant influence of inconsistency issues in the training process and improving generalization ability became critical problems faced by the EEG analysis.

Self-supervised learning (SSL) has shown its superior performance in solving the challenges mentioned above, which leverages the intrinsic structure and information within data to train models without labels. Self-supervised learning designed a series of pretext tasks different from the final modeling target that generate the \emph{pseudo-label} directly from the unlabeled samples to train the model \cite{re48}. In the Computer Vision (CV) and Natural Language Process (NLP), self-supervised learning has achieved tremendous success. In CV, SSL structure helps the model to learn the effect image representation through the pretext tasks such as image rotation \cite{re52}, jigsaw \cite{re53}, and reconstruction \cite{re55}, which significantly improved the downstream task performance, sample efficiency and mitigate the overfitting problem \cite{re56}. In NLP, the mask-reconstruction (MAE) and the prompt answering pretext tasks help the language model to comprehensively understand textual context, enabling a series of functions of machine translation and conversation system \cite{re57}. Therefore, the strong representation ability and low-labeled sample requirements of the self-supervised learning paradigm demonstrate its potential as an effective training method, which offers new insights and tools for addressing various complex problems in different domains.  

Implementing the SSL frameworks in the EEG field is gaining more and more attraction among various researchers \cite{re40}. Figure~\ref{fig_1} illustrates a typical SSL-integrated EEG analysis method. There have been certain studies investigating the combination of SSL with EEG analysis, which conduct the preliminary exploration of SSL to deal with temporal physiological signal-based tasks. Accordingly, this paper comprehensively reviews the utilization of self-supervised learning for EEG analysis, which provides an in-depth exploration of the taxonomy, the pros and cons, and the development potential of the EEG-based SSL frameworks. The main contributions of this paper are listed as follows:

\textbf{(1) Comprehensive review}. This paper provides a comprehensive up-to-date review of the self-supervised learning integrated EEG analysis methods. We analyze the technique details of different SSL approaches for EEG signals, including the type of pretext tasks, the mathematical description, the performance of the SSL, and some simple summaries. By comparing different methods, we outline the general process and characteristics of the EEG-based SSL methods.

\textbf{(2) Systematic and reasonable taxonomy}. Following the classical taxonomy of the traditional self-supervised learning methods, we rigorously categorize existing studies on self-supervised learning in EEG into four major classes: the prediction-based method, the generation-based method, the contrastive-based method, and the graph-based method.

\textbf{(3) Future potential directions}. We also analyze the pros and cons of various methods, identify the limitations of current works, and take into account the inherent characteristics of EEG data to indicate the potential directions for developing SSL-based EEG analysis.

\section{Preliminary}
This section provides a concise overview of traditional supervised EEG-analysis methods. In addition, we outline the form definition and mathematical description of the self-supervised learning frameworks proposed in other fields (CV, NLP), which serves as a preliminary of the EEG-based self-supervised method. 

\subsection{Supervised EEG Analysis}

EEG signals have been widely studied to decode brain activity for addressing various real-world tasks. For instance, EEG have been used to recognize specific emotion \cite{re44}, detect seizure \cite{re61}, classify sleep stage \cite{re62}, recognize motor imagery \cite{re63}, decode visual or auditory information \cite{re64}, etc. Machine learning and deep learning supervised methods have been widely adopted to analyze EEG signals, extract features, and complete specific tasks. Existing studies can be classified into two categories \cite{re60}: the feature-driven and the model-driven methods, where the feature-driven methods combine the handcrafted features with traditional machine learning classifiers to interpret EEG signals, and the model-driven methods construct end-to-end deep learning models to automated extract task-related EEG features. 

\textbf{Feature-driven methods}. The feature-driven methods use specific features extracted from EEG signals to guide the analysis process. In general, the feature-driven methods select handcrafted features that have been proven effective for the task according to the previous research by neuroscientists \cite{re68}. By leveraging the selected features through traditional machine learning classifiers, the models can uncover patterns, relationships, and insights in understanding EEG signals and brain activity. Various handcrafted features fed into different models are extensively applied to multiple tasks. For example, time domain features like the Hjorth parameter \cite{re69}, the high order crossing \cite{re70}, the statistical analysis features \cite{re71}, etc; the frequency domain features like different independent frequency bands \cite{re43} generated through Fast Fourier Transfer and differential entropy \cite{re45}, etc; the temporal frequency domain features which combined the frequency features with the time window to introduce the variation of frequency features overtime \cite{re71}. Utilizing these manually engineered features as input, machine learning have demonstrated a dependable performance in tasks such as emotion recognition, sleep stage classification, and motor imagery classification \cite{re27}.

\textbf{Model-driven methods}. Model-driven methods refer to approaches that incorporate deep end-to-end models to interpret and analyze temporal EEG raw data or the high dimensional EEG features. Deep models can capture specific spatial-temporal information to infer underlying brain dynamics, quantify brain activity, and complete complex EEG-based classification or regression tasks \cite{re73}. The existing model-driven methods are typical examples of supervised deep learning approaches that rely extensively on a substantial volume of training samples. Owing to the powerful learning capabilities of deep learning and the assistance of extensively labeled samples, the efficacy of models has been further heightened across diverse complex EEG tasks.

\begin{figure}[t]
  \centering
  \includegraphics[width=0.8\linewidth]{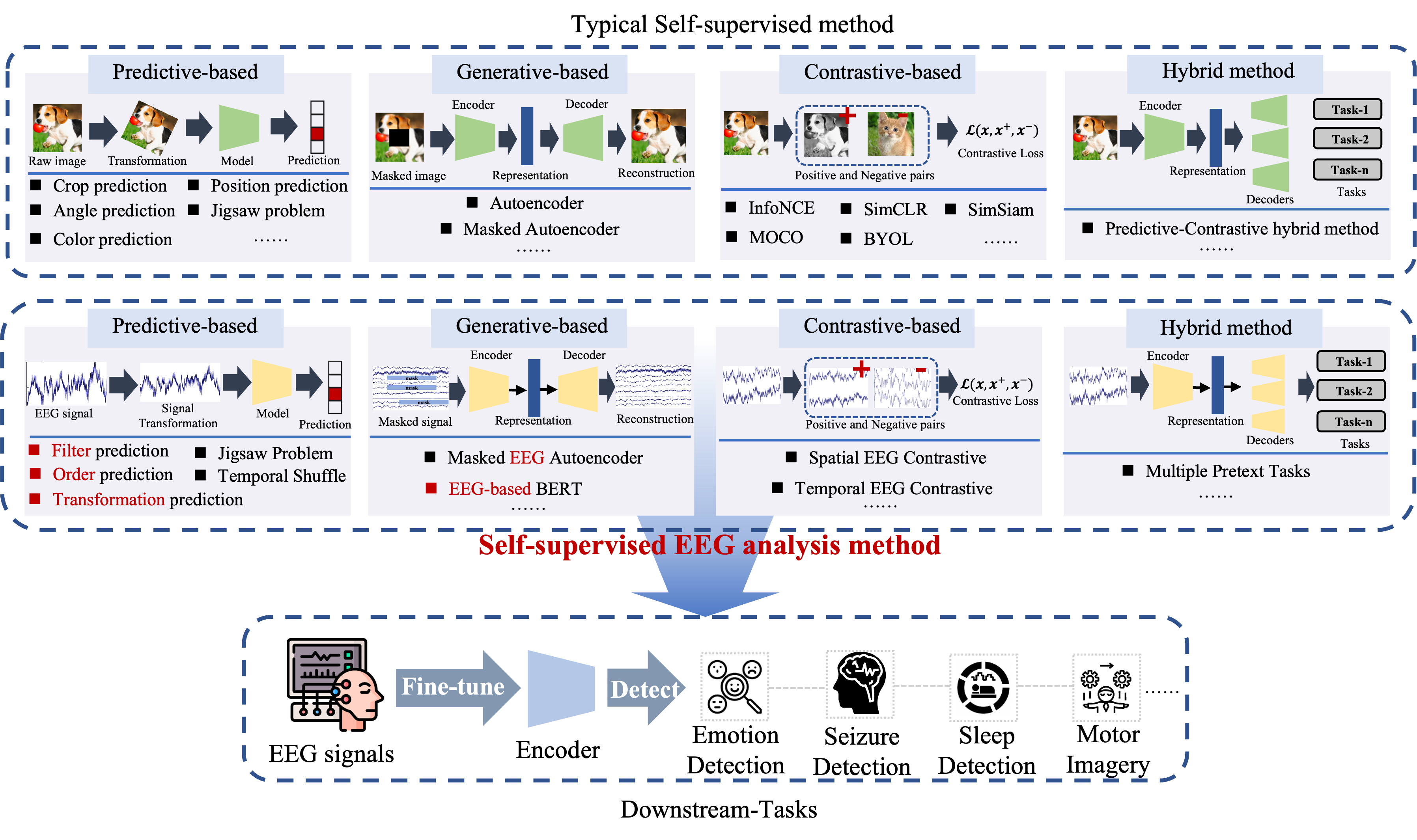}
  \caption{The taxonomy of the typical self-supervised learning methods and self-supervised EEG analysis methods}
  \label{fig_3}
\vspace{-\baselineskip}
\end{figure}
\subsection{Overview of Self-supervised Learning}
Self-supervised learning can extract effective representation from unlabeled samples instead of directly training end-to-end models through labeled samples, which has shown its superior performance in learning spatial images and sequential context representation in the fields of CV and NLP. In this part, we outline the mathematical definition of the SSL, explain the terms of essential concepts, and briefly divide the existing SSL frameworks into four distinct categories based on the variation in pretext tasks. 

\textbf{Term explanation}. We provide important definitions of terms to help further understand self-supervised learning. 
\begin{itemize}
    \item \textbf{Pretext task}. The pretext tasks $T=\{t_1,t_2,...,t_n\}$ refers to the learning objective or task designed to leverage the content or structure within unlabeled data to help the model learn knowledge and effective representations. The learned representations can then be transferred to downstream tasks with limited labeled data. 
    \item \textbf{Pseudo-label}. The pseudo-label $Y_p$ is the artificial label created based on the pretext tasks to train the model. The pseudo-labels serve as a form of supervision that guides the self-supervised learning process to extract specific features from unlabeled samples.
    \item \textbf{Downstream task}. The definition of downstream task $d_t$ is the target or final task to be performed using the features or representations learned from the previous phase of training (by pretext tasks). The downstream task typically requires labeled samples to fine-tune the previous model to transfer the representation model to become more specific and task-focused toward the downstream task.
    \item \textbf{Human-label}. The human-label refers to the labels for the downstream samples annotated by human experts.
\end{itemize}
\textbf{Mathematical definition}. The objective of self-supervised learning is to learn a function $\mathcal{F}(x) \rightarrow \mathbb{R}^{d}$ that maps the input instances $X$ to a $d$-dimensional representation space $\mathbb{R}^{d}$ capturing essential features from unlabeled samples. The frameworks of self-supervised learning are generally regarded as the \textbf{encoder-decoder} structure encompassing an encoder $f_{\theta}$ to generate representation and several decoders $g$ to decode the representation to complete different tasks: \textbf{pretext task decoder} $g_{\delta}^p $ cascade with the encoder to accomplish pretext task and pre-train the model without external labels, and \textbf{downstream task decoder} $g_{\xi}^d$ can recognize specific patterns in the representation to fine-tune the model adopted to complete downstream tasks. In general, the training paradigms of SSL can be summarized into three categories: 1. pre-train mode; 2. joint-train mode; 3. unsupervised-train mode. The first mode is the pre-train mode, which uses pretext tasks to pre-train the representation model and downstream tasks to fine-tune the encoder and downstream task decoder to transfer the model for addressing specific tasks. The process can be expressed as follows:
\begin{equation}
    \begin{split}
        \theta,\delta=\mathop{\arg\min} \limits_{\theta, \delta} \mathcal{L}_{pt}(g_{\delta}^p(f_{\theta}^p(X)),Y_p)  \\
        \theta,\xi=\mathop{\arg\min} \limits_{\theta, \xi} \mathcal{L}_{ft}(g_{\xi}^d(f_{\theta}^p(X)),Y)
    \end{split}
\end{equation}
where $\mathcal{L}_{pt}$ and $\mathcal{L}_{ft}$ represent the loss of pretext task and downstream task, respectively. The encoder is trained by the pretext task and fine-tuned by the downstream task to first generate effective representation and then transfer the learned knowledge into the specific task. The downstream task decoder is trained by the downstream task loss to fully leverage the representation for target task completion.

The second mode is the co-train mode, where a joint loss function is constructed to leverage pretext and downstream tasks to jointly train the model. The pretext task collaboratively explores the relevant knowledge for the downstream task and also serves as the regularization term to constrain the gradient during training, thereby mitigating the overfitting problem. This mode can be expressed as follows:
\begin{equation}
    \theta,\xi=\mathop{\arg\min} \limits_{\theta, \delta, \xi} \alpha\mathcal{L}(g_{\delta}^p(f_{\theta}^p(X)),Y_p)+\beta \mathcal{L}(g_{\xi}^d(f_{\theta}^p(X)),Y)
\end{equation}
where $\alpha$ and $\beta$ are the hyper-parameter to balance different losses.

The third mode is the unsupervised-train mode, which is similar to the pre-train mode, but the parameters of the encoder are frozen during the fine-tuning stage. This mode only fine-tunes the downstream task decoder to verify the generated representations' effectiveness. The process can be formulated as follows:
\begin{equation}
    \begin{split}
        \theta,\delta=\mathop{\arg\min} \limits_{\theta, \delta} \mathcal{L}_{pt}(g_{\delta}^p(f_{\theta}^p(X)),Y_p)  \\
        \theta,\xi=\mathop{\arg\min} \limits_{\xi} \mathcal{L}_{ft}(g_{\xi}^d(f_{\theta}^p(X)),Y)
    \end{split}
\end{equation}

Within the pretext task based taxonomy of SSL, we can categorize the SSL method into four types: predictive-based, generative-based, contrastive-based, and hybrid SSL method. Figure~\ref{fig_3} demonstrates the general taxonomy of SSL, and the detailed explanations are as follows:

\textbf{Predictive-based SSL method}. The predictive-based SSL method creates classification pretext tasks to predict discrete pseudo labels generated from unlabeled data to learn effective features. For instance, pretext tasks like predicting image rotation angles \cite{re52} and pixel colors \cite{re76} can force the model to extract spatial features and object boundaries in the images beneficial for the downstream tasks such as object detection; pretext tasks like predicting next sentence \cite{re57} can help language model understand contextual correlation. Due to their simpler execution nature, predictive-based SSL methods are mostly easy to combine with traditional deep models, and the proficient performance in prediction tasks signifies that the model has mastered specific knowledge for downstream tasks.

\begin{figure}[t]
  \centering
  \includegraphics[width=\linewidth]{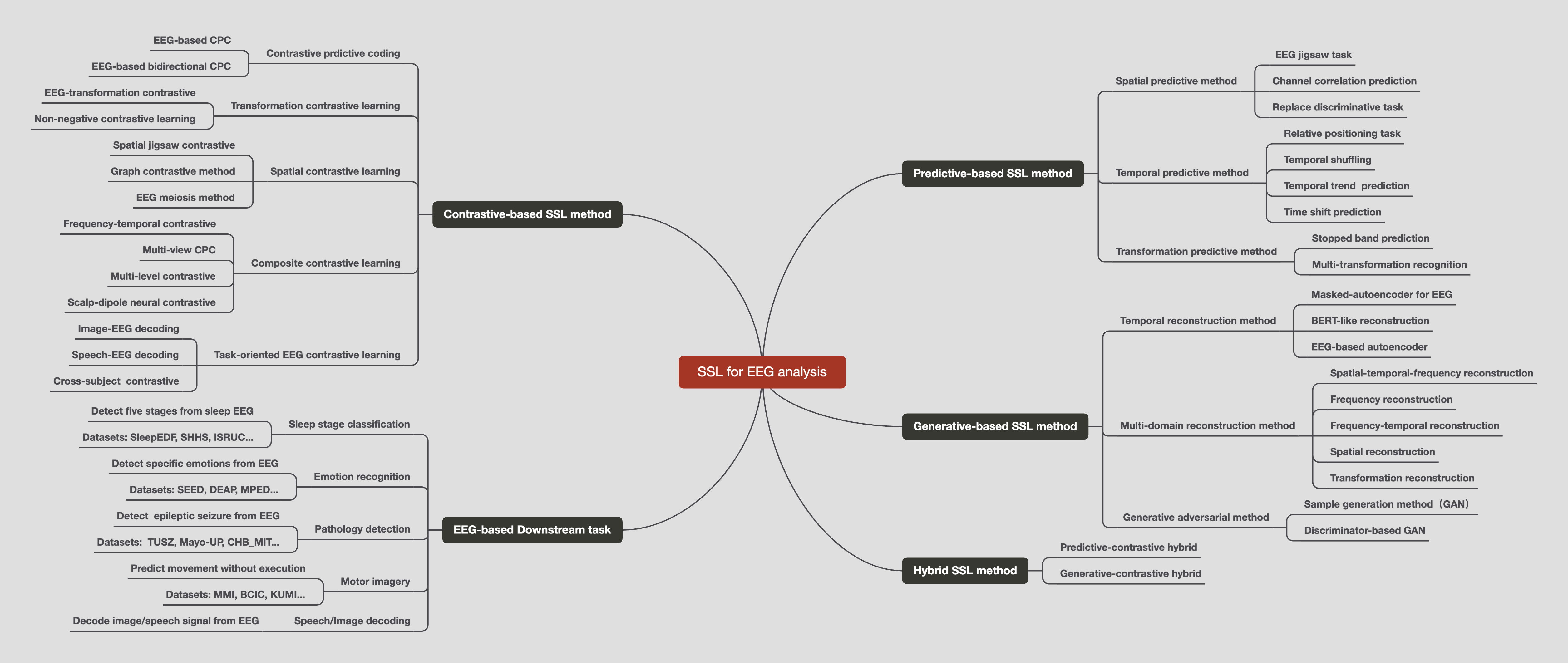}
  \caption{Categories of Self-supervised learning for EEG analysis}
  \label{structure}
\vspace{-\baselineskip}
\end{figure}

\textbf{Generative-based SSL method}. The generative-based SSL method designs generation or reconstruction pretext tasks to capture contextual features and correlations to generate effective representations. The most widely used generative-based pretext task is the reconstruction task. This task begins by encoding the input sample into a distinctive representation, followed by a decoding process to reconstruct the original input. By making the input and output as similar as possible, the encoder can learn significant features to reconstruct the input, which are highly effective for target downstream tasks. For example, typical methods like autoencoder \cite{re77} have been investigated to extract representations from image and textual data. Recently, the mask-reconstruction pretext task has supplanted traditional reconstruction tasks to extract the contextual information from unlabeled samples. This task masks part of the input samples and reconstructs the masked data through the contextual data, where the encoder is responsible for extracting features and generating representations, and the decoder is responsible for reconstructing the masked data. In vision tasks, the masked autoencoder \cite{re55} can extract spatial contextual features from unlabeled samples for downstream classification and segmentation. In language tasks, the BERT model captures token-level context correlation information, which greatly improves the performance of subsequent tasks such as machine translation and text generation.

\textbf{Contrastive-based SSL method}. The contrastive-based SSL method adopts the 'comparison' technique, which encourages similar data points to be closer in the representation space while pushing dissimilar data points apart. Augmentation methods are important in the constrastive-based SSL: input samples are augmented to create negative and positive sample pairs, where the positive pairs represent the similar samples, and negative pairs refer to the vastly dissimilar samples \cite{re79}. By optimizing the designed contrastive loss, the model minimizes the distance between positive pairs and maximizes the distance between negative pairs to extract identical features and transferable representations. Based on the theory of information bottleneck \cite{re81} and mutual information \cite{re80}, InfoNCE loss \cite{re82} is proposed to efficiently learn representations where positive pairs are closer together in the feature space compared to negative pairs. Besides, SimCLR \cite{re83}, MoCo \cite{re84}, and other contrastive learning methods have become important frameworks driving the development of computer vision.

\textbf{Hybrid SSL method}. The hybrid SSL method combines multiple SSL techniques or tasks to create a powerful framework for learning representations. The main idea is to leverage the strengths of different pretext tasks to capture diverse and informative features from unlabeled samples. The weighted fusion of losses from multiple pretext tasks enables the model to grasp multi-dimensional knowledge. It is particularly valuable when data is heterogeneous or a single pretext task may not capture all the relevant information in the unlabeled samples \cite{re74}.

Following the taxonomy of typical SSL frameworks in vision and language fields, this paper categorizes self-supervised EEG analysis methods into predictive, generative, contrastive, and hybrid frameworks. Comprehensive summary for different methods are provided from Section \ref{predictive} to Section \ref{hybrid}. The structure of this survey can be visualized in Figure~\ref{structure}.

\begin{figure}[t]
\centering
\subfloat[The framework of the spatial predictive method to predict channel augmentation techniques applied to EEG.]{
\includegraphics[width=0.48\linewidth]{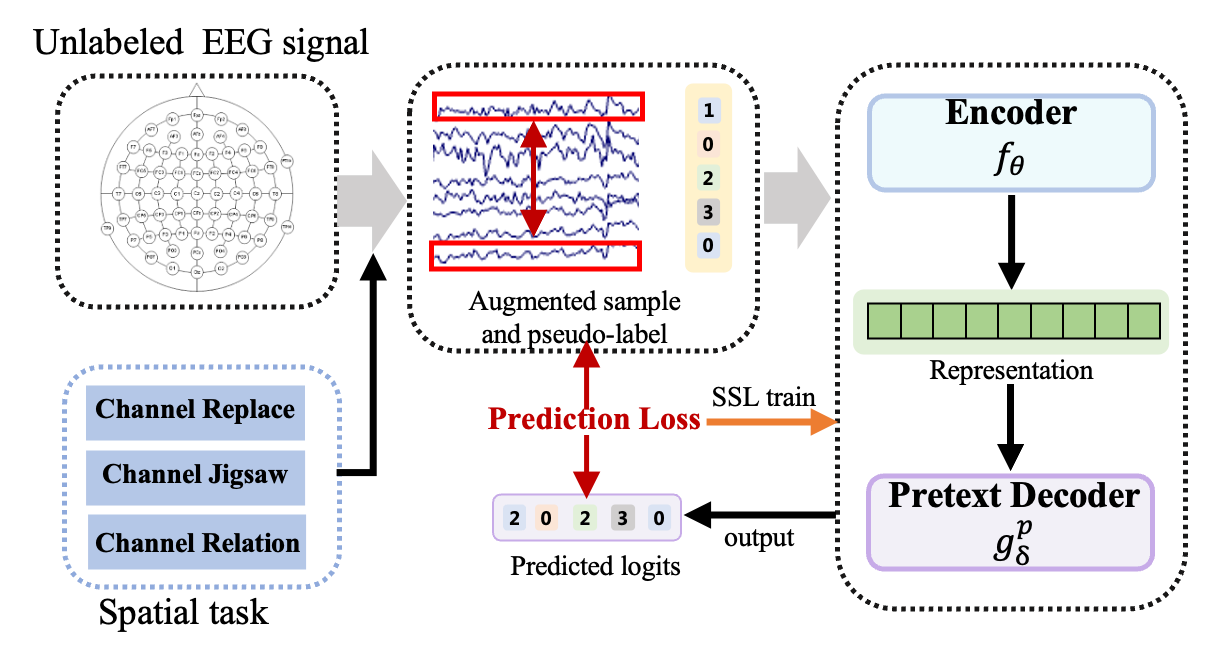}
\label{p1}
}
\hfill
\subfloat[The framework of the temporal predictive method to predict different temporal augmentation techniques applied to EEG.]{
\includegraphics[width=0.48\linewidth]{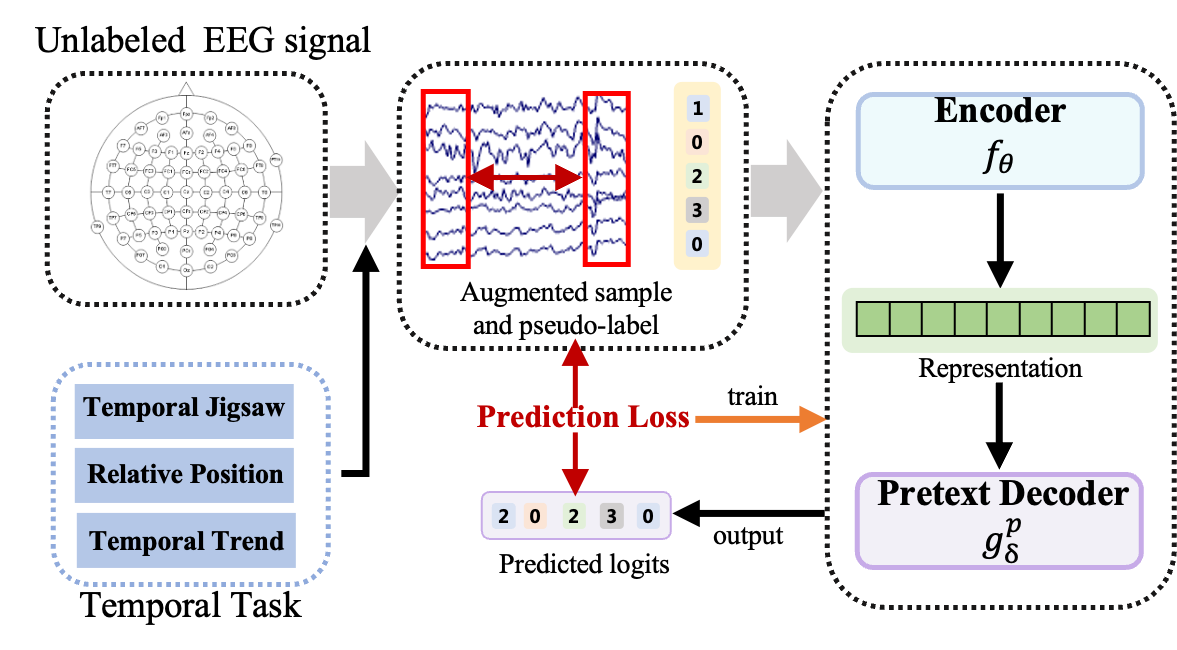}
\label{p2}
}

\centering
\caption{The comparison of spatial predictive and temporal predictive SSL EEG analysis methods }
\label{predictive_Fig1}
\vspace{-\baselineskip}
\end{figure}

\section{predictive-based SSL EEG analysis method}\label{predictive}

The predictive-based SSL EEG analysis method aims to design and execute classification to acquire domain-specific knowledge beneficial for various downstream tasks. Multi-channel EEG signals present distinctive characteristics, including high temporal density, pronounced temporal dependencies, and intricate inter-channel correlations, indicating the presence of critical features within the temporal, frequency, and spatial domains of EEG data. Sequentially, pretext tasks are implemented to distinguish EEG samples that are augmented through temporal, frequency, and spatial processing to acquire features from different domains. Therefore, we can categorize the existing studies into three sub-categories: (1) \textit{spatial predictive methods}, (2) \textit{temporal predictive methods}, and (3) \textit{transformation predictive methods}. The typical frameworks of three kinds of methods are demonstrated in Figure~\ref{predictive_Fig1} and Figure~\ref{pred_2}, and the summary of existing works is listed in Table~\ref{tab_1}.

\subsection{Spatial Predictive Method}

The spatial predictive method draws inspiration from SSL in the image domain, establishing local or global spatial-structure-related pretext tasks to help the model comprehend spatial contextual information. Figure~\ref{p1} shows the typical spatial predictive framework for EEG analysis, and different methods have been investigated to extract channel correlation and brain structure, which are listed as follows:

\textbf{EEG jigsaw task}\cite{re85} is analogous to the image jigsaw pretext task in CV. EEG jigsaw task involves the random shuffling of EEG channels, followed by an expectation that the model can reconstruct the original sequence of the scrambled channels or predict the order in which the channels were shuffled. For example, assuming the raw EEG data $X_{sp} \in {\Bbb{R}}^{c \times t}$, where $c$ is channel numbers and $t$ is the number of sampling points. The random shuffling operation then produces a permuted EEG matrix $X^*_{sp} \in {\Bbb{R}}$, where the temporal information remains unchanged, but the channel order has been shuffled. The loss function of the jigsaw task can be described as follows:
\begin{equation}
 \mathcal{L}(X^*_{sp},Y^*)=-\sum_{i=1}^{N} Y^*_i \log(g_\delta^p(f_{\theta}(X^*_{sp})))
\end{equation}
where $Y^*$ represents the one-hot pseudo-label (channel order), and $N$ represents the batch size. The loss function calculates the cross-entropy between the predicted order of the shuffled EEG sample and its corresponding label. By minimizing this pretext loss, researchers believed the model can capture spatial features related to the distribution of multi-channel EEG signals across the cortical regions of the brain, which are closely related to downstream tasks such as emotion recognition and seizure detection.

\textbf{Channel correlation prediction}\cite{re86} is designed to realize the spatial correlation between different channels. Researchers proposed that the time delay exists in the propagation of EEG signals between channels in distinct brain regions. EEG signals will experience delays when propagating from one region to a distant one, and exploring these features enables the model to understand the activation modes and information exchange patterns of brain activity. In this task, pseudo labels are generated from signal correlation between channels, which can be calculated as follows:
\begin{align}
    Y(i,j,t_1,t_2)= \left\{
    \begin{aligned}
        1, cossim(X_i(t_1),X_j(t_2)) \geq \kappa \\
        0, cossim(X_i(t_1),X_j(t_2)) < \kappa
    \end{aligned}
    \right.
\end{align}
This function calculates the cosine similarity between the $i$-th channel and the $j$-th channel at time slices $t_1$ and $t_2$. The $cossim$ represents the cosine similarity, and the $\kappa$ is the predefined threshold to determine whether the two slices are correlated and assign the pseudo label. The binary cross-entropy loss can be used as loss prediction loss to pre-train the model, where the predictions are generated by the encoder-pretext task decoder structure:$g_\delta^p(f_\theta(X_1,X_2))$. 

\textbf{Replace discriminative task}\cite{re86} is the binary classification task to extract channel-specific differential features by identifying distinct components from different channels. In this task, a random replacement is performed to replace a certain percentage $p_r \%$ of original EEG signal $X_i$ with signal $\overline{X_i}$ sampled at any channels and time slices. The pseudo labels are constructed to indicate whether the current samples have been replaced, which can be described as follows:
\begin{align}
    Y(X_i)= \left\{
    \begin{aligned}
        1, f_I(X_i,i)=0  \\
        0, f_I(X_i,i)\ne 0
    \end{aligned}
    \right.
\end{align}
where $f_I(X_i,i)$ is the function to judge whether the signal $X_i$ has been replaced or not. Subsequently, by minimizing the binary cross-entropy pretext task loss, the model learns the distinctive spatial features of different channels and retains essential information beneficial for various downstream tasks.

\subsection{Temporal Predictive Method}

The temporal predictive methods aim to capture the temporal correlation and sequential dependencies in EEG signals. As the temporal physiological signal, temporal characteristics play an important role in various EEG-based tasks. Figure~\ref{p2} shows a typical framework of temporal predictive SSL for EEG analysis, and different temporal predictive pretext tasks have been proposed to investigate potential temporal information, which are summarized as follows:

\textbf{Relative positioning task} is the temporal predictive method to distinguish whether two different EEG segments are close or distinct in time dimension \cite{re87}. This task firstly constructs an EEG pair  $X_{t_i},X_{t^{'}_i} \in {\Bbb{R}}^{c \times t}$ represent two sampled EEG segments. Representation of EEG signals should change slowly over time, which means EEG segments proximate in the time dimension convey similar information, and those further apart exhibit significant dissimilarities \cite{re87}. The duration parameter $\tau_{pos}$ controls the duration of positive context. For two EEG segments$X_{t_i}$ and $X_{t^{'}_i}$, the temporal interval $|t_i-t^{'}_i|\leq \tau_{pos}$ indicates that these segments are within positive duration, sharing common underlying labels. Therefore, the pseudo labels of the relative positioning task can be constructed as follows:
\begin{align}
    Y(X_{t_i},X_{t^{'}_i})= \left\{
    \begin{aligned}
        1, |t_i-t^{'}_i|\leq \tau_{pos} \\
        -1, |t_i-t^{'}_i|> \tau_{pos}
    \end{aligned}
    \right.
\end{align}
where $-1$ and $1$ represent samples of negative and positive duration. The training sample $S_N=\{(X_{t_i},X_{t^{'}_i}),Y((X_{t_i},X_{t^{'}_i})\}$ can be used to train the model with binary classification loss to capture temporal information. This method has been widely used for continuous EEG classification tasks such as the sleep stage classification \cite{re87}.

\textbf{Temporal Shuffling} is considered as the variation of the relative positioning task \cite{re87}. The temporal shuffling task first samples two different EEG segments $X_{t_i},X_{t^{'}_i}$ from positive duration, and then samples another EEG segment $X_{t^{''}_i}$ between the first two segments or in the negative duration. Three different segments form the triplet $(X_{t_i},X_{t^{'}_i},X_{t^{''}_i})$. The shuffling operation is performed to permute the order of the segments in the triplet randomly. The pseudo labels indicating whether the triplet has been shuffled can be constructed: 0 for the shuffled triplet and 1 for the normal triplet. Then, the model learns to distinguish whether the triplet has been shuffled through the concatenated differential features between segments, which can be calculated as follows:
\begin{equation}
    D(X_{t_i},X_{t^{'}_i},X_{t^{''}_i})=concat(|f_\theta(X_{t_i})-f_\theta(X_{t^{'}_i})|,|f_\theta(X_{t^{'}_i})-f_\theta(X_{t^{''}_i})|)
\end{equation}
where $cancat$ is the vector concatenation operation. The model conducts shuffling classification by utilizing differential encoded information from different segments as features, which can help to comprehend temporal dependencies within EEG signals. Besides, another temporal shuffling method proposed by \cite{re88} divides the EEG slice into three equidistant sequences, then randomly shuffles the order of sequences to form the shuffled sample. The model is asked to predict the order of input shuffled samples to capture temporal correlation. Therefore, for shuffled EEG signals, both binary classification (predict whether they have been shuffled ) and multi-class classification (predict the order of shuffled signals) can serve as the pretext task to extract temporal features of physiological signals at different granularities.

\textbf{Temporal trend prediction}\cite{re89} is a task to identify the potential trends of EEG to capture short-term and long-term dynamic patterns. This task divides the EEG signal into three categories according to its temporal characteristics: stationary, trendstationary, and cyclostationary. By learning how to identify temporal trends, the model can comprehend the temporal relationships within signals and capture both global and local essential waveform information to generate the temporally enriched representations, which can benefit a variety of downstream tasks, like sleep stage classification.

\textbf{Time shift prediction}\cite{re90} is a task to predict the time shift performed to the EEG signals by contrasting the differences in features between the raw EEG signal and shifted signals. In this task, the raw EEG signal $X_{t_i}$ and augmented EEG signal $X_{t_i+\rho}$ resulting from $\rho$-step time shifts applied to the raw EEG signal. The raw signal and shifted signal are encoded into representations, and the pretext task uses a classification method to analyze the difference between the two representations and classify how much the raw signal was shifted. By minimizing the classification loss, the encoder can learn the temporal-aligned features and dependencies within EEG signals, generating the representation containing rich time information significantly beneficial for long-term EEG tasks like clinical monitoring.

\subsection{Transformation Predictive Method}\label{transformation}
Figure~\ref{pred_2} shows the general process of the transformation predictive method. This task aims to predict specific transformations applied to the EEG signals to learn signal-related features in the time-frequency domain. Different EEG transformation techniques employed to augment EEG samples to be recognized in this task can be listed as follows:

\begin{figure}[t]
  \centering
  \includegraphics[width=0.48\linewidth]{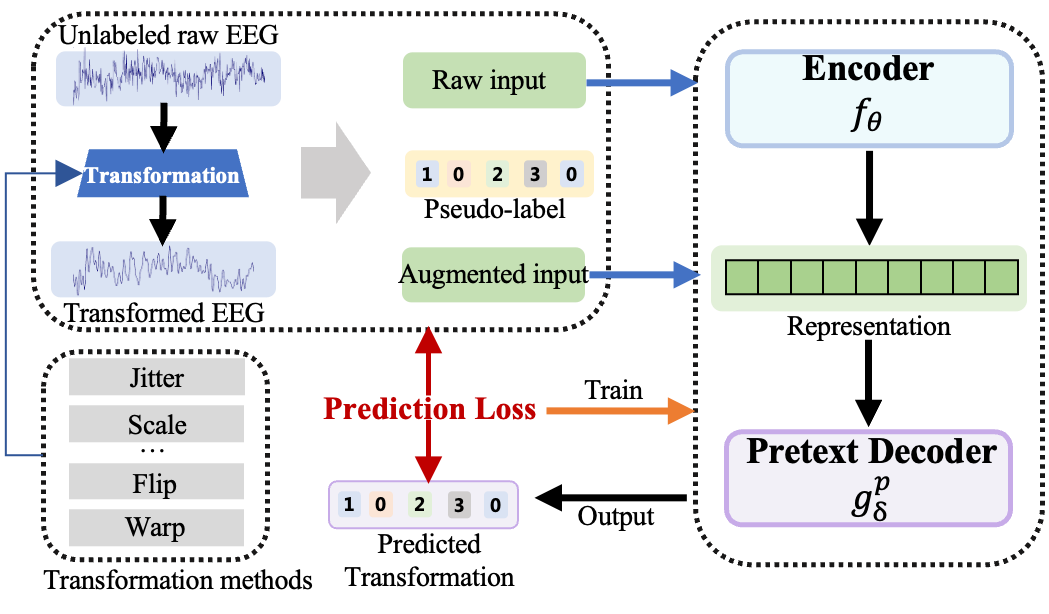}`
  \caption{The general process of transformation predictive method for EEG analysis. Different signal transformation techniques are applied to EEG signals to generate augmented samples and pseudo labels. The model can capture critical signal-level features for downstream tasks by correctly predicting the transformation method. }
  \label{pred_2}
\vspace{-\baselineskip}
\end{figure}

\textbf{Stopped band prediction} randomly removes specific frequency bands in EEG signals and forces the model to predict the index of the removed channel to learn frequency-related features \cite{re91}. EEG signals comprise information from multiple frequency bands, with essential information concentrated within the frequency range of 1 to 50 Hz, encompassing five independent frequency bands: $\delta$ (0.5-4Hz), $\theta$ (4-8Hz), $\alpha$ (8-12Hz), $\beta$ (12-30Hz) and  $\gamma$ (30-50Hz). This task transforms the EEG signal from time to the frequency domain and remaps the signal to the time domain after the random removal of a specific frequency band, and the pseudo labels $Y(x_i) \in [0,1,2,3,4]$ are set representing the index of the removed band. By forcing the model to predict the stopped band through encoded representation $f_\theta(X_i)$, the encoder $f_\theta$ can learn efficient frequency correlation and features to form the temporal-frequency representation.

\textbf{Multi-transformation recognition} aims to predict the transformation technique used to augment EEG signals to extract fine-grained signal features and form effective representations \cite{re92}. In this task, EEG signals are augmented through one transformation technique, and the model is asked to recognize the transformation methods. The common encoder $f_\theta$ extracts features from augmented EEG signals and encodes them into representation, with multiple binary classifiers to recognize different transformation methods. Each classifier corresponds to a specific transformation method to determine its occurrence. Six transformation methods are proposed to be recognized: 

(1)\textit{Noise Adding}. Adding random noise generated by Gaussian distribution $N~(\mu,\sigma^2)$. The noise $NS_i$ directly added to the original signal to $X_i$, resulting in a noise-augmented signal $X^{ns}_i$.

(2)\textit{Scale transformation} alters the waveform of the EEG signals. The amplitude of EEG signal is stretched or telescoped through a scale factor $\alpha$, where the scale-augmented signal can be expressed as $X^{st}_i=\alpha * X_i$.  

(3)\textit{Horizontal flipping}. This transformation method directly flips EEG signal horizontally. The horizontal-augmented signal can be expressed as $X^h_i=-X_i$.

(4)\textit{Vertical flipping} flips EEG signal (each sample) vertically. The vertical-augmented signal can be described as $X^{v}_i=flip(X_i)$, where the $flip$ represents the segment's vertical symmetric flip.

(5)\textit{Temporal dislocation} is consistent with the temporal shuffling in predictive methods. This method divides EEG segments into sub-segments and randomly shuffles these sub-segments to form the dislocation-augmented signal $X^{td}_i$. 

(6)\textit{Time warping} method randomly stretches and compresses sub-segments to form the augmented samples. This method randomly selects sub-segments to stretch and compress, with the recombination method to reassemble all the sub-segments and construct the warping-augmented signal $X^{tw}_i$ with the same dimension as the origin EEG signal.

By recognizing different transformation techniques, the model can generate the representation that captures temporal dependencies, frequency correlation, and time-frequency correspondences within EEG signals from unlabeled samples.

\begin{table}\scriptsize
\centering
\begin{center}
  \caption{The summarization of predictive-based EEG analysis self-supervised learning method. In the "training mode" column, "PT" represents pre-training and fine-tune mode, "UT" represents unsupervised training mode, and "CT" represents joint-training mode.}
  \begin{tabularx}{\linewidth}{p{2.5cm} c  p{3cm} c p{3cm} c}
    \toprule \centering
    \textbf{Approach}&\textbf{\makecell[c]{Type of \\ pretext method}}&\textbf{\makecell[c]{Detailed \\ method}}& \textbf{Backbone} & \textbf{\makecell[c]{Downstream \\ Tasks}} & \textbf{\makecell[c]{Training \\ Mode}} \\
    \midrule

    \makecell[c]{ EEG scaling SSL \cite{re94}}&\makecell[c]{transformation predictive}&\makecell[c]{scaling prediction}& SVM &\makecell[c]{epileptic classification}&\makecell[c]{PT } \\ \hline

    \makecell[c]{ Transformation SSL \cite{re92}}&\makecell[c]{transformation predictive}&\makecell[c]{multi-transformation}& CNN &\makecell[c]{emotion recognition}&\makecell[c]{UT} \\ \hline

    \makecell[c]{ Task-agnostic SSL \cite{re95}}&\makecell[c]{transformation predictive}&\makecell[c]{multi-transformation }& CNN &\makecell[c]{seizure \& motor imagery}&\makecell[c]{UT} \\ \hline

    \makecell[c]{ Temporal EEG SSL \cite{re96}}&\makecell[c]{temporal predictive}&\makecell[c]{relative position}& -&\makecell[c]{sleep \& pathology prediction}&\makecell[c]{PT} \\ \hline

    \makecell[c]{ SSL-EED AD \cite{re98}}&\makecell[c]{transformation predictive}&\makecell[c]{noise classification}& CNN,SVM&\makecell[c]{pathology prediction}&\makecell[c]{UT} \\ \hline

    \makecell[c]{Speech-EEG SSL \cite{re90}}&\makecell[c]{temporal predictive}&\makecell[c]{temporal-shift}& CNN &\makecell[c]{speech decoding}&\makecell[c]{UT} \\ \hline

    \makecell[c]{ SSL MI-EEG \cite{re88}}&\makecell[c]{temporal predictive}&\makecell[c]{temporal shuffling}& CNN &\makecell[c]{motor imagery}&\makecell[c]{PT} \\ \hline

    \makecell[c]{ MtCLSS \cite{re97}}&\makecell[c]{transformation predictive}&\makecell[c]{multi-transformation}& CNN&\makecell[c]{pediatric sleep classification}&\makecell[c]{CT} \\ \hline

    \makecell[c]{ MM Emotion \cite{rr1}} &\makecell[c]{transformation predictive}&\makecell[c]{multi-transformation}& CNN&\makecell[c]{Emotion recognition}&\makecell[c]{PT} \\ \hline

    \makecell[c]{ SSTSC \cite{rr2}} &\makecell[c]{transformation predictive}&\makecell[c]{Relative position}& CNN&\makecell[c]{Seizure detection}&\makecell[c]{PT} \\ \hline

    \makecell[c]{Clinical EEG SSL \cite{re87}}&\makecell[c]{temporal predictive}&\makecell[c]{relative position \\ temporal shuffling}& CNN &\makecell[c]{sleep classification \\ pathology classification }&\makecell[c]{PT} \\ \hline
    
    \makecell[c]{SSL for sleep EEG \cite{re93}}&\makecell[c]{temporal predictive}&\makecell[c]{relative position \\ temporal shuffling}& CNN &\makecell[c]{sleep classification}&\makecell[c]{PT} \\ \hline

    \makecell[c]{ EEG-oriented SSL \cite{re89}}&\makecell[c]{transformation predictive \\ temporal predictive}&\makecell[c]{band-stop prediction \\ temporal-trend}& CNN &\makecell[c]{sleep \& Pathology \\motor imagery}&\makecell[c]{PT } \\ \hline

    \makecell[c]{Robust EEG SSL \cite{re91}}&\makecell[c]{transformation predictive \\ temporal predictive}&\makecell[c]{band-stop prediction \\ temporal-trend}& CNN &\makecell[c]{sleep \& Pathology}&\makecell[c]{PT } \\ \hline

    \makecell[c]{ MBrain \cite{re86}}&\makecell[c]{Spatial predictive}&\makecell[c]{channel correlation \\ replace discriminative}& CNN,LSTM &\makecell[c]{Seizure detection}&\makecell[c]{PT} \\

  \bottomrule
\end{tabularx}
\label{tab_1}
\end{center}
\vspace{-\baselineskip}
\end{table}

\subsection{Section Discussion}
This section extensively reviews the predictive-based EEG analysis methods. In this section, we categorize the predictive methods into three sub-categories: spatial predictive, temporal predictive, and transformation predictive method. The spatial predictive tasks focus on exploring the channel-correlation features. In contrast, the temporal predictive tasks involve incorporating rich temporal dependency features, time-correlation features, and consistent temporal information into the representation. The transformation predictive task can help the model to extract temporal-frequency aligned features by recognizing typical signal transformation techniques. Those pretext tasks are simple to accomplish, where the encoder $g_theta$ can be CNN or LSTM, and the pretext task decoder $g_\delta^p$ can be simple forward neural networks or the traditional machine learning classifiers. The predictive tasks only require a few parameters and complex network architectures but may need help to learn general representations for downstream tasks.

\section{Generative-based SSL EEG analysis method}\label{generative}
Different from predictive methods, generative-based SSL EEG analysis methods are more complex and challenging. The critical terms of this method are "\textit{Reconstruction}" and "\textit{Generation}," where the fine-grained correlation and features can be captured through this pretext task. "Reconstruction" means reconstructing the masked or transformed samples to learn effective representation, and "Generation" means generating specific context to train the model to learn specific knowledge. In the EEG analysis, the generative-based SSL method adopts signal reconstruction and
\begin{figure}[t]
\centering
\subfloat[The typical framework of temporal reconstruction method, where EEG signals are randomly masked and the model is required to reconstruct raw EEG signals to extract signal contextual features]{
\includegraphics[width=0.44\linewidth]{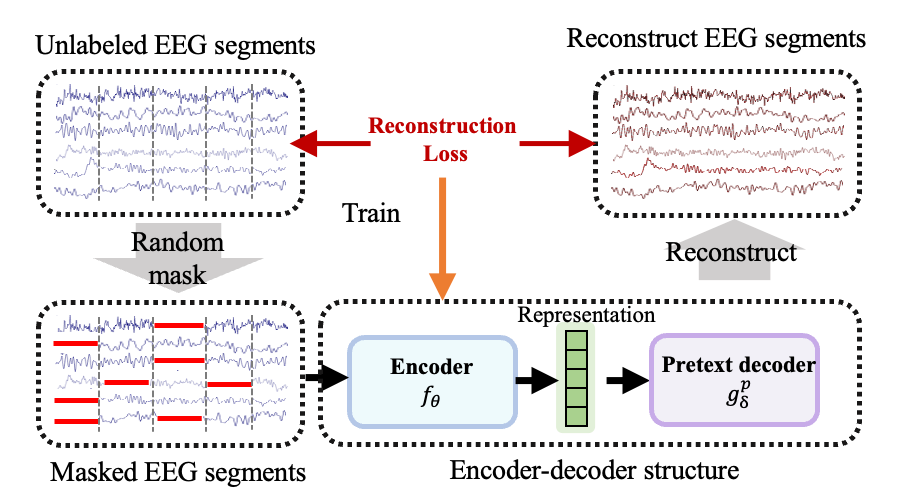}
\label{r1}
}
\hfill
\subfloat[The typical framework of multi-domain  reconstruction method, where the frequency-temporal features of EEG signals are randomly masked and the model is required to reconstruct the features to capture multi-domain correlation in EEG signals]{
\includegraphics[width=0.50\linewidth]{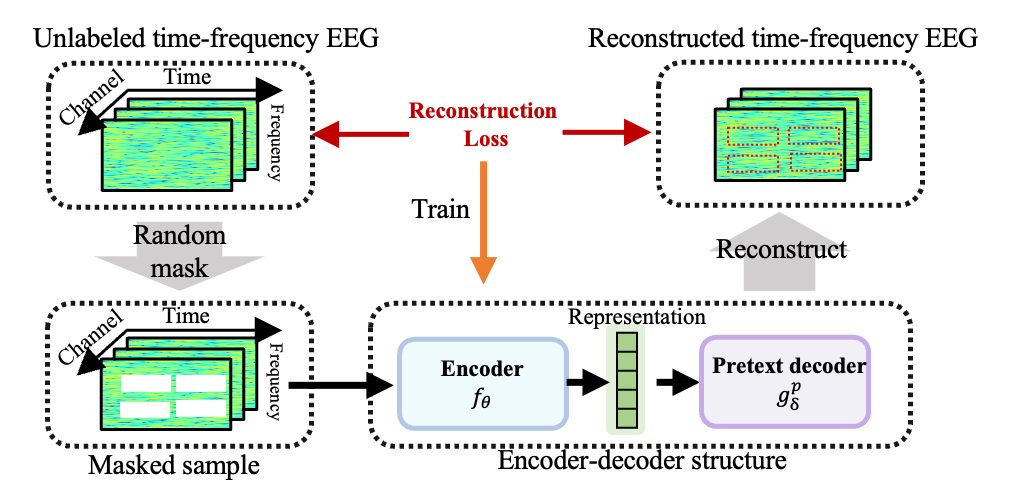}
\label{r2}
}

\centering
\caption{The frameworks of temporal reconstruction and multi-domain reconstruction SSL EEG analysis methods }
\label{reconstruction_Fig1}
\vspace{-\baselineskip}
\end{figure}
generative-adversarial task as the pretext tasks, which can be categorize into three independent sub-categories according to the task target: (1) \textit{Temporal reconstruction task}, (2) \textit{Multi-domain reconstruction task}, and (3) \textit{Generative adversarial task}. The typical frameworks of three kinds of methods are demonstrated in Figure~\ref{reconstruction_Fig1} and Figure~\ref{GAN}, and the summary of existing works is listed in Table \ref{tab_2}.

\subsection{Temporal Reconstruction Task}

The framework of the temporal reconstruction task is shown in Figure~\ref{r1}, which is inspired by the autoencoder method \cite{re77} to reconstruct the input data to capture contextual features without the need for human-labeled sample. The reconstruction task enables the encoder to learn fine-grained input correlation, which can generate representations containing rich contextual information. The EEG signal is the serialized temporal physiological data, which is applicable for conducting the temporal reconstruction pretext task \cite{re57} to learn signal contextual correlation, enhance the understanding of temporal dependencies, and provide effective representation for various EEG-based downstream tasks. Different temporal reconstruction tasks are listed as follows:

\textbf{EEG-based autoencoder} \cite{re105} is the adaptation of autoencoder for EEG analysis \cite{re106}. In this method, EEG signals are encoded into low-dimensional representation by the encoder $f_\theta$, and the low-dimensional representation is then used to reconstruct the original signal through the pretext task decoder $g_\delta^p$ symmetrical to the encoder. The encoder is responsible for preserving critical EEG signal information, while the decoder is responsible for reconstructing the EEG signal from the generated representation. The reconstruction loss can be calculated as follows:
\begin{equation}
    \mathcal{L}(X,X^*)=\frac{1}{len(X)}\Vert X-X^* \Vert_1
\end{equation}
where $len(X)$ represents the length of input signal and $\Vert \Vert_1$ represents the L1-norm. $X$ is the EEG signal and $X^*$ is the reconstructed signal. By minimizing the difference between the original and reconstructed signals, the encoder preserves critical information necessary for signal recovery, which can be considered signal compression.

\textbf{Signal-level mask-reconstruction (signal MAE)} is the typical mask-reconstruction method to capture temporal signal correlation for reconstructing the masked segments \cite{re101}. In this framework, multi-channel EEG signals are first encoded into temporal embeddings through the 1D convolution block similar to the famous wave2vec \cite{re102,re103} algorithm. The high-dimensional EEG signals $X$ are downsampled and compressed into low-dimensional feature embeddings $Z=\{z_1,z_2,z_3,...,z_k\}$ arranged in temporal order, and the stride of convolution block determines the number of input time-step to the encoder. The mask $M_i$ is generated to randomly replace parts of the information in embedding $Z$, creating the masked embedding $Z^m$ with local information dropout. The transformer encoders are then applied to extract bidirectional temporal correlation between input slices and output the signal representation $Re=\{re_1,re_2,...,re_k\}$. The convolution block and Transformer encoder are cascade as the encoder $f_\theta$ to generate signal representation, and followed by the linear and convolution layer as the pretext task decoder $g_\delta^p$ to reconstruct the raw signal $X$ through masked signal representation $Re$, the training process can be described by minimizing cosine similarity loss shown as follows:
\begin{equation}
    \mathcal{L}(X_m,X)=1-\frac{X_m \cdot X}{|X_m||X|}
\end{equation}
where $X_m$ is the reconstructed EEG signal. This method reconstruct raw signal through the masked signal representation, where the original EEG signals serve as the pseudo-labels. In this architecture, encoder $f_\theta$ is responsible for mining temporal correlation and preserving critical signal information while the pretext decoder $g_\delta^p$ is responsible for reconstructing original EEG signals. Therefore, this framework forces the encoder to generate representations containing fine-grained correlation and signal critical information, which exhibits strong expressiveness, generalization, and applicability across various EEG-based tasks.

\textbf{Embedding-level mask-reconstruction (embedding MAE)} is another mask-reconstruction method that is inspired by the BERT model\cite{re57} in the language domain to fuse the contextual relationship into the representation \cite{re104}. Similar to the MAE framework mentioned above, EEG signals are first transformed into temporal embeddings through the convolution block and then encoded by the transformer encoder to generate signal representations, followed by the pretext task decoder to accomplish the reconstruction task. However, different from the signal MAE, this task is based on the embedding-level reconstruction: the transformed EEG embedding $Z$ is randomly masked by the generated mask vector, where $z^*$ represents the randomly selected embeddings to be masked. The pretext task decoder is required to \textbf{reconstruct the masked embedding rather than EEG raw signals}. The contrastive loss function is designed to make the reconstructed embedding $z^*_{pre}$ to be as similar as possible to the original unmasked embedding $z^*$ while keeping it as dissimilar as possible to the remaining embeddings, which can be calculated as follows:
\begin{equation}
    \mathcal{L}(z^*,z^*_{pre})=- \log \frac{exp(sim(z^*_{pre},z^*)/\eta)}{\sum_{z_{r_i} \in {Z}} exp(sim(z^*_{pre},z_{r_i})/\eta)}
\end{equation}
where $z_{r_i}$ is the negative sample obtained by random sampling from the contextual embeddings, $sim$
represents cosine similarity to measure the distance between the reconstructed and original embeddings, and $\eta$ is the temperature parameter to control the contrastive loss. Compared with the MAE to reconstruct the original signals, the embedding-level reconstruction is simpler, with fewer parameters to capture critical contextual information precisely and understand EEG embedding temporal relationships. However, it may also lead to losing some original signal information. The combination of the transformation encoder can generate representations for various downstream tasks.

\subsection{Multi-domain Reconstruction Method}
The multi-domain reconstruction method extends the EEG-based MAE to multiple domains (signal, spatial, and frequency), which can be shown in Figure~\ref{r2}. Different from the temporal reconstruction methods, this method achieves collaborative and mutual reconstruction across different domains to extract spatial-temporal-frequency aligned and complementary features in the EEG signal, generating more powerful and general representations adapted to different tasks. Detailed explanations of multi-domain reconstruction methods are as follows:

\textbf{Spatial-temporal-frequency reconstruction (STF MAE)} conducts the synergistic reconstruction task in the temporal-frequency-spatial domain to extract integrated EEG features \cite{re107}. The idea of synergistic reconstruction task is inspired by the time-frequency analysis method \cite{re108}: temporal analysis method \cite{re109,re110} investigates the patterns of EEG amplitude changes over time, while frequency analysis \cite{re111} method studies the frequency energy distribution within EEG signals. The time-frequency analysis method utilizes the sliding time window to investigate temporal changes in frequency spectral features \cite{re112}. Based on the time-frequency analysis method, this task constructs a 3D matrix as the feature of the EEG signal. Through the continuous wavelet transform (CWT) \cite{re113}, EEG signals are transformed into 3D frequency-spatial-temporal matrix $X \in \Bbb{R}^{c \times t_n \times f_r}$, where $c$ is the channel number, $t_n$ is the number of temporal slices, and $f_r$ represents frequency feature resolution. This 3D matrix can be considered time-frequency features (2D image) with multiple channels. Inspired by the image MAE, the EEG feature matrix is divided into different patches and randomly masked with the mask patch $m_p$ to generate the masked matrix $X^*$, the encoder-decoder structure utilizing vision-transformer (ViT) \cite{re114} as the backbone is designed to reconstruct the EEG feature matrix. The mean squared error (MSE) can be used to train the model, which is defined as follows:
\begin{equation}
    \mathcal{L}(X^m,X^{pre})=E(X^{pre}-X^m)^2=\frac{1}{n_m} \sum_{i=1}^{n_m} (X_i^{pre}-X_i^{m})^2
\end{equation}
where $n_m$ is the dimension of the masked features $X^m$, and $X^{pre}$ represents the reconstructed features generated by the encoder and pretext task decoder: $X^{pre}=g_\delta^p(f_\theta(X^*))$. By minimizing the MSE loss, encoder $f_\theta$ fuses spatial-temporal-frequency contextual correlation into representation, and decoder $g_\delta^p$ is learned to reconstruct the original EEG feature matrix based on the representation. The generated representations contain multi-domain correlation and features, which exhibit greater expressive ability and a wider range of applications for downstream tasks.

\textbf{Frequency mask-reconstruction (frequency MAE)} conducts mask-reconstruction task in different frequency bands to capture frequency features, long-term dependencies, and critical time-frequency correlated information \cite{re132}. Initially, the EEG signal undergoes two distinct transformations: 1. The EEG signal is directly embedded into the patch sequence through division, linear projection, and flattening operation, representing the EEG temporal patch sequence. 2. The EEG signal is transformed into six independent frequency bands (0-4Hz, 4-8Hz, 8-18Hz, 16-32Hz, 32-64Hz, and other frequencies), representing the EEG frequency patch sequences. $10\%$ of the frequency patch sequences are randomly masked, followed by six independent ViT-based encoders to generate representations for all frequency bands (one encoder corresponds to one frequency band). Six independent ViT-based decoders are sequentially used to reconstruct the frequency patch sequences. Differently, the target of the pretext task is to minimize the difference between the summation of all reconstructed frequency sequences and the temporal patch sequence, which can be calculated similarly to equation (12). This task can reconstruct the temporal information by the masked frequency patch sequences, which can help the model to align temporal-frequency information in the EEG signal and understand its correlation, generating high-dimensional representations with rich time-frequency coherent features of the signals, and  providing valuable EEG features for various EEG-based tasks.

\textbf{Frequency-temporal reconstruction (FT MAE)} is the framework to reconstruct the masked EEG representations in the frequency and time domain \cite{re116}. This framework transforms EEG signals into discrete patches through the non-overlapping 1D-CNN, with some patches randomly masked with ratio $r$. The ViT-based encoder is subsequently employed to generate representations, followed by a symmetric decoder to reconstruct the masked patches. Two reconstruction methods are proposed in the framework: the first is the \textbf{spatiotemporal domain reconstruction}, where the decoder reconstructs the masked patches directly, with the MSE loss function to train the model and capture the temporal correlations in EEG signal. The second is the \textbf{Fourier domain reconstruction} to reconstruct masked patches in the frequency domain. Through the Discrete Fourier Transform (DFT) \cite{re117}, EEG signals can be transformed from the time domain to the frequency domain:
\begin{equation}
    x^f_k=\sum_{j=1}^{n} x_j * \cos (\frac{2\pi}{n}jk)-\mathbf{i}*\sin(\frac{2\pi}{n}jk) 
\end{equation}
where $k \in (0,n)$, $n$ is the number of sampling points for the EEG segments, $x_j$ is the temporal amplitude at sampling point $j$, and $\mathbf{i}$ represents the imaginary unity. $x^f_k$ represents the generated spectrum features at sampling point $k$. The first term in this equation represents the \textit{Real} part of the result, and the second term represents the \textit{imagery} part. Then, the magnitude and phase of the frequency signal can be calculated as follows:
\begin{align}
    \left\{
    \begin{aligned}
        magnitude_k=\frac{1}{n}\sqrt{Re(x^f_k)^2+Im(x^f_k)^2} \\
        phase_k=atan2(Re(x^f_k)^2,Im(x^f_k)^2)
    \end{aligned}
    \right.
\end{align}
where $Re$ and $Im$ represent the real and imagery part of the spectrum feature, and $atan2$ represents the arctangent function with two arguments. Researchers believe that the study of both magnitude and phase is important: For EMG signals, the magnitude and phase are highly correlated with muscle movement. Muscles move both longitudinally and transversely according to the direction of the fibers. As a result, the biological impedance of the motion units changes, leading to variations in amplitude and phase responses. Therefore, analyzing magnitude and phase can help the model capture muscle contraction patterns as part of the representation learning process \cite{re116}. In the EEG signal, the magnitude and phase are highly related to the phase synchronization information between neurons, which can help reveal the synchrony and information transmission between different brain regions. The Fourier domain reconstruction task predicts the magnitude and phase sequence of masked EEG patches, which are then reconstructed through the inverse Fourier transform. The mean squared error can measure the difference between the original patches and those reconstructed by magnitude and phase. Encoder $f_\theta$ can understand the correlation between spectrum features and temporal signal and capture critical neuron activity knowledge through this task.

\textbf{Spatial reconstruction (Spatial MAE)} aims to learn the spatial correlation between different channels in EEG signal \cite{re118}. In this framework, the correlation between EEG channels can be defined using a graph structure $\mathcal{G}=(\mathcal{A},\mathcal{X})$, where $\mathcal{X} \in \Bbb{R}^{c \times n}$ represents the node feature matrix in the graph (each channel corresponds to a node on the graph) and  $\mathcal{A} \in \Bbb{R}^{c \times c}$ is the adjacency matrix representing the connectivity between nodes. The graph structure can be calculated through channel spatial distance and correlation. In the framework, the sub-graph $\mathcal{G}^s$ is sampled containing $n_s$ nodes and their connectivity graph structure. For the sampled sub-graph, the feature of a random node is masked and then reconstructed by the model through adjacent node features and graph structure, which can train the model to capture the spatial correlations. The graph neural network (GNN) \cite{re119} is used as the backbone for the encoder and decoder to deal with topological graph data, and the MSE is the loss to measure the node reconstruction performance. By reconstructing the graph node, the generated representation enables a deeper exploration of spatial features and channel correlation, which is valuable for tasks that require high spatial resolution in EEG (such as visual decoding).

\textbf{Transformation reconstruction} aims to reconstruct EEG signals after different signal transformations to preserve critical signal-related information \cite{re120}. The model reconstructs the EEG signal after the following signal transformations: 1. \textit{signal jitter}, where EEG samples are added with random noise:$x_s=x+s$. 2. \textit{Random sample}, where some points in the temporal EEG signal are replaced by the average value of neighbor points. This transformation can be considered as the \textit{smoothing} operation. 3. \textit{Channel removal}, where a specific channel in the EEG signal is removed to be reconstructed. 4. \textit{Window replace}, where EEG signals in randomly selected time windows are replaced by dummy value zero. 5. \textit{Jitter in time windows}, where the signal in the randomly selected time window is corrupted by noise. The model can fuse temporal correlation, spatial correlation, and transformation features into representation for downstream tasks by reconstructing raw signals from various transformations in the pre-training process.

\subsection{Generative Adversarial Method}

The generative adversarial method encompasses two pretext tasks: the generation task to generate fake EEG samples continually, and the adversarial task strives to distinguish real and fake samples (shown in Figure~\ref{GAN}) \cite{re121,re122}. Through self-supervised the adversarial training of the generator and discriminator in the framework, the model can generate enhanced EEG samples. In the field of EEG, two kinds of generative adversarial networks (GANs) have been investigated, which are listed as follows:
\begin{figure}[t]
  \centering
  \includegraphics[width=0.6\linewidth]{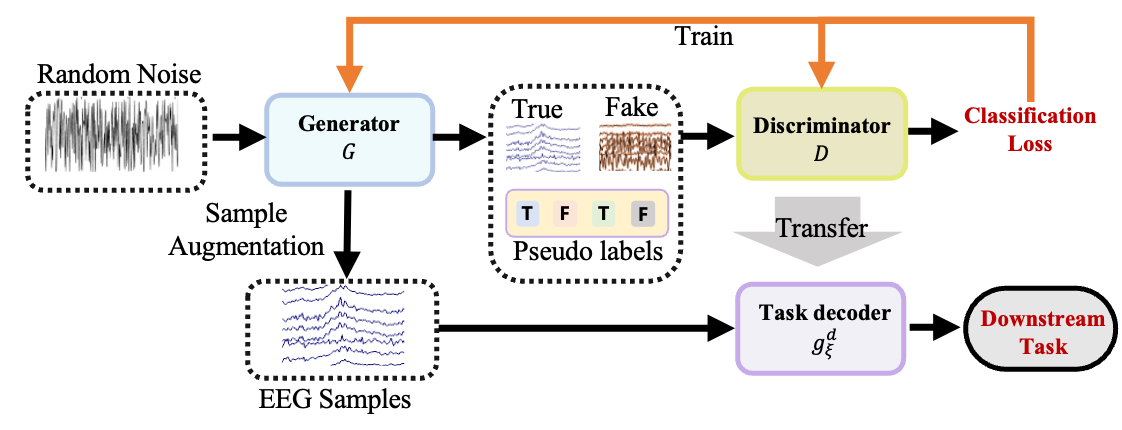}`
  \caption{The general framework of generative adversarial network (GAN) for EEG analysis }
  \label{GAN}
\vspace{-\baselineskip}
\end{figure}

\textbf{Sample generation method} aims to produce new EEG samples through the generation and adversarial pretext tasks \cite{re123}. This framework uses the generator $G$ and discriminator $D$ to accomplish the generation adversarial task. The input of $G$ are augmented EEG signals (e.g. masked signal) or random noise, and the output of $G$ are the generated fake EEG samples. The input of $D$ are the sample pairs $(x^n,x)$, where $x^n$ is the generated fake sample and $x$ is the true EEG signal. The generator aims to produce pseudo-samples highly similar to real EEG samples, while the discriminator attempts to distinguish between real and fake samples accurately. Through adversarial training, the generator can produce highly believed EEG samples for training, which can help alleviate EEG collection and labeling issues.

\textbf{Discriminator-based GAN} is another generative adversarial method to extract discriminative representations from EEG signal \cite{re124}. Discriminator-based GAN focuses on the discriminator to extract efficient features. By distinguishing real samples from fake ones, the discriminator can learn critical invariant and discriminative features of EEG signals. Through adversarial training, the discriminator can be considered as the encoder that can extract pre-trained EEG features and generate representations for downstream tasks.

\subsection{Section Discussion}

This section reviews the generative-based SSL EEG analysis methods, which conduct complex generative pretext tasks to train the encoder to capture effective signal features for downstream tasks. The existing methods are categorized into three sub-categories: 1) The temporal reconstruction task that masks part of the temporal signal and requires the model to reconstruct. 2) The multi-domain reconstruction task that masks temporal-frequency features and requires the model to reconstruct. 3) The adversarial generative task that generates pseudo sample by generator and requires the discriminator to distinguish real and fake samples. Compared with the predictive tasks, generation tasks are more challenging and need more trainable parameters and complex structures to accomplish, they can learn more efficient features in representation. Emulating the MAE, BERT, and other generative SSL methods in the vision and language field, generative SSL methods for EEG signals have achieved significant success in various downstream tasks.

\begin{table}\scriptsize
\centering
\begin{center}
  \caption{The summarization of generative-based EEG analysis self-supervised learning method. "PT" represents pre-training and fine-tune mode, "UT" represents unsupervised training mode, "CT" represents joint-training mode, and "SA" represents the sample augmentation.}
  \begin{tabularx}{\linewidth}{p{2cm} c  p{3cm} c p{3cm} c}
    \toprule \centering
    
    \textbf{Approach}&\textbf{\makecell[c]{Sub-category}}&\textbf{\makecell[c]{Detailed \\ method}}& \textbf{Backbone} & \textbf{\makecell[c]{Downstream \\ Tasks}} & \textbf{\makecell[c]{Training \\ Mode}} \\
    \midrule
    
    \makecell[c]{BENDR\cite{re104}}&\makecell[c]{Temporal reconstruction}&\makecell[c]{Embedding MAE}& CNN\&Transformer &\makecell[c]{Multiple tasks}&\makecell[c]{PT\&UT} \\ \hline

    \makecell[c]{GANSER[\cite{re125}}&\makecell[c]{Generative adversarial }&\makecell[c]{Sample generation}& CNN(U-NET) &\makecell[c]{Emotion recognition}&\makecell[c]{SA} \\ \hline

    \makecell[c]{EEG-CGS\cite{re118}}&\makecell[c]{Multi-domain reconstruction}&\makecell[c]{Spatial MAE}& GNN &\makecell[c]{Seizure analysis}&\makecell[c]{PT} \\ \hline

    \makecell[c]{Eeg2vec\cite{re127}}&\makecell[c]{Temporal reconstruction}&\makecell[c]{Embedding MAE}& CNN\&Transformer &\makecell[c]{Speech decoding}&\makecell[c]{UT} \\ \hline

    \makecell[c]{MAEEG\cite{re101}}&\makecell[c]{Temporal reconstruction}&\makecell[c]{Signal MAE}& Transformer &\makecell[c]{Sleep classification}&\makecell[c]{PT\&UT} \\ \hline

    \makecell[c]{Cognitive MAE\cite{re128}}&\makecell[c]{Temporal reconstruction}&\makecell[c]{Embedding reconstruction}& CNN\&Transformer &\makecell[c]{Cognitive-load classification}&\makecell[c]{PT\&UT} \\ \hline

    \makecell[c]{SSLAPP\cite{re129}}&\makecell[c]{Generative adversarial}&\makecell[c]{Sample augmentation}& Transformer &\makecell[c]{Sleep classification}&\makecell[c]{SA} \\ \hline
    
    \makecell[c]{MV-SSTMA\cite{re130}}&\makecell[c]{Multi-domain reconstruction}&\makecell[c]{STF MAE}& CNN\&Transformer &\makecell[c]{Emotion recognition}&\makecell[c]{PT} \\ \hline

    \makecell[c]{EpilepsyNet\cite{re105}}&\makecell[c]{Temporal reconstruction}&\makecell[c]{EEG-based autoencoder}& CNN &\makecell[c]{Epileptic classification}&\makecell[c]{JT} \\ \hline

    \makecell[c]{EEGMAE\cite{re107}}&\makecell[c]{Multi-domain reconstruction}&\makecell[c]{STF MAE}& ViT &\makecell[c]{ASD classification}&\makecell[c]{UT\&PT} \\ \hline

    \makecell[c]{brain2vec\cite{re131}}&\makecell[c]{Temporal reconstruction}&\makecell[c]{Embedding MAE}& CNN\&Transformer &\makecell[c]{Speech decoding}&\makecell[c]{PT\&UT} \\ \hline

    \makecell[c]{ Wavelet2vec \cite{re132}}&\makecell[c]{Multi-domain reconstruction}&\makecell[c]{FT MAE}& ViT &\makecell[c]{Seizure detection}&\makecell[c]{PT} \\ \hline

    \makecell[c]{CRT \cite{re133}}&\makecell[c]{Multi-domain reconstruction}&\makecell[c]{STF MAE}& Transformer &\makecell[c]{Sleep classification}&\makecell[c]{UT} \\ \hline

    \makecell[c]{Neuro2vec \cite{re116}}&\makecell[c]{Multi-domain reconstruction}&\makecell[c]{FT MAE}& Transformer &\makecell[c]{Seizure \& sleep}&\makecell[c]{PT\&UT} \\ \hline

    \makecell[c]{SDCAN\cite{re124}}&\makecell[c]{Generative adversarial}&\makecell[c]{Discriminator-based}& CNN &\makecell[c]{Stress classification}&\makecell[c]{JT} \\ \hline

    \makecell[c]{WGAN-GP\cite{re123}}&\makecell[c]{Generative adversarial}&\makecell[c]{Sample generation}& CNN &\makecell[c]{Emotion recognition}&\makecell[c]{SA} \\ \hline

    \makecell[c]{CWGAN\cite{re134}}&\makecell[c]{Generative adversarial}&\makecell[c]{Sample generation}& LSTM &\makecell[c]{Sleep classification}&\makecell[c]{SA} \\ \hline

    \makecell[c]{SAE-EEG\cite{re135}}&\makecell[c]{Temporal reconstruction}&\makecell[c]{EEG-based autoencoder}& CNN &\makecell[c]{Emotion recognition}&\makecell[c]{PT} \\ \hline

    \makecell[c]{AE-CDNN\cite{re136}}&\makecell[c]{Temporal reconstruction}&\makecell[c]{EEG-based autoencoder}& CNN &\makecell[c]{Seizure detection}&\makecell[c]{UT} \\ \hline

    \makecell[c]{MI-AE\cite{re137}}&\makecell[c]{Temporal reconstruction}&\makecell[c]{EEG-based autoencoder}& CNN &\makecell[c]{Motor imagery}&\makecell[c]{UT} \\

  \bottomrule
\end{tabularx}
\label{tab_2}
\end{center}
\vspace{-\baselineskip}
\end{table}

\section{contrastive-based SSL EEG analysis method}\label{contrastive}

Contrastive learning is the most widely used SSL technique in EEG analysis. Contrastive learning framework combined with EEG augmentation methods have been investigated to generate representation that integrates invariant features between positive pairs while eliminating irrelevant features between negative pairs. The target of contrastive learning is to encourage the model to pull positive pairs (similar samples) closer together and push negative samples apart in the representation space, which is defined as follows:
\begin{equation}
    \mathcal{L}_{con} \overset{\mathrm{def}}{=} max(d(x^+,x)-d(x^-,x)+\alpha,0)
\end{equation}
This loss function is the triplet loss \cite{re142} that trains the model to achieve $d(x^+,x)>d(x^-,x)+\alpha$, and $\alpha$ is a small positive number to avoid clustering overfitting. Different augmentation methods are applied to EEG signals to form positive and negative pairs. According to the type of augmentation methods for generating positive and negative sample pairs, we can categorize the contrastive-based SSL EEG analysis method into five sub-categories: (1) \textit{Contrastive predictive coding}, (2) \textit{transformation contrastive learning}, (3) \textit{spatial contrastive learning}, (4) \textit{composite contrastive learning}, and (5) \textit{task-oriented contrastive learning}. The typical frameworks of different kinds of methods are demonstrated in Figure~\ref{CPC} to Figure~\ref{contrastive_5}, and the summary of existing works is listed in Table \ref{tab_3}.

\subsection{Contrastive Predictive Coding}
Contrastive Predictive Coding (CPC) is a self-supervised learning technique used in NLP and CV for learning high-level representations \cite{re143,re144}. In CPC, data are divided into overlapping context windows, which are used to generate positive and negative pairs. The main idea of CPC is to generate the representation of the context window that can accurately predict the representation of future windows to extract shared invariant features. In the EEG field, two different CPC methods have been investigated:

\begin{figure}[t]
  \centering
  \includegraphics[width=0.48\linewidth]{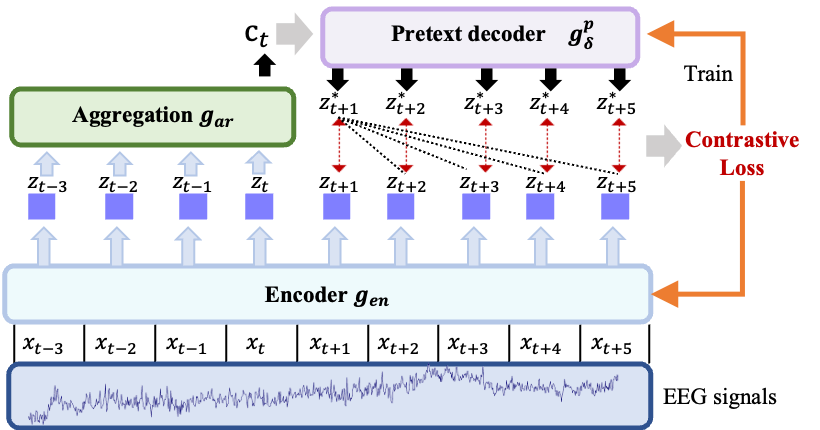}
  \caption{The framework of contrastive predictive coding  (CPC) for EEG analysis. }
  \label{CPC}
\vspace{-\baselineskip}
\end{figure}

\textbf{EEG-based CPC} extends the CPC for EEG analysis \cite{re87}, which can be shown in Figure~\ref{CPC}. This method divides EEG signals into time slices through the sliding windows. The context window $X_c$ contains $N_c$ samples is defined as $X_c=\{x_{t_i-N_c+1},...,x_{t_i}\}$, where $t_i$ is the temporal index. Similarly, the following predictive window $X_p$ is defined as $X_p=\{x_{t_i+1},...,x_{t_i+N_p}\}$, where $N_p$ is the length of the prediction window. The encoder $g_{en}$ calculates representation $z_t=g_{en}(x_t)$ for context window and prediction window, generating context and prediction representation sequence $Z_c=\{z_{t_i-N_c+1},...,z_{t_i}\}$ and $Z_p=\{z_{t_i+1},...,z_{t_i+N_p}\}$, separately. The integrated feature $c_{t_i}$ is calculated by a GRU-based regression encoder $g_{ar}$ that summarizes the information of representations within the context window. $c_{t_i}$ is used to predict future representations in the prediction window through weight $W_k, k \in [1,N_p]$, where $W_kc_t$ is the prediction for $z_{t_i+k}$ through the contextual feature $c_t$. Positive and negative pairs are then constructed: the predicted representation $W_kc_{t_i}$ forms positive pairs with the corresponding original representation $z_{t_i+k}$, while forming negative pairs with the remaining representations. The loss function is described as follows:
\begin{equation}
    \mathcal{L}_{CPC}=-\frac{1}{|\mathcal{B}|}\sum_{t_i \in \mathcal{B} }\sum_{k=1}^{N_p} log\frac{exp(s(x_{t_i+k},W_kc_{t_i}))}{exp(s(x_{t_i+k},W_kc_{t_i}))+\sum_{j \in N_e} exp(s(x_j,W_kc_{t_i}))} 
\end{equation}
where $\mathcal{B}$ is the sample batch and $|\mathcal{B}|$ is the batch size. The $N_e$ indexes the negative samples of $W_kc_{t_i}$, where $j \ne k+t_i $. By minimizing the contrastive loss, the model can extract invariant temporal features from EEG signals and integrate long-term temporal dependencies within EEG signals to form representations, which can maximize the correlation between representation and EEG raw signal to preserve critical signal information for EEG-based downstream tasks.

\textbf{EEG-based bidirectional contrastive predictive coding (BCPC)} is the extension of CPC to extract bidirectional temporal correlation in EEG signals \cite{re145}. Unlike CPC, the BCPC method adds an additional backward prediction window in the framework, representing the EEG signal prior to the context window in the time dimension. The contextual feature $c_{t_i}$ is used to predict the representation in the prediction window and the backward prediction window to construct the positive and negative pairs. By adding the backward prediction window to introduce the reverse negative and positive sample pairs for contrastive learning, the bidirectional model can capture the contextual features with temporal semantic information from both directions in the EEG signal.

\subsection{Transformation Contrastive Learning}

The transformation contrastive learning method is inspired by the typical contrastive learning framework such as SimCLR \cite{re83}and MoCo \cite{re84} in CV. EEG signals are augmented into negative and positive sample pairs through the signal transformation methods designed according to the characteristics of temporal physiological signals. The framework is shown in Figure~\ref{contrastive_2}. Multiple transformation contrastive learning methods have been studied to solve different downstream tasks, the typical frameworks are listed as follows:
\begin{figure}[t]
  \centering
  \includegraphics[width=0.6\linewidth]{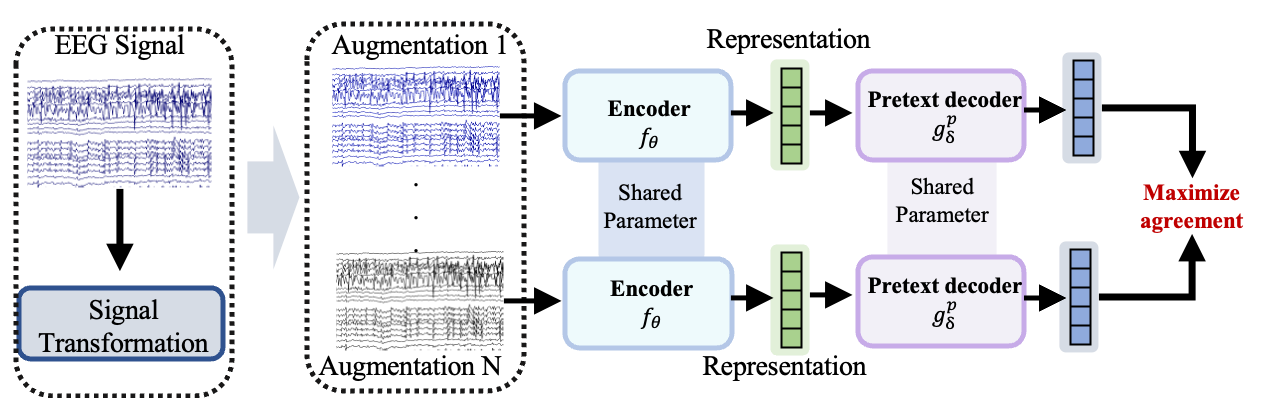}
  \caption{The framework of transformation contrastive method to capture invariant signal temporal features from unlabeled signals. }
  \label{contrastive_2}
\end{figure}

\textbf{Signal-transformation contrastive} emulates the typical framework SimCLR to conduct EEG contrastive learning \cite{re146}. For the random selected EEG sample $x_{t}$, different transformation methods are employed to generate augmentations $T_1(x_{t})$ and $T_2(x_{t})$. This method leverages the concept that augmentations applied to the same sample yield similar information, forming positive pairs, while augmentations from distinct samples exhibit significant dissimilarity, constituting negative pairs. The encoder $f_\theta$ generates representations and pretext task decoder $g_\delta$ maps the representation into loss space to calculate the contrastive loss. For batch with size $|B|$, the loss function is defined as follows:
\begin{equation}
    \mathcal{L}=-\frac{1}{|\mathcal{B}|}\sum_{t=0}^{|\mathcal{B}|} log \frac{exp(sim(z_{t_1},z_{t_2})/\tau)}{\sum_{i=1}^{2k} \mathbbm{1}_{[i \ne t ]} exp(sim(z_{t_1},z_i)/\tau) }
\end{equation}
where $z_{t_1}$ and $z_{t_2}$ is generated by $g_\delta^p(f_\theta(T_1(x_t)))$ and $g_\delta^p(f_\theta(T_2(x_t)))$ respectively, indicating the representations of augmentations from the same sample. $\mathbbm{1}_{[i \ne t ]} \in \{0,1\}$ is the indicator function, and $\tau$ is the temperature parameter. By minimizing this loss function, the model can optimize the representation space to capture discriminative representations. In the EEG analysis domain, EEG augmentation methods can be applied using signal transformation methods mentioned in Section \ref{transformation}, and other EEG signal augmentation methods are listed as follows:

(1) \textit{Cutout \& resize} divides EEG signals into different segments, and one segment is randomly discarded, representing the "cut out" operation. The remaining segments are then concatenated and resized to the length of the original sample. (2) \textit{Crop \& resize} divides EEG signals into different segments, and one segment is randomly chosen and resized to the length of the original sample. (3) \textit{Average filter}, regarded as the smoothing operation, replaces some points in the signal with the value of several neighbor points. (4) \textit{Amplitude scaling}. This method scales the temporal amplitude of the original EEG signal. The scale value should be between 0.5 and 2, suggested by prior research \cite{re146}. (5) \textit{Time shift} method shifts the EEG segments along the time dimension, representing the horizontal offset in temporal sampling. (6) \textit{Direct-current shift} method shifts the EEG segments along the voltage dimension, representing the magnitude offset in temporal sampling. The model can learn invariant EEG features and understand its latent knowledge by conducting contrastive learning on those transformation-augmented EEG signals.

\textbf{Non-negative EEG contrastive} is the contrastive framework without negative samples \cite{re147}. In traditional contrastive learning, the quantity and quality of negative samples play a crucial role in determining the effectiveness and quality of contrastive learning. In this framework, $z_i$ and $z_j$ represent the anchor and its positive samples through augmentation. To reduce the impact of negative pairs, this method proposes the \textit{world representation} $z_w$ representing the average information of EEG signal, where $z_w=E_{k\sim p(.)}[z_k]$ is generated by random representation $z_k$ and the distribution $p(\cdot)$. Based on the idea that the similarity between positive pairs should be greater than the similarity between anchor sample and global representation, the loss function is designed as follows:
\begin{equation}
    l(i,j)=s(z_i,z_w)+\epsilon-s(z_i,z_j)
\end{equation}
where $\epsilon$ is the empirical margin, and $s( \cdot , \cdot)$ is the Gaussian kernel to measure the similarity between input representations. By minimizing the loss, the model makes the similarity between the anchor sample and positive sample greater than the world representation to learn consistent EEG information between samples without human labels. Besides, some EEG analysis methods integrated with non-negative contrastive frameworks in CV like Barlow Twins \cite{re155} and BYOL \cite{re156} to conduct EEG-based non-negative contrastive learning, where all the augmented samples form the positive pairs with anchor sample and the well-designed loss function can extract invariant features from only positive pairs. Those methods are also non-negative contrastive frameworks without global representation.  

\subsection{Spatial Contrastive Learning}
\begin{figure}[t]
  \centering
  \includegraphics[width=0.6\linewidth]{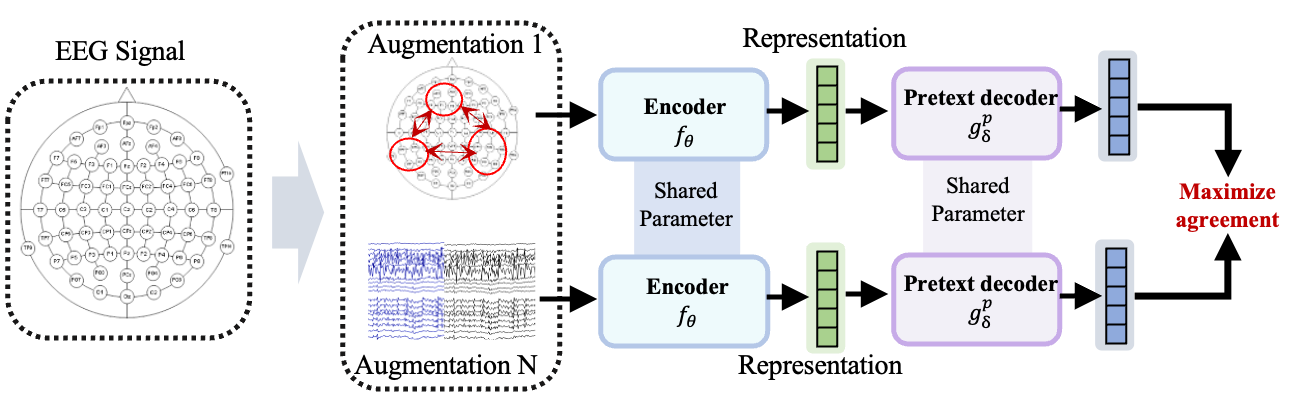}
  \caption{The framework of spatial contrastive method. In this framework, various channel-level spatial augmentation methods (e.g., channel shuffle, channel meiosis) are used to construct positive and negative sample pairs for contrastive learning, where the model can capture invariant spatial features and channel correlations from unlabeled EEG samples. }
  \label{contrastive_3}
\end{figure}

The spatial contrastive learning method shown in Figure~\ref{contrastive_3} focuses on spatial information and utilizes channel-level spatial augmentation techniques (e.g., jigsaw, meiosis) on EEG signals to construct positive and negative sample pairs, from which the model can integrate efficient spatial features and channel correlation into representation. Typical methods are listed as follows:

\textbf{Spatial shuffle contrastive} method conducts the channel-shuffling technique to construct positive and negative pairs \cite{re85}. In this method, EEG signal $x \in \Bbb{R}^{c \times t}$ is augmented through spatial shuffle: different EEG channels are categorized into different brain regions based on their spatial positions, generating $X^B=\{X^1,X^2,...,X^M\}$, where $M$ is the number of brain regions and each region in $X^B$ contains features from multiple channels. $X^B$ is randomly shuffled and reassembled into the augmented EEG sample $X^*\in\Bbb{R}^{c \times t}$. Each sample generates two shuffling augmentations to form positive sample pairs, and shuffling augmentations from different samples form negative pairs. InfoNCE loss described in equation (16) serves as the loss function for model training. The model can understand relationships between spatial channel location and signal features by contrasting the shuffling augmented EEG samples.

\textbf{Graph contrastive method} mines the relationship between channels using the graph structure \cite{re118,re148}. In this framework, EEG signals are embedded into node features in the graph, and the edges between nodes are calculated by the channel correlation or the spatial distance. Assuming $\mathcal{G}$ is the generated graph, $\mathcal{V}$ is the node-set, and $\mathcal{E}$ is the edge set, two augmentation methods are employed for contrastive learning: (1) \textit{Node dropping method}. For a sample $\mathcal{G}_t$, two augmented samples $\mathcal{G}^1_t$ and $\mathcal{G}^2_t$ are generated by randomly dropping nodes and their edges according to the dropping rate $r\%$. The augmentations from the same sample form the positive pairs, and from different samples form the negative pairs. (2) \textit{Sub-graph augmentation}. For each node $v_i$ in the sample $\mathcal{G}_t$, two positive and one negative samples are constructed: The random walk with restart algorithm is used to generate positive sub-graph $\mathcal{G}^+_{i,1}$ and $\mathcal{G}^+_{i,2}$ centered at the selected node $v_i$ with the radius parameter $ra$ to control the size, and generates negative sub-graph $\mathcal{G}^-_{i}$ centered at the farthest node from the selected node. In the positive sub-graph, the features of target nodes are masked with zero to avoid interference from target node information. For different sub-graphs, the representations are encoded by the trainable weight $W_e$ through the GNN, where $re^+_{i,1}$ and $re^+_{i,2}$ are the representations for positive samples, and $re^-_{i,1}$ is the representation of negative sample. The embedding of the selected node is also calculated by $W_e$, where $e_i=ReLU(v_iW_e)$. A trainable score matrix $W_s$ is then designed to quantify the similarity between the selected node and its sub-graphs, which is described as follows:
\begin{equation}
    S^+_{i,j}=\sigma(e_iW_s{re}^+_{i,j})
\end{equation}
where $\sigma$ represents the logistic function. The contrastive loss is designed to maximize the correlation between the embedding of nodes and positive samples, which makes the representation of channels in latent space closer to similar channels. The loss function is defined as:
\begin{equation}
    \mathcal{L}=-\frac{1}{2c|\mathcal{B}|}\sum_{j=1}^2 \sum_{i=1}^c (log(S^+_{i,j}+log(1-S^-_{i,1})))
\end{equation}
where $c$ is the number of channels and $|\mathcal{B}|$ is the batch size. By minimizing this loss, the model focuses on channel-level spatial features, which confers the model with the robust ability to comprehend high spatial resolution in EEG, leading to superior performance in downstream tasks involving multiple channels and complex channel configurations.

\textbf{EEG meiosis contrastive} method conducts meiosis augmentation technique and contrastive learning framework to integrate invariant channel features into representation \cite{re149}. The meiosis data augmentation technique is used to generate contrastive pairs: two EEG samples are randomly sampled into the group $X^g_i=\{A_i,B_i\}$, where the format of samples $A_i $ and $B_i$ are $\Bbb{R}^{c \times t}$, $c$ is the channel number and $t$ is the number of sampling points. For the group, $A_i=\{a_1,a_2,a_3,...,a_t\}$ and $B_i=\{b_1,b_2,b_3,...,b_t\}$ are engaged into meiosis with each other, which signifies data exchange between $A^p$ and $B^p$, generating the augmented sample $V^1_i=\{a_1,a_2,a_3,...,a_i,b_{i+1},b_{i+2},...,b_c\}$ and $V^2_i=\{b_1,b_2,b_3,...,b_i,a_{i+1},a_{i+2},...,a_c\}$. EEG samples $A_i$ and $B_i$ are under the same stimulus/event to increase contrasting complexity. All training samples are augmented and transformed into $V^1_i$ and $V^2_i$ for contrast, and the sample feature representation can be generated through encoder $f_\theta$ and projector (pretext task decoder) $g_\delta^p$ by $z^1_i=g^p_\delta(f_\theta(V^1_i))$. In the framework, $z^1_i$ and $z^2_i$ form the positive pair, indicating samples exchanged EEG signal with each other, while $z^1_i$ and $z^2_j$ form the negative pair ($i \ne j$). The loss function is then defined as follows:
\begin{equation}
\begin{aligned}
        L=-\frac{1}{2}(\frac{1}{|\mathcal{B}|}\sum_{i=0}^{|\mathcal{B}|}log\frac{exp(s(z^1_i,z^2_i)/\tau)}{\sum_{j=0}^{|\mathcal{B}|}\mathbbm{1}_{[j\ne i]}(s(z^1_i,z^1_j)/\tau)+\sum_{j=0}^{|B|}(s(z^1_i,z^2_j)/\tau)}\\ +\frac{1}{|\mathcal{B}|}\sum_{i=0}^{|\mathcal{B}|}log\frac{exp(s(z^1_i,z^2_i)/\tau)}{\sum_{j=0}^{|\mathcal{B}|}\mathbbm{1}_{[j\ne i]}(s(z^2_i,z^2_j)/\tau)+\sum_{j=0}^{|\mathcal{B}|}(s(z^1_j,z^2_i)/\tau)})
\end{aligned}
\end{equation}
where $s(\cdot ,\cdot)$ is the function to measure the similarity between representations, and $\mathbbm{1}_{[j \ne i]} \in \{0,1\}$ is the indicator function that equals 0 when $i = j $. The proposed contrastive loss aims to minimize the distance between mutually coupled sample pairs $(V^1_i,V^2_i)$, and maximize the distance between other sample pairs without mutual coupling:$(V^1_i,V^1_j)$, $(V^2_i,V^2_j)$, and $(V^1_i,V^2_j)$, where $i \ne j$. By minimizing the loss function, the model is trained to comprehend specific and coherent channel features and can discriminate homologous EEG channel data, which can be regarded as the model capturing the EEG channel distribution knowledge, proving highly beneficial for EEG-based tasks. 
 
\subsection{Composite Contrastive Learning}
\begin{figure}[t]
  \centering
  \includegraphics[width=0.6\linewidth]{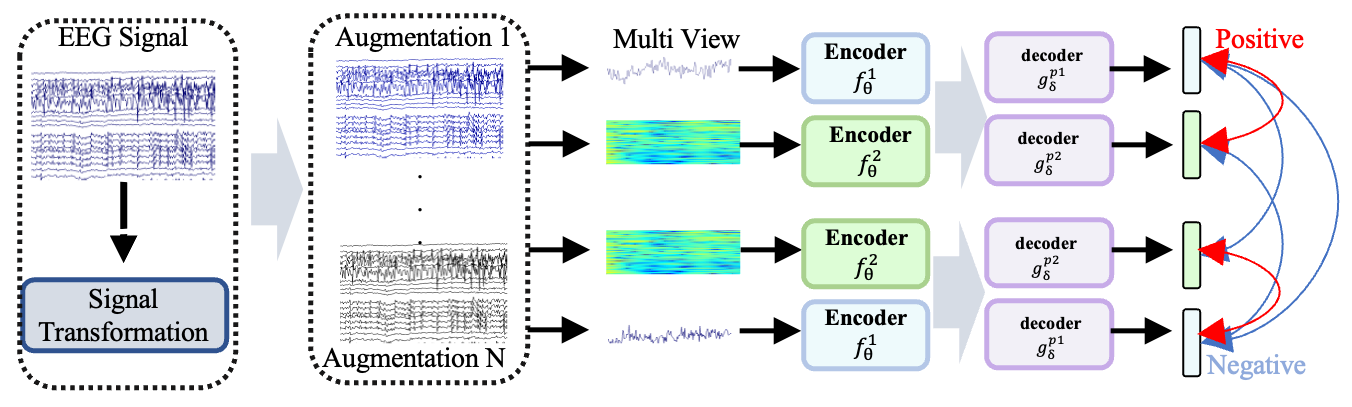}
  \caption{An example of the composite contrastive learning framework (temporal-frequency contrastive method).}
  \label{contrastive_4}
\vspace{-\baselineskip}
\end{figure}

Composite contrastive learning is the complex framework that augments EEG signals in multiple views or domains and conducts cross-view and cross-domain and contrastive learning to extract more expressive and complex representations integrating specific signal knowledge. Figure~\ref{contrastive_4} shows an example of the typical framework, and existing composite EEG contrastive learning frameworks are listed as follows:

\textbf{Frequency-temporal contrastive} method conducts contrastive learning on temporal and frequency domain. Two different frequency-temporal contrastive strategies have been investigated: 

(1) \textit{Complementary strategy} conducts cross-view contrastive learning to avoid the ignorance of complementary information in different views \cite{re151}. EEG signal $x_i$ is augmented into $x_{i,1}$ and $x_{i,2}$ through signal transformations, which are then mapped into the temporal and spectral domain independently, generating temporal components $x^t_{i,1}$, $x^t_{i,1}$ and spectrum components $x^s_{i,1}$, $x^s_{i,2}$. Different augmentations are subsequently processed through the temporal encoder $f_\theta^t$ and spectrum encoder $f_\theta^s$ to construct the representations $z^s_{i,1}$ and $z^t_{i,1}$ from augmentation $x_{i,1}$, and $z^s_{i,2}$ and $z^t_{i,2}$ from augmentation $x_{i,2}$. Four losses are combined to train the model: 1. temporal contrastive loss, denoted as
$\mathcal{L}_{tt}$. The temporal representations generated from same augmentation form positive pairs $\{z^t_{i,1},z^t_{i,2}\}$, and from different augmentation form negative pairs $\{z^t_{i,1},z^t_{i,2}\}, i \ne j$, with the infoNCE serves as the loss function. 2. Spectrum contrastive loss, denoted as $\mathcal{L}_{ss}$ calculated by spectral augmented representations similar to temporal contrastive loss. 3. Mixing contrastive loss, denoted as $\mathcal{L}_{gg}$. The spectrum and temporal augmented representations are concatenated to form the mixing augmented representation, where $z^g_{i,1}=cat(z^t_{i,1},z^s_{i,1})$, $cat$ represents the concatenation operation. This loss can be calculated similarly to the first two losses. 4. Complementary loss, denoted as $\mathcal{L}_d$. The above losses may narrow the distance between representations, losing complementary features in each view. Therefore, the complementary loss is designed to pull corresponding augmented samples in the same view closer while pushing away the corresponding augmented samples in different views. Assuming $z_i=\{z^t_{i,1},z^t_{i,2},z^s_{i,1},z^t_{i,2}\}$, complementary loss is defined as:
 \begin{align}
    \left\{
    \begin{aligned}
        &l_d(z_i,j,k)=-log\frac{exp(s(z_i[j],z_i[k])/\tau)}{\sum_{q=1}^4 \mathbbm{1}_{[q \ne j]}exp(s(z_i[j],z_i[q])/\tau)} \\
        &\mathcal{L}_D =\frac{1}{4|\mathcal{B}|}\sum_{i=0}^{|\mathcal{B}|} l_d(z_i,1,2)+l_d(z_i,2,1)+l_d(z_i,3,4)+l_d(z_i,4,3)
    \end{aligned}
    \right.
\end{align}
where $s()$ is the similarity function. Multiple loss functions are combined to train the model to extract multi-domain features and preserve the domain-specific and complementary features, which can be described as follows:
\begin{equation}
\mathcal{L}_{con}=\lambda_1(\mathcal{L}_{tt}+\mathcal{L}_{ss}+\mathcal{L}_{gg})+\lambda_2\mathcal{L}_D
\end{equation}
where $\lambda_1$ and $\lambda_2$ are the hyperparameters to balance different contrastive losses.

(2) \textit{Consistent strategy} focuses on extracting consistent information between temporal and frequency representations through contrastive learning \cite{re152}. Different from the complementary strategy, this strategy aims to maximize the mutual information between temporal and frequency representation to align different representations in a latent feature space to extract multi-domain coherent features. The consistent loss function is described as follows:
\begin{equation}
    \mathcal{L}_c=\frac{1}{|\mathcal{B}|}\sum_{i=1}^{|\mathcal{B}|} \sum_{Sim^*} (Sim^{t_1,f_1}-Sim^*+\delta), Sim^* \in \{Sim^{t_1,f_2}, Sim^{t_2,f_1}, Sim^{t_2,f_2}\}
\end{equation}
where $\delta$ is the hyperparameter. In this loss function, the $Sim^{t_1,f_1}_i=d(z^t_{i,1},z^f_{i,1})$ is defined to measure the representation similarity, where $z^t_{i,1}$ and $z^f_{i,1}$ are the temporal and frequency representations generated by sample $x_i$, and $z^t_{i,2}$ and $z^f_{i,2}$ are generated by augmented sample $x^*_i$. By minimizing this loss, the frequency and temporal representations can be pulled closer for a sample in the latent space to mine for multi-domain consistent features.

\textbf{Multi-view CPC} extends CPC from the single view to multiple views for exploring complex EEG features \cite{re153}. In this method, the weak and strong augmentation methods are designed to construct two views: jitter-and-scale strategy is used to construct weak augmentations $x^1_i$, while permutation-and-jitter generates complex strong augmentation $x^2_i$. According to the definition of CPC, the integrated features $c^1_i$ and $c^2_i$ of context windows from two views are generated. Cross-view prediction strategy is implemented, where $c^1_i$ is used to predict future windows $z^2_{i+1}$ and $c^2_i$ is used to predict future windows $z^1_{i+1}$, generating CPC losses $\mathcal{L}_{1,2}$ and $\mathcal{L}_{2,1}$. Besides, the cross-view contextual contrastive strategy is designed to extract discriminative features: $c_i^1$ and $c_i^2$ generated from the same sample but in different views form the positive pairs, while other representations form negative pairs. For the samples in batch $\mathcal{B}$, one given sample can construct 1 positive pair and $2|\mathcal{B}|-2$ negative pairs. The loss function is defined as follows:
\begin{equation}
    \mathcal{L}_{CC}=-\sum_{i=1}^{|\mathcal{B}|}\frac{exp(s(c_i^1,c_i^2)/\tau)}{\sum_{j=1}^{|\mathcal{B}|}\sum_{q=1}^2 \mathbbm{1}_{[j \ne i]}exp(s(c_i^1,c_j^q)/\tau)}
\end{equation} 
Different loss functions are combined as $\mathcal{L}=\lambda_1(\mathcal{L}_{1,2}+\mathcal{L}_{2,1})+\lambda_2\mathcal{L}_{CC}$ through the weight $\lambda_1$ and $\lambda_2$ to balance different losses. The multi-view CPC method can mine complex temporal features and understand the aligned representation.

\textbf{Multi-level contrastive} method conducts contrastive learning at multiple levels to capture complex signal features \cite{re154}. In this framework, EEG sample is divided into $n_l$ segments: $X_i=\{x_1,x_2,x_3,...,x_{(n_l)}\}$, with CNN encoder $f_\theta^1$ to generate local representation $Z_i=\{z_1,z_2,z_3,...,z_{n_l}\}$ and transformer encoder $f_\theta^2$ to generate contextual representation $R_i=\{r_1,r_{1+k},r_{1+2k}...\}$, where contextual representation $r_j$ is generated by the integration of $k$ neighbor local representations. The positive and negative sample pairs are constructed according to the filter with EEG rhythm rules. InfoNCE loss is used at local and contextual levels to extract multi-granular features. The fusion of multi-level contrastive learning can integrate different aspects of temporal features of EEG signals into representation, making representation more efficient and expressive.

\textbf{Scalp-dipole neural contrastive} is a knowledge-based cross-view contrastive method to generate general neural representation \cite{re150}. In this framework, two views are constructed according to the neural source \cite{re4} of EEG signal $x_i$: the scalp view $sc_i$ is constructed by spatial matrix, indicating the distribution of EEG voltage across the scalp; the dipole view $dp_i$ is constructed by undirected graph, indicating the inner correlation of dipoles (activated pyramidal cells) that produce the EEG signals. EEG signals are augmented through mask and jigsaw also transformed into two views. The CNN encoder $f_\theta^c$ and graph convolutional encoder $f_\theta^g$ are designed to generate scalp and dipole representation from different views, and two contrastive
\begin{figure}[t]
  \centering
  \includegraphics[width=0.5\linewidth]{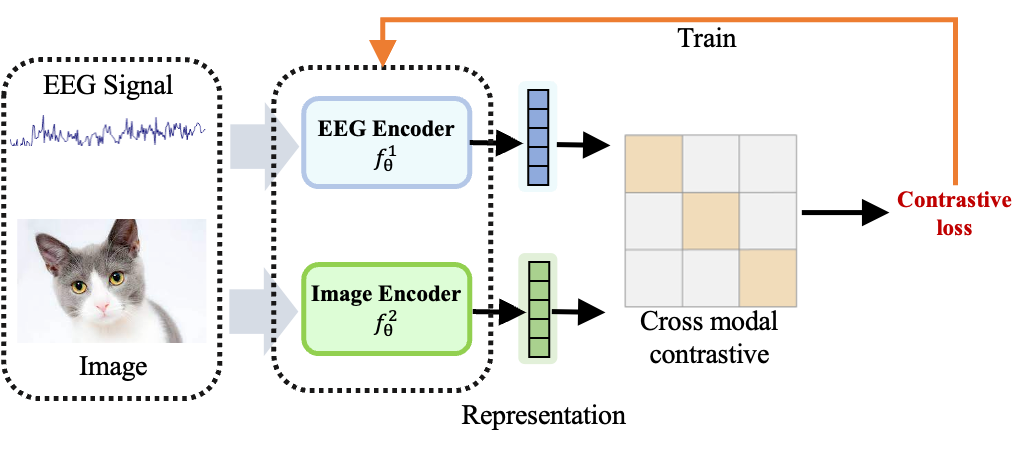}
  \caption{An example of task-oriented EEG contrastive learning: where the contrastive sample pairs are constructed from EEG signal and image data to align the features of EEG signal with image feature.}
  \label{contrastive_5}
\vspace{-\baselineskip}
\end{figure}
strategies are proposed: (1) \textit{Inner-view contrastive} aims to extract invariant features in each view. Augmented samples in a specific view are considered as positive pairs and the Barlow twins loss is implemented to minimize their distance in the representation space and capture invariant information between augmented samples; (2) \textit{Cross-view contrastive} is based on the theory that EEG representations for different views are homogenous and contain similar neural information. Therefore, augmentations in two views generated by the same sample construct positive pairs and augmentations in two views construct negative pairs, with InfoNCE loss to train the model. Combining the inner-view and cross-view losses can extract both view-specific features and latent neural information, generating general representations effective for different EEG-based tasks.

\subsection{Task-oriented EEG Contrastive Learning}
Task-oriented contrastive learning is an idiosyncratic framework set up to solve specific tasks. Some specific contrastive frameworks are designed for particular tasks: (1) \textbf{Image-EEG contrastive}\cite{re157}, where the image and the corresponding EEG signals elicited by viewing this image can form positive pairs, and the image with other EEG signals can form negative pairs. The process of image-EEG contrastive learning can be shown in Figure~\ref{contrastive_5}. The mutual information between the image and its corresponding EEG signal is maximized by image-EEG contrastive, which can solve the task of \textit{EEG image decoding}. (2) \textbf{Speech-EEG contrastive learning} is inspired by the contrastive language-image pre-training (CLIP)\cite{re158} method that forms the EEG-speech sample pairs to extract correlation and solve the \textit{EEG speech decoding} task \cite{re159}. (3) \textbf{Cross-subject contrastive learning} aims to address the individual variability issue in EEG signals. For example, \textit{age contrastive learning} selects anchor samples at different age groups, and samples with a slight difference in age from the anchor sample are used as positive samples to construct positive pairs, while those with a large difference in age from them are used to form positive pairs. By contrasting negative and positive pairs, the model can capture the age-related brain features to improve the generalizability of generated representations \cite{re160}. 
\begin{table}\scriptsize
\centering
\begin{center}
  \caption{The summarization of contrastive-based EEG analysis SSL method. "CPC" represents contrastive predictive coding, "CL" represents contrastive learning, "Tfm" represents the transformer model, "PT" represents pre-training and fine-tune mode, "UT" represents unsupervised training mode, and "CT" represents joint-training mode.}
  \begin{tabularx}{\linewidth}{p{2cm} c  p{3cm} c p{3cm} c}
    \toprule \centering
    
    \textbf{Approach}&\textbf{\makecell[c]{Sub-category}}&\textbf{\makecell[c]{Detailed \\ method}}& \textbf{Backbone} & \textbf{\makecell[c]{Downstream \\ Tasks}} & \textbf{\makecell[c]{Training \\ Mode}} \\
    \midrule

    \makecell[c]{Clinical EEG SSL \cite{re87}}&\makecell[c]{Contrastive predictive coding}&\makecell[c]{EEG-based CPC}& CNN &\makecell[c]{Sleep \& pathology classification}&\makecell[c]{PT} \\ \hline

    \makecell[c]{SSL for EEG \cite{re93}}&\makecell[c]{Contrastive predictive coding}&\makecell[c]{EEG-based CPC}& CNN &\makecell[c]{Sleep classification}&\makecell[c]{PT} \\ \hline

    \makecell[c]{ContrastWR\cite{re147}}&\makecell[c]{Transformation-based contrastive}&\makecell[c]{Non-negative CL}& CNN &\makecell[c]{Sleep \& pathology classification}&\makecell[c]{UT} \\ \hline

    \makecell[c]{SSCL for EEG\cite{re162}}&\makecell[c]{Transformation-based contrastive}&\makecell[c]{Signal transformation CL}& CNN &\makecell[c]{Sleep classification}&\makecell[c]{PT\&UT} \\ \hline

    \makecell[c]{EEG-CGS\cite{re118}}&\makecell[c]{Spatial contrastive}&\makecell[c]{Spatial shuffle CL}& GNN &\makecell[c]{Seizure analysis}&\makecell[c]{PT} \\ \hline

    \makecell[c]{GMSS \cite{re85}}&\makecell[c]{Spatial contrastive}&\makecell[c]{Graph-based CL}& GNN &\makecell[c]{Emotion recognition}&\makecell[c]{PT\&UT\&JT} \\ \hline

    \makecell[c]{SleepDPC \cite{re163}}&\makecell[c]{Contrastive predictive coding}&\makecell[c]{EEG-based CPC}& CNN\&LSTM &\makecell[c]{Sleep classification}&\makecell[c]{UT} \\ \hline

    \makecell[c]{Seq-SimCLR \cite{re146}}&\makecell[c]{Transformation-based contrastive}&\makecell[c]{Signal transformation CL}& CNN\&GRU &\makecell[c]{Multiple tasks}&\makecell[c]{PT\&UT} \\ \hline

    \makecell[c]{Domain-guide CL\cite{re160}}&\makecell[c]{Task-oriented contrastive}&\makecell[c]{Cross-subject CL}& CNN &\makecell[c]{Multiple tasks}&\makecell[c]{PT} \\ \hline

    \makecell[c]{Multivariate CL\cite{re164}}&\makecell[c]{Spatial contrastive}&\makecell[c]{Graph-based CL}& GNN &\makecell[c]{Sleep classification}&\makecell[c]{PT} \\ \hline

    \makecell[c]{DS-AGC\cite{re148}}&\makecell[c]{Spatial contrastive}&\makecell[c]{Graph-based CL}& GNN &\makecell[c]{Emotion recognition}&\makecell[c]{PT} \\ \hline

    \makecell[c]{ME-MHAC\cite{re149}}&\makecell[c]{Spatial contrastive}&\makecell[c]{Meiosis-based CL}& CNN &\makecell[c]{Emotion recognition}&\makecell[c]{PT} \\ \hline

    \makecell[c]{MBrain\cite{re86}}&\makecell[c]{Contrastive predictive coding}&\makecell[c]{EEG-based CPC}& CNN\&LSTM &\makecell[c]{Seizure detection}&\makecell[c]{PT} \\ \hline

    \makecell[c]{BrainNet\cite{re145}}&\makecell[c]{Contrastive predictive coding}&\makecell[c]{Bidirectional CPC}& GNN &\makecell[c]{Seizure detection}&\makecell[c]{JT} \\ \hline

    \makecell[c]{SleepECL\cite{re154}}&\makecell[c]{Composite contrastive}&\makecell[c]{Multi-level CL}& Transformer &\makecell[c]{Sleep classification}&\makecell[c]{UT} \\ \hline

    \makecell[c]{TS-TCC\cite{re153}}&\makecell[c]{Composite contrastive}&\makecell[c]{Multi-view CPC}& Transformer &\makecell[c]{Sleep \& seizure detection}&\makecell[c]{PT\&UT} \\ \hline

    \makecell[c]{CoSleep\cite{re165}}&\makecell[c]{Contrastive predictive coding}&\makecell[c]{EEG-based CPC}& CNN &\makecell[c]{Sleep classification}&\makecell[c]{UT} \\ \hline

    \makecell[c]{SLAM-EEG\cite{re166}}&\makecell[c]{Transformation-based contrastive}&\makecell[c]{Transformation-based CL}& ViT &\makecell[c]{Seizure detection}&\makecell[c]{PT} \\ \hline

    \makecell[c]{SPP-EEGNET\cite{re167}}&\makecell[c]{Transformation-based contrastive}&\makecell[c]{Transformation-based CL}& CNN &\makecell[c]{Multiple tasks}&\makecell[c]{PT} \\ \hline

    \makecell[c]{DSSNet\cite{re168}}&\makecell[c]{Contrastive predictive coding}&\makecell[c]{EEG-based CPC}& CNN\&RNN &\makecell[c]{Sleep classification}&\makecell[c]{UT} \\ \hline

    \makecell[c]{TF-C\cite{re152}}&\makecell[c]{Composite contrastive}&\makecell[c]{Frequency-temporal CL}& CNN\&Tfm &\makecell[c]{Sleep classification}&\makecell[c]{UT} \\ \hline

    \makecell[c]{TS-MoCo\cite{re169}}&\makecell[c]{Transformation-based contrastive}&\makecell[c]{Transformation-based CL}& Transformer &\makecell[c]{Emotion recognition}&\makecell[c]{PT\&UT} \\ \hline

    \makecell[c]{MV-EEG\cite{re170}}&\makecell[c]{Composite contrastive}&\makecell[c]{Frequency-temporal CL}& Transformer &\makecell[c]{Pathology detection}&\makecell[c]{UT} \\ \hline

    \makecell[c]{PSN-Sleep\cite{re171}}&\makecell[c]{Transformation-based contrastive}&\makecell[c]{Non-negative CL}& CNN &\makecell[c]{Sleep classification}&\makecell[c]{UT} \\ \hline

    \makecell[c]{MulEEG\cite{re151}}&\makecell[c]{Composite contrastive}&\makecell[c]{Frequency-temporal CL}& CNN &\makecell[c]{Sleep classification}&\makecell[c]{UT} \\ \hline

    \makecell[c]{MI-SSLEEG\cite{re172}}&\makecell[c]{Transformation-based contrastive}&\makecell[c]{Transformation-based CL}& CNN &\makecell[c]{Motor imagery}&\makecell[c]{JT} \\ \hline

    \makecell[c]{SA-EEG\cite{re173}}&\makecell[c]{Transformation-based contrastive}&\makecell[c]{Transformation-based CL}& CNN &\makecell[c]{Motor imagery}&\makecell[c]{UT} \\ \hline

    \makecell[c]{MtCLSS\cite{re92}}&\makecell[c]{Transformation-based contrastive}&\makecell[c]{Transformation-based CL}& CNN &\makecell[c]{Sleep classification}&\makecell[c]{UT} \\ \hline

    \makecell[c]{Multi-channel CL\cite{re174}}&\makecell[c]{Transformation-based contrastive}&\makecell[c]{Transformation-based CL}& CNN &\makecell[c]{Sleep \& pathology classification}&\makecell[c]{UT} \\ \hline

    \makecell[c]{SGMC \cite{re175}}&\makecell[c]{Spatial contrastive}&\makecell[c]{Meiosis-based CL}& CNN &\makecell[c]{Emotion recognition}&\makecell[c]{PT} \\ \hline

    \makecell[c]{CLISA \cite{re161}}&\makecell[c]{Task-oriented contrastive}&\makecell[c]{Cross-subject CL}& CNN &\makecell[c]{Emotion recognition}&\makecell[c]{UT} \\ \hline

    \makecell[c]{KDC\cite{re150}}&\makecell[c]{Composite contrastive}&\makecell[c]{Scalp-dipole neural CL}& CNN\&GNN &\makecell[c]{Multiple tasks}&\makecell[c]{PT\&UT} \\ \hline

    \makecell[c]{NICE-EEG\cite{re157}}&\makecell[c]{Task-oriented contrastive}&\makecell[c]{Image-signal CL}& ViT\&GNN &\makecell[c]{Image-decoding}&\makecell[c]{UT} \\ \hline

    \makecell[c]{AAD\cite{re159}}&\makecell[c]{Task-oriented contrastive}&\makecell[c]{Speech-signal CL}& CNN\&LSTM &\makecell[c]{Speech-decoding}&\makecell[c]{UT} \\ 
  \bottomrule
\end{tabularx}
\label{tab_3}
\end{center}
\vspace{-\baselineskip}
\end{table}

\subsection{Section Discussion}
In this section, various contrastive-based EEG analysis methods are comprehensively reviewed. The contrastive-based frameworks are categorized into five sub-categories: 1. contrastive predictive coding method that integrates the prediction and contrastive tasks to capture temporal information. 2. Transformation contrastive learning to extract signal-related invariant features. 3. Spatial contrastive method to capture spatial channel correlation. 4. Composite contrastive method that conducts multi-view contrastive learning to extract spatial-temporal-spectral features. 5. Task-oriented contrastive method that constructs specialized framework towards specific tasks. Compared to other SSL for EEG analysis, contrastive-based tasks are the most effective, with fewer parameters and simpler tasks to generate representations with higher generation and information density. Contrastive methods rely on the augmentation techniques, where the well-designed sample pairs can help the model integrate critical neural knowledge and arbitrarily chosen sample pairs may yield counterproductive results.

\section{hybrid SSL EEG analysis method}\label{hybrid}

The hybrid SSL EEG analysis method combines various pretext tasks to jointly train the model to learn complex knowledge or information. The idea of multi-task learning\cite{re176,re177} has been applied in hybrid SSL methods: the common encoder $f_\theta$ is used to extract features and generate representation from EEG signal, with different pretext task decoders $\{g_\delta^{p_1}, g_\delta^{p_2},...\}$ are used to solve multiple pretext tasks. The losses from different tasks are fused to train the model, where the shared encoder can fully leverage the advantages of different tasks to obtain representation that encompasses more knowledge and exhibits stronger expressive capabilities. The combination of multi-task losses with weight $\lambda$ can be described as follows:
\begin{equation}
    \mathcal{L}_{mt}=\sum_{i=1}^{t_n} \lambda_i\mathcal{L}_i
\end{equation}

Tabel~\ref{tab_4} shows the existing hybrid EEG SSL methods. In the existing studies, different combinations of pretext tasks are used to generate representations: many methods combine the predictive and contrastive tasks \cite{re87,re93}, where the decoders predict the transformations and conduct negative and positive pairs for contrastive learning to capture critical discriminative information and invariant features; Another method combines the generative and contrastive tasks\cite{re118} to explore the local correlations and global coherent features of EEG signal. Although the hybrid SSL method can capture complex features through multiple tasks, the gradient interference caused by training various tasks may influence the effectiveness of the generated representation. Therefore, this method requires careful selection of correlated pretext tasks to avoid interference between tasks.

\begin{table}\scriptsize
\centering
\begin{center}
  \caption{The summarization of hybrid-based EEG analysis self-supervised learning method.}
  \begin{tabular}{c c c c c}
    \toprule \centering
    
    \textbf{Approach}&\textbf{\makecell[c]{Pretext-category}}& \textbf{Backbone} & \textbf{\makecell[c]{Downstream \\ Tasks}} & \textbf{\makecell[c]{Training \\ Mode}} \\
    \midrule

    Clinical EEG SSL\cite{re87} & Predictive task/contrastive task & CNN & Sleep and pathology classification&PT\\ 

    SSL for sleep EEG \cite{re93} & Predictive task/contrastive task & CNN & Sleep and pathology classification&PT \\ 

    EEG-CGS \cite{re118} & Generative task/contrastive task & CNN & Seizure analysis&PT\\ 

    GMSS\cite{re85} & Predictive task/contrastive task & GNN & Emotion recognition & PT\&UT \\

    MBrain\cite{re86} & Predictive task/contrastive task & CNN\&LSTM &Seizure detection &PT \\

    MtCLSS\cite{re97}& Predictive task/contrastive task& CNN & Sleep classification & UT\\
    
  \bottomrule
\end{tabular}
\label{tab_4}
\end{center}
\vspace{-\baselineskip}
\end{table}
\section{Practical downstream tasks}

SSL EEG analysis methods have been applied to various EEG-based tasks. Table \ref{tab_5} demonstrates the EEG-based downstream tasks and related datasets, the practical downstream tasks are listed as follows:

\textbf{Emotion recognition} is the task aims to decode emotional states from EEG signals collected by non-invasive electrodes. The traditional emotion recognition method combines machine learning with hand-crafted features to predict discrete emotions from EEG, while the recent emotion recognition method conducts end-to-end deep models to capture continuous emotion scores \cite{re60}. The labels of the training samples are derived from the subjective rating scales or the type of stimuli that elicited the signals, which may introduce significant bias into the model training process. By combining SSL with emotion recognition task, the issue of label shift can be mitigated and the representation can improve task performance in the low-label scenarios. 

\textbf{Motor imagery} is the task to decode the mental simulation without physically performing the movement\cite{rp6}. This task involves mentally rehearsing or imaging a specific motor action, such as imaging moving the left limb, right limb or executing a complex physical activity. The decoded imagined patterns can be applied as the control signal in the brain-computer interface (BCI). For example, controlling the exoskeleton for the disabled \cite{rn1}. EEG-based motor imagery recognition methods have been widely investigated. The challenges in motor imagery lie in difficult labeling and significant subject variability, which can be effectively addressed by combining different pretext tasks in SSL.

\textbf{Pathology detection} is the most crucial clinical tasks for EEG-based applications. This task aims to recognize the mental or neural diseases that occur in the brain from EEG signals. Deep models are used to detect seizure, autism spectrum disorder, and other disorders from EEG signals \cite{re107}. However, the clinical applications of EEG signals demand high-density training data and expert knowledge to label the samples, which introduces substantial challenges in data collection. The SSL framework can reduce the number of labeled EEG samples and combine medical knowledge pretext tasks, which holds significance for the development and improvement of EEG-based clinical detection.
\begin{table}\scriptsize
\centering
\begin{center}
  \caption{The summarization of datasets that have been used in SSL EEG analysis, where the symbol '-' represents the missing information for the dataset}
  \begin{tabularx}{\linewidth}{c c c c c c c}
    \toprule \centering
    
    \textbf{Dataset}&\textbf{Subject number}&\textbf{Sampling rates}& \textbf{EEG channels} & \textbf{Task} & \textbf{Label} & \textbf{Auxiliary data} \\
    \midrule

    Physionet Challenge 2018 \cite{rp1,rp11} & 1983 & 200 Hz & 6 &Sleep classification & Weak,N1,N2,N3,RAM  & EMG,EOG etc. \\

    TUH abnormal \cite{rp2} & 2329 & 250,256,512 Hz & 27 to 36 & Abnormal detection & Normal,Abnormal  & - \\
    
    Sleep EDFx \cite{rp10,rp3} & 83 & 100 Hz & 2 & Sleep classification & Weak,N1,N2,N3,RAM & Breathe,ERP \\

    MASS \cite{rp4} & 62 & 256Hz & 20 &Sleep classification&Weak,N1,N2,N3,RAM &EOG,EMG,ECG\\

    MMI \cite{rp5} & 105 & 160Hz & 64 & Motor imagery & Rest, MI(left), MI(right) & - \\

    BCIC \cite{rp6}& 9 & 250Hz & 22 & Motor imagery & MI(l),MI(r),MI(f),MI(t) & EOG \\

    Mayo-UPenn Seizure Dataset \cite{rp7}& 4 & 400Hz & 16 & Seizure detection & Normal,Abnormal & Dog signal \\
    
    SHHS dataset\cite{rp8} & - & 125Hz & 14 & Sleep classification & Weak,N1,N2,N3,RAM & EOG,Heart \\

    MGH Sleep \cite{rp9} & - & 200Hz & 6 & Sleep classification & Weak,N1,N2,N3,RAM & - \\

    Sleep EDF\cite{rp10,rp3} & 20 & 100Hz & 2 & Sleep classification &Weak,N1,N2,N3,RAM& EOG,EMG,ERP \\

    Dreem Open Dataset\cite{rp11}& 80 & 250Hz & 8,12 &  Sleep classification &Weak,N1,N2,N3,RAM& - \\

    DEAP \cite{rp12} & 32 & 512Hz & 32 & Emotion recognition & Arousal,Valance,Dominant & Video,EOG,EMG \\

    TUSZ \cite{rp13} & over 300 & - & 19 & Seizure detection & Different seizure types & - \\

    SEED \cite{re44,re45} & 15 & 200Hz & 62 & Emotion recognition & Negative,Neutral,Positive & - \\

    SEED-IV \cite{rp15} & 15 & 200Hz & 62 & Emotion recognition & Happy,Neutral,Sad,Fear & - \\

    MPED \cite{rp16} & 23 & 1000Hz & 62 & Emotion recognition & Different discrete emotions& ECG,ESR,RSP \\

    KU-MI \cite{rp17} & 52 & 1000Hz & 62 & Motor imagery & MI(left),MI(right) & EMG \\

    ISRUC \cite{rp18} & 100,8,10 & 200Hz & 6 & Sleep classification &Weak,N1,N2,N3,RAM & Multiple signals \\

    parrKULee \cite{rp19} & 85 & 8192Hz & 64 & Speech decoding & speech signal & - \\

    CHB-MIT \cite{rp20} & 24 & 256Hz & 24-26 & Seizure detection & Seizure,Non-seizure & - \\

    MPI-LEMON \cite{rp21} & 216 & 2500Hz & 62 & Non & Resting states & MRI,ECG etc. \\

    Visual object \cite{rp22} & 10 & 1000Hz &64 &Image decoding & Object label & - \\

    MAHNOB-HCI \cite{rp23} & 27 & 256Hz & 32&Emotion recognition &Arousal,Valance,Dominant& Multiple signals \\

    SEEG \cite{re145} & - & 1000 or 2000Hz & 52 to 124 & Seizure detection& Seizure, Non seizure & Ecog \\

    Epilepsy Dataset \cite{rp25} & 500 & 173Hz & 19 & Seizure detection &Five seizure labels & - \\

    AMIGOS \cite{rp26} & 40 & 128Hz & 14 & Emotion recognition & Arousal,Valance,Dominant & ECG,GRS \\

    DREAMER \cite{rp27} & 23 & 128Hz &14&Emotion recognition & Arousal,Valance,Dominant & ECG \\

    ASD dataset \cite{re107} & 4899 & 250Hz & 20 to 129 & AS-Disorder & ASD lables & - \\

    NMT sculp dataset \cite{rp28} & - & 250Hz & 19 & Pathology detection & Normal,Abnormal & - \\

    CUHZ \cite{rp29} & 25 & 500,698,1000Hz & 22 &Seizure datection& Different seizure types &- \\

  \bottomrule
\end{tabularx}
\label{tab_5}
\end{center}
\vspace{-\baselineskip}
\end{table}

\textbf{Sleep stage classification} is the task of classifying sleep EEG signals into different stages. The criteria for sleep stage classification are proposed by the American Academy of Sleep Medicine (AASM), dividing sleep EEG signals into five stages \cite{rn3}: \textit{W} stage is the weak stage, \textit{N1} and \textit{N2} stages (Non-REM stage) are the light sleep, \textit{N3} stage is deep sleep, and \textit{REM} stage is the Rapid Eye Movement sleep \cite{rn4}. Temporal models are used to capture the temporal correlation and difference between sleep stages to accurately identify the sleep stage to which the EEG sample belongs.

\textbf{Speech/Image decoding} is the complex task of decoding image or speech information from the EEG signals. This task involves translating brain activity patterns recorded by EEG into meaningful visual and speech information, which can help to understand the neural mechanisms of vision and audition in the brain \cite{re157}. Inspired by the visual question answering \cite{re158} that aligned the text and image patches through SSL to extract the semantic information, SSL can align the EEG signal and image, speech patches to capture the inner correlation and improve the task performance.

As the practical downstream tasks mentioned above, corresponding datasets have been proposed to train the model. For emotion recognition and motor imagery tasks, the existing datasets such as SEED \cite{re44} and MMI \cite{rp5} contain EEG signal with more than 30 channels, where the fine-grained spatial correlation can be extracted; On the contrary, the datasets for sleep stage classification task contain fewer channels but longer time windows, where the temporal information is critical for sleep classifications. Besides, the pathology datasets contain more subjects to ensure that the general EEG features can be extracted for clinical application.

\section{Future directions}

In the EEG analysis field, combining deep model and SSL frameworks can help improve the model performance on various EEG-based tasks through extra parameter training on unlabeled EEG samples with well-designed pretext tasks. In addition to the advantages of EEG-based SSL frameworks, we analyze the challenges in the existing EEG-based SSL studies and propose potential future directions for EEG-based SSL to address the challenges and problems.

\textbf{Signal-oriented pretext task}. Most existing pretext tasks are the straightforward extension of pretext tasks in CV and NLP, which treat EEG signals as 2D matrix and temporal vector like image or text patches to capture spatial and contextual correlations but ignore the intrinsic characteristics of EEG signal. Therefore, designing the EEG-oriented pretext task to extract the spatial-temporal-frequency EEG features is a feasible approach worth further exploration.

\textbf{Knowledge-driven SSL framework}. Although SSL frameworks have achieved significant success in various EEG-based tasks, the lack of theoretical foundation and neural knowledge for EEG signals leads to the generated representations lacking generalization and interpretability. Therefore, how to integrate the EEG-based neural knowledge with the SSL framework to construct the knowledge-driven interpretable EEG model is another important direction, which needs to design specific pretext tasks and augmentation techniques that can fuse explainable neural knowledge into representation. We believe that by integrating knowledge of EEG into the self-supervised framework, the models are expected to bring generalization and interpretability to representations.

\textbf{Graph-based SSL}. Deep learning models like CNN and RNN have been widely used to extract spatial-temporal features from EEG signals for different tasks. However, most existing methods ignore the inherent topological connections among electrodes. EEG signals are generated from the activity of neurons that are topologically connected inner the brain. Graph neural networks can explore the inherent connectivity patterns among neurons, and we believe that researching GNN-based EEG SSL methods can integrate richer latent brain information into representations, offering a new perspective for information expression.

\textbf{SSL for Heterogeneous EEG}. The ultimate goal of SSL for EEG analysis is to generate general representations for various downstream tasks. However, EEG signals are collected from multiple scenarios which encompass variations in channel, device, sampling rate, task, subject, and distribution. The significant differences between EEG signals from different sources make it challenging for self-supervised training collaboratively. Therefore, constructing SSL framework tailored for heterogeneous EEG data is an important direction for future development. Exploration in this direction can utilize heterogeneous EEG samples from multiple sources to jointly pre-train the model to fully utilize existing differentiated EEG datasets to mine universal representations for different downstream tasks.

\textbf{Multimodal SSL}. SSL for EEG signals is the unimodal approach aiming to extract neural information from unlabeled EEG samples. However, the features mined from  EEG signals are difficult to adapt to some complex downstream tasks, which require other brain or physiological signals to provide more abundant information. Therefore, the EEG-based multimodal self-supervised learning method needs to be further studied to extract integrated and aligned features from unlabeled multimodal signals (ECG, EMG, EOG, etc) for challenging downstream tasks.

\section{Conclusion}

This paper is a comprehensive review of self-supervised learning for EEG analysis, including the reasonable taxonomy, different kinds of existing EEG-based SSL methods, downstream EEG tasks, and the available training datasets, offering detailed guidelines for researchers interested in deep learning combined with EEG analysis. We first review typical SSL frameworks and pretext tasks in the CV and NLP and introduce traditional supervised EEG analysis methods as the preliminary, to illustrate the drawbacks of supervised EEG analysis and underscore the necessity of introducing SSL for EEG analysis. We then provide a detailed exposition on four categories of SSL frameworks for EEG analysis, elucidating the technical details of representative methods to extract spatial-temporal-frequency features from EEG signals. Subsequently, we enumerate EEG-based downstream tasks effective for SSL frameworks and present relevant EEG datasets suitable for pre-training or downstream task fine-tuning. Finally, we discuss the challenges in the existing studies and propose new insights and potential future directions that warrant exploration, which can help generate a more general explainable representation to solve various complex downstream tasks.


\bibliographystyle{ACM-Reference-Format}
\bibliography{manuscript}


\begin{thebibliography}{163}


\ifx \showCODEN    \undefined \def \showCODEN     #1{\unskip}     \fi
\ifx \showDOI      \undefined \def \showDOI       #1{#1}\fi
\ifx \showISBNx    \undefined \def \showISBNx     #1{\unskip}     \fi
\ifx \showISBNxiii \undefined \def \showISBNxiii  #1{\unskip}     \fi
\ifx \showISSN     \undefined \def \showISSN      #1{\unskip}     \fi
\ifx \showLCCN     \undefined \def \showLCCN      #1{\unskip}     \fi
\ifx \shownote     \undefined \def \shownote      #1{#1}          \fi
\ifx \showarticletitle \undefined \def \showarticletitle #1{#1}   \fi
\ifx \showURL      \undefined \def \showURL       {\relax}        \fi
\providecommand\bibfield[2]{#2}
\providecommand\bibinfo[2]{#2}
\providecommand\natexlab[1]{#1}
\providecommand\showeprint[2][]{arXiv:#2}

\bibitem[Accou et~al\mbox{.}(2023)]%
        {re90}
\bibfield{author}{\bibinfo{person}{Bernd Accou}, \bibinfo{person}{Tom Francart}, {et~al\mbox{.}}} \bibinfo{year}{2023}\natexlab{}.
\newblock \showarticletitle{Self-supervised enhancement of stimulus-evoked brain response data}.
\newblock \bibinfo{journal}{\emph{arXiv preprint arXiv:2302.01924}} (\bibinfo{year}{2023}).
\newblock


\bibitem[Aguiar-Conraria and Soares(2014)]%
        {re113}
\bibfield{author}{\bibinfo{person}{Lu{\'\i}s Aguiar-Conraria} {and} \bibinfo{person}{Maria~Joana Soares}.} \bibinfo{year}{2014}\natexlab{}.
\newblock \showarticletitle{The continuous wavelet transform: Moving beyond uni-and bivariate analysis}.
\newblock \bibinfo{journal}{\emph{Journal of Economic Surveys}} \bibinfo{volume}{28}, \bibinfo{number}{2} (\bibinfo{year}{2014}), \bibinfo{pages}{344--375}.
\newblock


\bibitem[Al-Quraishi et~al\mbox{.}(2018)]%
        {re20}
\bibfield{author}{\bibinfo{person}{Maged~S Al-Quraishi}, \bibinfo{person}{Irraivan Elamvazuthi}, \bibinfo{person}{Siti~Asmah Daud}, \bibinfo{person}{S Parasuraman}, {and} \bibinfo{person}{Alberto Borboni}.} \bibinfo{year}{2018}\natexlab{}.
\newblock \showarticletitle{EEG-based control for upper and lower limb exoskeletons and prostheses: A systematic review}.
\newblock \bibinfo{journal}{\emph{Sensors}} \bibinfo{volume}{18}, \bibinfo{number}{10} (\bibinfo{year}{2018}), \bibinfo{pages}{3342}.
\newblock


\bibitem[Alotaiby et~al\mbox{.}(2014)]%
        {re61}
\bibfield{author}{\bibinfo{person}{Turkey~N Alotaiby}, \bibinfo{person}{Saleh~A Alshebeili}, \bibinfo{person}{Tariq Alshawi}, \bibinfo{person}{Ishtiaq Ahmad}, {and} \bibinfo{person}{Fathi~E Abd El-Samie}.} \bibinfo{year}{2014}\natexlab{}.
\newblock \showarticletitle{EEG seizure detection and prediction algorithms: a survey}.
\newblock \bibinfo{journal}{\emph{EURASIP Journal on Advances in Signal Processing}}  \bibinfo{volume}{2014} (\bibinfo{year}{2014}), \bibinfo{pages}{1--21}.
\newblock


\bibitem[Altaheri et~al\mbox{.}(2023)]%
        {re63}
\bibfield{author}{\bibinfo{person}{Hamdi Altaheri}, \bibinfo{person}{Ghulam Muhammad}, \bibinfo{person}{Mansour Alsulaiman}, \bibinfo{person}{Syed~Umar Amin}, \bibinfo{person}{Ghadir~Ali Altuwaijri}, \bibinfo{person}{Wadood Abdul}, \bibinfo{person}{Mohamed~A Bencherif}, {and} \bibinfo{person}{Mohammed Faisal}.} \bibinfo{year}{2023}\natexlab{}.
\newblock \showarticletitle{Deep learning techniques for classification of electroencephalogram (EEG) motor imagery (MI) signals: A review}.
\newblock \bibinfo{journal}{\emph{Neural Computing and Applications}} \bibinfo{volume}{35}, \bibinfo{number}{20} (\bibinfo{year}{2023}), \bibinfo{pages}{14681--14722}.
\newblock


\bibitem[Andrzejak et~al\mbox{.}(2001)]%
        {rp25}
\bibfield{author}{\bibinfo{person}{Ralph~G Andrzejak}, \bibinfo{person}{Klaus Lehnertz}, \bibinfo{person}{Florian Mormann}, \bibinfo{person}{Christoph Rieke}, \bibinfo{person}{Peter David}, {and} \bibinfo{person}{Christian~E Elger}.} \bibinfo{year}{2001}\natexlab{}.
\newblock \showarticletitle{Indications of nonlinear deterministic and finite-dimensional structures in time series of brain electrical activity: Dependence on recording region and brain state}.
\newblock \bibinfo{journal}{\emph{Physical Review E}} \bibinfo{volume}{64}, \bibinfo{number}{6} (\bibinfo{year}{2001}), \bibinfo{pages}{061907}.
\newblock


\bibitem[Aserinsky and Kleitman(1953)]%
        {rn4}
\bibfield{author}{\bibinfo{person}{Eugene Aserinsky} {and} \bibinfo{person}{Nathaniel Kleitman}.} \bibinfo{year}{1953}\natexlab{}.
\newblock \showarticletitle{Regularly occurring periods of eye motility, and concomitant phenomena, during sleep}.
\newblock \bibinfo{journal}{\emph{Science}} \bibinfo{volume}{118}, \bibinfo{number}{3062} (\bibinfo{year}{1953}), \bibinfo{pages}{273--274}.
\newblock


\bibitem[Babayan et~al\mbox{.}(2018)]%
        {rp21}
\bibfield{author}{\bibinfo{person}{A Babayan}, \bibinfo{person}{M Erbey}, \bibinfo{person}{D Kumral}, \bibinfo{person}{JD Reinelt}, \bibinfo{person}{AMF Reiter}, \bibinfo{person}{J R{\"o}bbig}, \bibinfo{person}{H Lina~Schaare}, \bibinfo{person}{M Uhlig}, \bibinfo{person}{A Anwander}, \bibinfo{person}{PL Bazin}, {et~al\mbox{.}}} \bibinfo{year}{2018}\natexlab{}.
\newblock \bibinfo{title}{Data descriptor: a mind-brain-body dataset of MRI, EEG, cognition, emotion, and peripheral physiology in young and old adults. Sci. Data 6, 180308}.
\newblock
\newblock


\bibitem[Baevski et~al\mbox{.}(2020)]%
        {re103}
\bibfield{author}{\bibinfo{person}{Alexei Baevski}, \bibinfo{person}{Yuhao Zhou}, \bibinfo{person}{Abdelrahman Mohamed}, {and} \bibinfo{person}{Michael Auli}.} \bibinfo{year}{2020}\natexlab{}.
\newblock \showarticletitle{wav2vec 2.0: A framework for self-supervised learning of speech representations}.
\newblock \bibinfo{journal}{\emph{Advances in neural information processing systems}}  \bibinfo{volume}{33} (\bibinfo{year}{2020}), \bibinfo{pages}{12449--12460}.
\newblock


\bibitem[Bagchi and Mitra(2012)]%
        {re117}
\bibfield{author}{\bibinfo{person}{Sonali Bagchi} {and} \bibinfo{person}{Sanjit~K Mitra}.} \bibinfo{year}{2012}\natexlab{}.
\newblock \bibinfo{booktitle}{\emph{The nonuniform discrete Fourier transform and its applications in signal processing}}. Vol.~\bibinfo{volume}{463}.
\newblock \bibinfo{publisher}{Springer Science \& Business Media}.
\newblock


\bibitem[Balconi and Lucchiari(2006)]%
        {re109}
\bibfield{author}{\bibinfo{person}{Michela Balconi} {and} \bibinfo{person}{Claudio Lucchiari}.} \bibinfo{year}{2006}\natexlab{}.
\newblock \showarticletitle{EEG correlates (event-related desynchronization) of emotional face elaboration: a temporal analysis}.
\newblock \bibinfo{journal}{\emph{Neuroscience letters}} \bibinfo{volume}{392}, \bibinfo{number}{1-2} (\bibinfo{year}{2006}), \bibinfo{pages}{118--123}.
\newblock


\bibitem[Banville et~al\mbox{.}(2019)]%
        {re93}
\bibfield{author}{\bibinfo{person}{Hubert Banville}, \bibinfo{person}{Isabela Albuquerque}, \bibinfo{person}{Aapo Hyv{\"a}rinen}, \bibinfo{person}{Graeme Moffat}, \bibinfo{person}{Denis-Alexander Engemann}, {and} \bibinfo{person}{Alexandre Gramfort}.} \bibinfo{year}{2019}\natexlab{}.
\newblock \showarticletitle{Self-supervised representation learning from electroencephalography signals}. In \bibinfo{booktitle}{\emph{2019 IEEE 29th International Workshop on Machine Learning for Signal Processing (MLSP)}}. IEEE, \bibinfo{pages}{1--6}.
\newblock


\bibitem[Banville et~al\mbox{.}(2021)]%
        {re87}
\bibfield{author}{\bibinfo{person}{Hubert Banville}, \bibinfo{person}{Omar Chehab}, \bibinfo{person}{Aapo Hyv{\"a}rinen}, \bibinfo{person}{Denis-Alexander Engemann}, {and} \bibinfo{person}{Alexandre Gramfort}.} \bibinfo{year}{2021}\natexlab{}.
\newblock \showarticletitle{Uncovering the structure of clinical EEG signals with self-supervised learning}.
\newblock \bibinfo{journal}{\emph{Journal of Neural Engineering}} \bibinfo{volume}{18}, \bibinfo{number}{4} (\bibinfo{year}{2021}), \bibinfo{pages}{046020}.
\newblock


\bibitem[Bhat and Hortal(2021)]%
        {re123}
\bibfield{author}{\bibinfo{person}{Sudhanva Bhat} {and} \bibinfo{person}{Enrique Hortal}.} \bibinfo{year}{2021}\natexlab{}.
\newblock \showarticletitle{Gan-based data augmentation for improving the classification of eeg signals}. In \bibinfo{booktitle}{\emph{The 14th pervasive technologies related to assistive environments conference}}. \bibinfo{pages}{453--458}.
\newblock


\bibitem[Biswal et~al\mbox{.}(2018)]%
        {rp9}
\bibfield{author}{\bibinfo{person}{Siddharth Biswal}, \bibinfo{person}{Haoqi Sun}, \bibinfo{person}{Balaji Goparaju}, \bibinfo{person}{M~Brandon Westover}, \bibinfo{person}{Jimeng Sun}, {and} \bibinfo{person}{Matt~T Bianchi}.} \bibinfo{year}{2018}\natexlab{}.
\newblock \showarticletitle{Expert-level sleep scoring with deep neural networks}.
\newblock \bibinfo{journal}{\emph{Journal of the American Medical Informatics Association}} \bibinfo{volume}{25}, \bibinfo{number}{12} (\bibinfo{year}{2018}), \bibinfo{pages}{1643--1650}.
\newblock


\bibitem[Bollens et~al\mbox{.}(2023)]%
        {rp19}
\bibfield{author}{\bibinfo{person}{Lies Bollens}, \bibinfo{person}{Bernd Accou}, \bibinfo{person}{Marlies Gillis}, \bibinfo{person}{Wendy Verheijen}, \bibinfo{person}{Tom Francart}, {et~al\mbox{.}}} \bibinfo{year}{2023}\natexlab{}.
\newblock \showarticletitle{SparrKULee: A Speech-evoked Auditory Response Repository of the KU Leuven, containing EEG of 85 participants}.
\newblock  (\bibinfo{year}{2023}).
\newblock


\bibitem[Boostani et~al\mbox{.}(2017)]%
        {re62}
\bibfield{author}{\bibinfo{person}{Reza Boostani}, \bibinfo{person}{Foroozan Karimzadeh}, {and} \bibinfo{person}{Mohammad Nami}.} \bibinfo{year}{2017}\natexlab{}.
\newblock \showarticletitle{A comparative review on sleep stage classification methods in patients and healthy individuals}.
\newblock \bibinfo{journal}{\emph{Computer methods and programs in biomedicine}}  \bibinfo{volume}{140} (\bibinfo{year}{2017}), \bibinfo{pages}{77--91}.
\newblock


\bibitem[Bos et~al\mbox{.}(2006)]%
        {re15}
\bibfield{author}{\bibinfo{person}{Danny~Oude Bos} {et~al\mbox{.}}} \bibinfo{year}{2006}\natexlab{}.
\newblock \showarticletitle{EEG-based emotion recognition}.
\newblock \bibinfo{journal}{\emph{The influence of visual and auditory stimuli}} \bibinfo{volume}{56}, \bibinfo{number}{3} (\bibinfo{year}{2006}), \bibinfo{pages}{1--17}.
\newblock


\bibitem[Br{\"u}sch et~al\mbox{.}(2023)]%
        {re164}
\bibfield{author}{\bibinfo{person}{Thea Br{\"u}sch}, \bibinfo{person}{Mikkel~N Schmidt}, {and} \bibinfo{person}{Tommy~S Alstr{\o}m}.} \bibinfo{year}{2023}\natexlab{}.
\newblock \showarticletitle{Multi-view self-supervised learning for multivariate variable-channel time series}. In \bibinfo{booktitle}{\emph{2023 IEEE 33rd International Workshop on Machine Learning for Signal Processing (MLSP)}}. IEEE, \bibinfo{pages}{1--6}.
\newblock


\bibitem[Cai et~al\mbox{.}(2023)]%
        {re86}
\bibfield{author}{\bibinfo{person}{Donghong Cai}, \bibinfo{person}{Junru Chen}, \bibinfo{person}{Yang Yang}, \bibinfo{person}{Teng Liu}, {and} \bibinfo{person}{Yafeng Li}.} \bibinfo{year}{2023}\natexlab{}.
\newblock \showarticletitle{MBrain: A Multi-channel Self-Supervised Learning Framework for Brain Signals}.
\newblock \bibinfo{journal}{\emph{arXiv preprint arXiv:2306.13102}} (\bibinfo{year}{2023}).
\newblock


\bibitem[Chang et~al\mbox{.}(2022)]%
        {re168}
\bibfield{author}{\bibinfo{person}{Shuohua Chang}, \bibinfo{person}{Zhihong Yang}, \bibinfo{person}{Yuyang You}, {and} \bibinfo{person}{Xiaoyu Guo}.} \bibinfo{year}{2022}\natexlab{}.
\newblock \showarticletitle{Dssnet: A deep sequential sleep network for self-supervised representation learning based on single-channel eeg}.
\newblock \bibinfo{journal}{\emph{IEEE Signal Processing Letters}}  \bibinfo{volume}{29} (\bibinfo{year}{2022}), \bibinfo{pages}{2143--2147}.
\newblock


\bibitem[Chen et~al\mbox{.}(2023)]%
        {re107}
\bibfield{author}{\bibinfo{person}{He Chen}, \bibinfo{person}{Ouyang Gaoxiang}, {and} \bibinfo{person}{Xiaoli Li}.} \bibinfo{year}{2023}\natexlab{}.
\newblock \showarticletitle{Extracting Temporal-Spectral-Spatial Representation of EEG Using Self-Supervised Learning for the Identification of Children with ASD}. In \bibinfo{booktitle}{\emph{2023 IEEE 13th International Conference on CYBER Technology in Automation, Control, and Intelligent Systems (CYBER)}}. IEEE, \bibinfo{pages}{1263--1266}.
\newblock


\bibitem[Chen et~al\mbox{.}(2022b)]%
        {re145}
\bibfield{author}{\bibinfo{person}{Junru Chen}, \bibinfo{person}{Yang Yang}, \bibinfo{person}{Tao Yu}, \bibinfo{person}{Yingying Fan}, \bibinfo{person}{Xiaolong Mo}, {and} \bibinfo{person}{Carl Yang}.} \bibinfo{year}{2022}\natexlab{b}.
\newblock \showarticletitle{Brainnet: Epileptic wave detection from seeg with hierarchical graph diffusion learning}. In \bibinfo{booktitle}{\emph{Proceedings of the 28th ACM SIGKDD Conference on Knowledge Discovery and Data Mining}}. \bibinfo{pages}{2741--2751}.
\newblock


\bibitem[Chen et~al\mbox{.}(2020)]%
        {re83}
\bibfield{author}{\bibinfo{person}{Ting Chen}, \bibinfo{person}{Simon Kornblith}, \bibinfo{person}{Mohammad Norouzi}, {and} \bibinfo{person}{Geoffrey Hinton}.} \bibinfo{year}{2020}\natexlab{}.
\newblock \showarticletitle{A simple framework for contrastive learning of visual representations}. In \bibinfo{booktitle}{\emph{International conference on machine learning}}. PMLR, \bibinfo{pages}{1597--1607}.
\newblock


\bibitem[Chen et~al\mbox{.}(2022a)]%
        {re42}
\bibfield{author}{\bibinfo{person}{Xun Chen}, \bibinfo{person}{Chang Li}, \bibinfo{person}{Aiping Liu}, \bibinfo{person}{Martin~J McKeown}, \bibinfo{person}{Ruobing Qian}, {and} \bibinfo{person}{Z~Jane Wang}.} \bibinfo{year}{2022}\natexlab{a}.
\newblock \showarticletitle{Toward open-world electroencephalogram decoding via deep learning: A comprehensive survey}.
\newblock \bibinfo{journal}{\emph{IEEE Signal Processing Magazine}} \bibinfo{volume}{39}, \bibinfo{number}{2} (\bibinfo{year}{2022}), \bibinfo{pages}{117--134}.
\newblock


\bibitem[Cheng et~al\mbox{.}(2020)]%
        {re173}
\bibfield{author}{\bibinfo{person}{Joseph~Y Cheng}, \bibinfo{person}{Hanlin Goh}, \bibinfo{person}{Kaan Dogrusoz}, \bibinfo{person}{Oncel Tuzel}, {and} \bibinfo{person}{Erdrin Azemi}.} \bibinfo{year}{2020}\natexlab{}.
\newblock \showarticletitle{Subject-aware contrastive learning for biosignals}.
\newblock \bibinfo{journal}{\emph{arXiv preprint arXiv:2007.04871}} (\bibinfo{year}{2020}).
\newblock


\bibitem[Chien et~al\mbox{.}(2022)]%
        {re101}
\bibfield{author}{\bibinfo{person}{Hsiang-Yun~Sherry Chien}, \bibinfo{person}{Hanlin Goh}, \bibinfo{person}{Christopher~M Sandino}, {and} \bibinfo{person}{Joseph~Y Cheng}.} \bibinfo{year}{2022}\natexlab{}.
\newblock \showarticletitle{MAEEG: Masked Auto-encoder for EEG Representation Learning}.
\newblock \bibinfo{journal}{\emph{arXiv preprint arXiv:2211.02625}} (\bibinfo{year}{2022}).
\newblock


\bibitem[Choi et~al\mbox{.}(2020)]%
        {rn1}
\bibfield{author}{\bibinfo{person}{Junhyuk Choi}, \bibinfo{person}{Keun~Tae Kim}, \bibinfo{person}{Ji~Hyeok Jeong}, \bibinfo{person}{Laehyun Kim}, \bibinfo{person}{Song~Joo Lee}, {and} \bibinfo{person}{Hyungmin Kim}.} \bibinfo{year}{2020}\natexlab{}.
\newblock \showarticletitle{Developing a motor imagery-based real-time asynchronous hybrid BCI controller for a lower-limb exoskeleton}.
\newblock \bibinfo{journal}{\emph{Sensors}} \bibinfo{volume}{20}, \bibinfo{number}{24} (\bibinfo{year}{2020}), \bibinfo{pages}{7309}.
\newblock


\bibitem[Cimtay and Ekmekcioglu(2020)]%
        {re43}
\bibfield{author}{\bibinfo{person}{Yucel Cimtay} {and} \bibinfo{person}{Erhan Ekmekcioglu}.} \bibinfo{year}{2020}\natexlab{}.
\newblock \showarticletitle{Investigating the use of pretrained convolutional neural network on cross-subject and cross-dataset EEG emotion recognition}.
\newblock \bibinfo{journal}{\emph{Sensors}} \bibinfo{volume}{20}, \bibinfo{number}{7} (\bibinfo{year}{2020}), \bibinfo{pages}{2034}.
\newblock


\bibitem[Craik et~al\mbox{.}(2019)]%
        {re73}
\bibfield{author}{\bibinfo{person}{Alexander Craik}, \bibinfo{person}{Yongtian He}, {and} \bibinfo{person}{Jose~L Contreras-Vidal}.} \bibinfo{year}{2019}\natexlab{}.
\newblock \showarticletitle{Deep learning for electroencephalogram (EEG) classification tasks: a review}.
\newblock \bibinfo{journal}{\emph{Journal of neural engineering}} \bibinfo{volume}{16}, \bibinfo{number}{3} (\bibinfo{year}{2019}), \bibinfo{pages}{031001}.
\newblock


\bibitem[Creswell et~al\mbox{.}(2018)]%
        {re121}
\bibfield{author}{\bibinfo{person}{Antonia Creswell}, \bibinfo{person}{Tom White}, \bibinfo{person}{Vincent Dumoulin}, \bibinfo{person}{Kai Arulkumaran}, \bibinfo{person}{Biswa Sengupta}, {and} \bibinfo{person}{Anil~A Bharath}.} \bibinfo{year}{2018}\natexlab{}.
\newblock \showarticletitle{Generative adversarial networks: An overview}.
\newblock \bibinfo{journal}{\emph{IEEE signal processing magazine}} \bibinfo{volume}{35}, \bibinfo{number}{1} (\bibinfo{year}{2018}), \bibinfo{pages}{53--65}.
\newblock


\bibitem[Das et~al\mbox{.}(2022)]%
        {re120}
\bibfield{author}{\bibinfo{person}{Sudip Das}, \bibinfo{person}{Pankaj Pandey}, {and} \bibinfo{person}{Krishna~Prasad Miyapuram}.} \bibinfo{year}{2022}\natexlab{}.
\newblock \showarticletitle{Improving self-supervised pretraining models for epileptic seizure detection from EEG data}.
\newblock \bibinfo{journal}{\emph{arXiv preprint arXiv:2207.06911}} (\bibinfo{year}{2022}).
\newblock


\bibitem[D{\'e}fossez et~al\mbox{.}(2023)]%
        {re159}
\bibfield{author}{\bibinfo{person}{Alexandre D{\'e}fossez}, \bibinfo{person}{Charlotte Caucheteux}, \bibinfo{person}{J{\'e}r{\'e}my Rapin}, \bibinfo{person}{Ori Kabeli}, {and} \bibinfo{person}{Jean-R{\'e}mi King}.} \bibinfo{year}{2023}\natexlab{}.
\newblock \showarticletitle{Decoding speech perception from non-invasive brain recordings}.
\newblock \bibinfo{journal}{\emph{Nature Machine Intelligence}} (\bibinfo{year}{2023}), \bibinfo{pages}{1--11}.
\newblock


\bibitem[Dement and Kleitman(1957)]%
        {rn3}
\bibfield{author}{\bibinfo{person}{William Dement} {and} \bibinfo{person}{Nathaniel Kleitman}.} \bibinfo{year}{1957}\natexlab{}.
\newblock \showarticletitle{Cyclic variations in EEG during sleep and their relation to eye movements, body motility, and dreaming}.
\newblock \bibinfo{journal}{\emph{Electroencephalography and clinical neurophysiology}} \bibinfo{volume}{9}, \bibinfo{number}{4} (\bibinfo{year}{1957}), \bibinfo{pages}{673--690}.
\newblock


\bibitem[Devlin et~al\mbox{.}(2018)]%
        {re57}
\bibfield{author}{\bibinfo{person}{Jacob Devlin}, \bibinfo{person}{Ming-Wei Chang}, \bibinfo{person}{Kenton Lee}, {and} \bibinfo{person}{Kristina Toutanova}.} \bibinfo{year}{2018}\natexlab{}.
\newblock \showarticletitle{Bert: Pre-training of deep bidirectional transformers for language understanding}.
\newblock \bibinfo{journal}{\emph{arXiv preprint arXiv:1810.04805}} (\bibinfo{year}{2018}).
\newblock


\bibitem[Dosovitskiy et~al\mbox{.}(2020)]%
        {re114}
\bibfield{author}{\bibinfo{person}{Alexey Dosovitskiy}, \bibinfo{person}{Lucas Beyer}, \bibinfo{person}{Alexander Kolesnikov}, \bibinfo{person}{Dirk Weissenborn}, \bibinfo{person}{Xiaohua Zhai}, \bibinfo{person}{Thomas Unterthiner}, \bibinfo{person}{Mostafa Dehghani}, \bibinfo{person}{Matthias Minderer}, \bibinfo{person}{Georg Heigold}, \bibinfo{person}{Sylvain Gelly}, {et~al\mbox{.}}} \bibinfo{year}{2020}\natexlab{}.
\newblock \showarticletitle{An image is worth 16x16 words: Transformers for image recognition at scale}.
\newblock \bibinfo{journal}{\emph{arXiv preprint arXiv:2010.11929}} (\bibinfo{year}{2020}).
\newblock


\bibitem[Du et~al\mbox{.}(2015)]%
        {re28}
\bibfield{author}{\bibinfo{person}{Nguyen~Duy Du}, \bibinfo{person}{Nguyen~Hoang Huy}, {and} \bibinfo{person}{Nguyen~Xuan Hoai}.} \bibinfo{year}{2015}\natexlab{}.
\newblock \showarticletitle{The impact of high dimensionality on SVM when classifying ERP data-a solution from LDA}. In \bibinfo{booktitle}{\emph{Proceedings of the 6th International Symposium on Information and Communication Technology}}. \bibinfo{pages}{32--37}.
\newblock


\bibitem[Duan et~al\mbox{.}(2013)]%
        {re45}
\bibfield{author}{\bibinfo{person}{Ruo-Nan Duan}, \bibinfo{person}{Jia-Yi Zhu}, {and} \bibinfo{person}{Bao-Liang Lu}.} \bibinfo{year}{2013}\natexlab{}.
\newblock \showarticletitle{Differential entropy feature for {EEG}-based emotion classification}. In \bibinfo{booktitle}{\emph{6th International IEEE/EMBS Conference on Neural Engineering (NER)}}. IEEE, \bibinfo{pages}{81--84}.
\newblock


\bibitem[Eldele et~al\mbox{.}(2021)]%
        {re153}
\bibfield{author}{\bibinfo{person}{Emadeldeen Eldele}, \bibinfo{person}{Mohamed Ragab}, \bibinfo{person}{Zhenghua Chen}, \bibinfo{person}{Min Wu}, \bibinfo{person}{Chee~Keong Kwoh}, \bibinfo{person}{Xiaoli Li}, {and} \bibinfo{person}{Cuntai Guan}.} \bibinfo{year}{2021}\natexlab{}.
\newblock \showarticletitle{Time-series representation learning via temporal and contextual contrasting}.
\newblock \bibinfo{journal}{\emph{arXiv preprint arXiv:2106.14112}} (\bibinfo{year}{2021}).
\newblock


\bibitem[Ericsson et~al\mbox{.}(2022)]%
        {re48}
\bibfield{author}{\bibinfo{person}{Linus Ericsson}, \bibinfo{person}{Henry Gouk}, \bibinfo{person}{Chen~Change Loy}, {and} \bibinfo{person}{Timothy~M Hospedales}.} \bibinfo{year}{2022}\natexlab{}.
\newblock \showarticletitle{Self-supervised representation learning: Introduction, advances, and challenges}.
\newblock \bibinfo{journal}{\emph{IEEE Signal Processing Magazine}} \bibinfo{volume}{39}, \bibinfo{number}{3} (\bibinfo{year}{2022}), \bibinfo{pages}{42--62}.
\newblock


\bibitem[Fahimi et~al\mbox{.}(2019)]%
        {re23}
\bibfield{author}{\bibinfo{person}{Fatemeh Fahimi}, \bibinfo{person}{Zhuo Zhang}, \bibinfo{person}{Wooi~Boon Goh}, \bibinfo{person}{Kai~Keng Ang}, {and} \bibinfo{person}{Cuntai Guan}.} \bibinfo{year}{2019}\natexlab{}.
\newblock \showarticletitle{Towards EEG generation using GANs for BCI applications}. In \bibinfo{booktitle}{\emph{2019 IEEE EMBS International Conference on Biomedical \& Health Informatics (BHI)}}. IEEE, \bibinfo{pages}{1--4}.
\newblock


\bibitem[Fu et~al\mbox{.}(2022)]%
        {re124}
\bibfield{author}{\bibinfo{person}{Ruiqi Fu}, \bibinfo{person}{Yi-Feng Chen}, \bibinfo{person}{Yongqi Huang}, \bibinfo{person}{Shuping Chen}, \bibinfo{person}{Feiyan Duan}, \bibinfo{person}{Jiewei Li}, \bibinfo{person}{Jianhui Wu}, \bibinfo{person}{Dongmei Jiang}, \bibinfo{person}{Junling Gao}, \bibinfo{person}{Jason Gu}, {et~al\mbox{.}}} \bibinfo{year}{2022}\natexlab{}.
\newblock \showarticletitle{Symmetric convolutional and adversarial neural network enables improved mental stress classification from EEG}.
\newblock \bibinfo{journal}{\emph{IEEE Transactions on Neural Systems and Rehabilitation Engineering}}  \bibinfo{volume}{30} (\bibinfo{year}{2022}), \bibinfo{pages}{1384--1400}.
\newblock


\bibitem[Gao et~al\mbox{.}({[n.\,d.]})]%
        {re174}
\bibfield{author}{\bibinfo{person}{Wei Gao}, \bibinfo{person}{Zhengqing Hu}, \bibinfo{person}{Yu Lei}, \bibinfo{person}{Changming Wang}, \bibinfo{person}{Fangbing Qiu}, \bibinfo{person}{Yanqing Liu}, {and} \bibinfo{person}{Lin Han}.} \bibinfo{year}{[n.\,d.]}\natexlab{}.
\newblock \showarticletitle{A Multi-Channel Sleep Staging Method Based on Self-Supervised Learning}.
\newblock \bibinfo{journal}{\emph{Available at SSRN 4580453}} (\bibinfo{year}{[n.\,d.]}).
\newblock


\bibitem[Ge et~al\mbox{.}(2021)]%
        {re41}
\bibfield{author}{\bibinfo{person}{Wendong Ge}, \bibinfo{person}{Jin Jing}, \bibinfo{person}{Sungtae An}, \bibinfo{person}{Aline Herlopian}, \bibinfo{person}{Marcus Ng}, \bibinfo{person}{Aaron~F Struck}, \bibinfo{person}{Brian Appavu}, \bibinfo{person}{Emily~L Johnson}, \bibinfo{person}{Gamaleldin Osman}, \bibinfo{person}{Hiba~A Haider}, {et~al\mbox{.}}} \bibinfo{year}{2021}\natexlab{}.
\newblock \showarticletitle{Deep active learning for interictal ictal injury continuum EEG patterns}.
\newblock \bibinfo{journal}{\emph{Journal of neuroscience methods}}  \bibinfo{volume}{351} (\bibinfo{year}{2021}), \bibinfo{pages}{108966}.
\newblock


\bibitem[Ghassemi et~al\mbox{.}(2018)]%
        {rp1}
\bibfield{author}{\bibinfo{person}{Mohammad~M Ghassemi}, \bibinfo{person}{Benjamin~E Moody}, \bibinfo{person}{Li-Wei~H Lehman}, \bibinfo{person}{Christopher Song}, \bibinfo{person}{Qiao Li}, \bibinfo{person}{Haoqi Sun}, \bibinfo{person}{Roger~G Mark}, \bibinfo{person}{M~Brandon Westover}, {and} \bibinfo{person}{Gari~D Clifford}.} \bibinfo{year}{2018}\natexlab{}.
\newblock \showarticletitle{You snooze, you win: the physionet/computing in cardiology challenge 2018}. In \bibinfo{booktitle}{\emph{2018 Computing in Cardiology Conference (CinC)}}, Vol.~\bibinfo{volume}{45}. IEEE, \bibinfo{pages}{1--4}.
\newblock


\bibitem[Gidaris et~al\mbox{.}(2018)]%
        {re52}
\bibfield{author}{\bibinfo{person}{Spyros Gidaris}, \bibinfo{person}{Praveer Singh}, {and} \bibinfo{person}{Nikos Komodakis}.} \bibinfo{year}{2018}\natexlab{}.
\newblock \showarticletitle{Unsupervised representation learning by predicting image rotations}.
\newblock \bibinfo{journal}{\emph{arXiv preprint arXiv:1803.07728}} (\bibinfo{year}{2018}).
\newblock


\bibitem[Gifford et~al\mbox{.}(2022)]%
        {rp22}
\bibfield{author}{\bibinfo{person}{Alessandro~T Gifford}, \bibinfo{person}{Kshitij Dwivedi}, \bibinfo{person}{Gemma Roig}, {and} \bibinfo{person}{Radoslaw~M Cichy}.} \bibinfo{year}{2022}\natexlab{}.
\newblock \showarticletitle{A large and rich EEG dataset for modeling human visual object recognition}.
\newblock \bibinfo{journal}{\emph{NeuroImage}}  \bibinfo{volume}{264} (\bibinfo{year}{2022}), \bibinfo{pages}{119754}.
\newblock


\bibitem[Goldberger et~al\mbox{.}(2000)]%
        {rp11}
\bibfield{author}{\bibinfo{person}{Ary~L Goldberger}, \bibinfo{person}{Luis~AN Amaral}, \bibinfo{person}{Leon Glass}, \bibinfo{person}{Jeffrey~M Hausdorff}, \bibinfo{person}{Plamen~Ch Ivanov}, \bibinfo{person}{Roger~G Mark}, \bibinfo{person}{Joseph~E Mietus}, \bibinfo{person}{George~B Moody}, \bibinfo{person}{Chung-Kang Peng}, {and} \bibinfo{person}{H~Eugene Stanley}.} \bibinfo{year}{2000}\natexlab{}.
\newblock \showarticletitle{PhysioBank, PhysioToolkit, and PhysioNet: components of a new research resource for complex physiologic signals}.
\newblock \bibinfo{journal}{\emph{circulation}} \bibinfo{volume}{101}, \bibinfo{number}{23} (\bibinfo{year}{2000}), \bibinfo{pages}{e215--e220}.
\newblock


\bibitem[Goodfellow et~al\mbox{.}(2014)]%
        {re122}
\bibfield{author}{\bibinfo{person}{Ian Goodfellow}, \bibinfo{person}{Jean Pouget-Abadie}, \bibinfo{person}{Mehdi Mirza}, \bibinfo{person}{Bing Xu}, \bibinfo{person}{David Warde-Farley}, \bibinfo{person}{Sherjil Ozair}, \bibinfo{person}{Aaron Courville}, {and} \bibinfo{person}{Yoshua Bengio}.} \bibinfo{year}{2014}\natexlab{}.
\newblock \showarticletitle{Generative adversarial nets}.
\newblock \bibinfo{journal}{\emph{Advances in neural information processing systems}}  \bibinfo{volume}{27} (\bibinfo{year}{2014}).
\newblock


\bibitem[Gotman(1982)]%
        {re13}
\bibfield{author}{\bibinfo{person}{Jean Gotman}.} \bibinfo{year}{1982}\natexlab{}.
\newblock \showarticletitle{Automatic recognition of epileptic seizures in the EEG}.
\newblock \bibinfo{journal}{\emph{Electroencephalography and clinical Neurophysiology}} \bibinfo{volume}{54}, \bibinfo{number}{5} (\bibinfo{year}{1982}), \bibinfo{pages}{530--540}.
\newblock


\bibitem[Gramfort et~al\mbox{.}(2021)]%
        {re96}
\bibfield{author}{\bibinfo{person}{Alexandre Gramfort}, \bibinfo{person}{Hubert Banville}, \bibinfo{person}{Omar Chehab}, \bibinfo{person}{Aapo Hyv{\"a}rinen}, {and} \bibinfo{person}{Denis Engemann}.} \bibinfo{year}{2021}\natexlab{}.
\newblock \showarticletitle{Learning with self-supervision on EEG data}. In \bibinfo{booktitle}{\emph{2021 9th International Winter Conference on Brain-Computer Interface (BCI)}}. IEEE, \bibinfo{pages}{1--2}.
\newblock


\bibitem[Grill et~al\mbox{.}(2020)]%
        {re156}
\bibfield{author}{\bibinfo{person}{Jean-Bastien Grill}, \bibinfo{person}{Florian Strub}, \bibinfo{person}{Florent Altch{\'e}}, \bibinfo{person}{Corentin Tallec}, \bibinfo{person}{Pierre Richemond}, \bibinfo{person}{Elena Buchatskaya}, \bibinfo{person}{Carl Doersch}, \bibinfo{person}{Bernardo Avila~Pires}, \bibinfo{person}{Zhaohan Guo}, \bibinfo{person}{Mohammad Gheshlaghi~Azar}, {et~al\mbox{.}}} \bibinfo{year}{2020}\natexlab{}.
\newblock \showarticletitle{Bootstrap your own latent-a new approach to self-supervised learning}.
\newblock \bibinfo{journal}{\emph{Advances in neural information processing systems}}  \bibinfo{volume}{33} (\bibinfo{year}{2020}), \bibinfo{pages}{21271--21284}.
\newblock


\bibitem[Guo et~al\mbox{.}(2023)]%
        {re149}
\bibfield{author}{\bibinfo{person}{Yunfei Guo}, \bibinfo{person}{Tao Zhang}, {and} \bibinfo{person}{Wu Huang}.} \bibinfo{year}{2023}\natexlab{}.
\newblock \showarticletitle{Emotion recognition based on multi-modal electrophysiology multi-head attention Contrastive Learning}.
\newblock \bibinfo{journal}{\emph{arXiv preprint arXiv:2308.01919}} (\bibinfo{year}{2023}).
\newblock


\bibitem[Hallgarten et~al\mbox{.}(2023)]%
        {re169}
\bibfield{author}{\bibinfo{person}{Philipp Hallgarten}, \bibinfo{person}{David Bethge}, \bibinfo{person}{Ozan {\"O}zdcnizci}, \bibinfo{person}{Tobias Grosse-Puppendahl}, {and} \bibinfo{person}{Enkelejda Kasneci}.} \bibinfo{year}{2023}\natexlab{}.
\newblock \showarticletitle{TS-MoCo: Time-Series Momentum Contrast for Self-Supervised Physiological Representation Learning}. In \bibinfo{booktitle}{\emph{2023 31st European Signal Processing Conference (EUSIPCO)}}. IEEE, \bibinfo{pages}{1030--1034}.
\newblock


\bibitem[Han et~al\mbox{.}(2021)]%
        {re172}
\bibfield{author}{\bibinfo{person}{Jinpei Han}, \bibinfo{person}{Xiao Gu}, {and} \bibinfo{person}{Benny Lo}.} \bibinfo{year}{2021}\natexlab{}.
\newblock \showarticletitle{Semi-supervised contrastive learning for generalizable motor imagery eeg classification}. In \bibinfo{booktitle}{\emph{2021 IEEE 17th International Conference on Wearable and Implantable Body Sensor Networks (BSN)}}. IEEE, \bibinfo{pages}{1--4}.
\newblock


\bibitem[Harpale and Bairagi(2016)]%
        {re111}
\bibfield{author}{\bibinfo{person}{Varsha~K Harpale} {and} \bibinfo{person}{Vinayak~K Bairagi}.} \bibinfo{year}{2016}\natexlab{}.
\newblock \showarticletitle{Time and frequency domain analysis of EEG signals for seizure detection: A review}. In \bibinfo{booktitle}{\emph{2016 International Conference on Microelectronics, Computing and Communications (MicroCom)}}. IEEE, \bibinfo{pages}{1--6}.
\newblock


\bibitem[He et~al\mbox{.}(2022)]%
        {re55}
\bibfield{author}{\bibinfo{person}{Kaiming He}, \bibinfo{person}{Xinlei Chen}, \bibinfo{person}{Saining Xie}, \bibinfo{person}{Yanghao Li}, \bibinfo{person}{Piotr Doll{\'a}r}, {and} \bibinfo{person}{Ross Girshick}.} \bibinfo{year}{2022}\natexlab{}.
\newblock \showarticletitle{Masked autoencoders are scalable vision learners}. In \bibinfo{booktitle}{\emph{Proceedings of the IEEE/CVF conference on computer vision and pattern recognition}}. \bibinfo{pages}{16000--16009}.
\newblock


\bibitem[He et~al\mbox{.}(2020)]%
        {re84}
\bibfield{author}{\bibinfo{person}{Kaiming He}, \bibinfo{person}{Haoqi Fan}, \bibinfo{person}{Yuxin Wu}, \bibinfo{person}{Saining Xie}, {and} \bibinfo{person}{Ross Girshick}.} \bibinfo{year}{2020}\natexlab{}.
\newblock \showarticletitle{Momentum contrast for unsupervised visual representation learning}. In \bibinfo{booktitle}{\emph{Proceedings of the IEEE/CVF conference on computer vision and pattern recognition}}. \bibinfo{pages}{9729--9738}.
\newblock


\bibitem[Henaff(2020)]%
        {re144}
\bibfield{author}{\bibinfo{person}{Olivier Henaff}.} \bibinfo{year}{2020}\natexlab{}.
\newblock \showarticletitle{Data-efficient image recognition with contrastive predictive coding}. In \bibinfo{booktitle}{\emph{International conference on machine learning}}. PMLR, \bibinfo{pages}{4182--4192}.
\newblock


\bibitem[Hinton and Salakhutdinov(2006)]%
        {re106}
\bibfield{author}{\bibinfo{person}{Geoffrey~E Hinton} {and} \bibinfo{person}{Ruslan~R Salakhutdinov}.} \bibinfo{year}{2006}\natexlab{}.
\newblock \showarticletitle{Reducing the dimensionality of data with neural networks}.
\newblock \bibinfo{journal}{\emph{science}} \bibinfo{volume}{313}, \bibinfo{number}{5786} (\bibinfo{year}{2006}), \bibinfo{pages}{504--507}.
\newblock


\bibitem[Hjelm et~al\mbox{.}(2018)]%
        {re82}
\bibfield{author}{\bibinfo{person}{R~Devon Hjelm}, \bibinfo{person}{Alex Fedorov}, \bibinfo{person}{Samuel Lavoie-Marchildon}, \bibinfo{person}{Karan Grewal}, \bibinfo{person}{Phil Bachman}, \bibinfo{person}{Adam Trischler}, {and} \bibinfo{person}{Yoshua Bengio}.} \bibinfo{year}{2018}\natexlab{}.
\newblock \showarticletitle{Learning deep representations by mutual information estimation and maximization}.
\newblock \bibinfo{journal}{\emph{arXiv preprint arXiv:1808.06670}} (\bibinfo{year}{2018}).
\newblock


\bibitem[Ho and Armanfard(2023)]%
        {re118}
\bibfield{author}{\bibinfo{person}{Thi Kieu~Khanh Ho} {and} \bibinfo{person}{Narges Armanfard}.} \bibinfo{year}{2023}\natexlab{}.
\newblock \showarticletitle{Self-supervised learning for anomalous channel detection in EEG graphs: application to seizure analysis}. In \bibinfo{booktitle}{\emph{Proceedings of the AAAI Conference on Artificial Intelligence}}, Vol.~\bibinfo{volume}{37}. \bibinfo{pages}{7866--7874}.
\newblock


\bibitem[Hojjati(2023)]%
        {re170}
\bibfield{author}{\bibinfo{person}{Amirabbas Hojjati}.} \bibinfo{year}{2023}\natexlab{}.
\newblock \emph{\bibinfo{title}{A Multi-View Self-Supervised Approach to Learn Representations of EEG Data for Downstream Prediction Tasks}}.
\newblock \bibinfo{thesistype}{Master's\ thesis}. \bibinfo{school}{NTNU}.
\newblock


\bibitem[Hosseini et~al\mbox{.}(2020)]%
        {re27}
\bibfield{author}{\bibinfo{person}{Mohammad-Parsa Hosseini}, \bibinfo{person}{Amin Hosseini}, {and} \bibinfo{person}{Kiarash Ahi}.} \bibinfo{year}{2020}\natexlab{}.
\newblock \showarticletitle{A review on machine learning for EEG signal processing in bioengineering}.
\newblock \bibinfo{journal}{\emph{IEEE reviews in biomedical engineering}}  \bibinfo{volume}{14} (\bibinfo{year}{2020}), \bibinfo{pages}{204--218}.
\newblock


\bibitem[Huang et~al\mbox{.}(2023)]%
        {re105}
\bibfield{author}{\bibinfo{person}{Baichuan Huang}, \bibinfo{person}{Renato Zanetti}, \bibinfo{person}{Azra Abtahi}, \bibinfo{person}{David Atienza}, {and} \bibinfo{person}{Amir Aminifar}.} \bibinfo{year}{2023}\natexlab{}.
\newblock \showarticletitle{Epilepsynet: Interpretable self-supervised seizure detection for low-power wearable systems}. In \bibinfo{booktitle}{\emph{2023 IEEE 5th International Conference on Artificial Intelligence Circuits and Systems (AICAS)}}. IEEE, \bibinfo{pages}{1--5}.
\newblock


\bibitem[Jackson and Bolger(2014)]%
        {re4}
\bibfield{author}{\bibinfo{person}{Alice~F Jackson} {and} \bibinfo{person}{Donald~J Bolger}.} \bibinfo{year}{2014}\natexlab{}.
\newblock \showarticletitle{The neurophysiological bases of EEG and EEG measurement: A review for the rest of us}.
\newblock \bibinfo{journal}{\emph{Psychophysiology}} \bibinfo{volume}{51}, \bibinfo{number}{11} (\bibinfo{year}{2014}), \bibinfo{pages}{1061--1071}.
\newblock


\bibitem[Jaiswal et~al\mbox{.}(2020)]%
        {re56}
\bibfield{author}{\bibinfo{person}{Ashish Jaiswal}, \bibinfo{person}{Ashwin~Ramesh Babu}, \bibinfo{person}{Mohammad~Zaki Zadeh}, \bibinfo{person}{Debapriya Banerjee}, {and} \bibinfo{person}{Fillia Makedon}.} \bibinfo{year}{2020}\natexlab{}.
\newblock \showarticletitle{A survey on contrastive self-supervised learning}.
\newblock \bibinfo{journal}{\emph{Technologies}} \bibinfo{volume}{9}, \bibinfo{number}{1} (\bibinfo{year}{2020}), \bibinfo{pages}{2}.
\newblock


\bibitem[Jiang et~al\mbox{.}(2021)]%
        {re162}
\bibfield{author}{\bibinfo{person}{Xue Jiang}, \bibinfo{person}{Jianhui Zhao}, \bibinfo{person}{Bo Du}, {and} \bibinfo{person}{Zhiyong Yuan}.} \bibinfo{year}{2021}\natexlab{}.
\newblock \showarticletitle{Self-supervised contrastive learning for EEG-based sleep staging}. In \bibinfo{booktitle}{\emph{2021 International Joint Conference on Neural Networks (IJCNN)}}. IEEE, \bibinfo{pages}{1--8}.
\newblock


\bibitem[Jiao et~al\mbox{.}(2020)]%
        {re134}
\bibfield{author}{\bibinfo{person}{Yingying Jiao}, \bibinfo{person}{Yini Deng}, \bibinfo{person}{Yun Luo}, {and} \bibinfo{person}{Bao-Liang Lu}.} \bibinfo{year}{2020}\natexlab{}.
\newblock \showarticletitle{Driver sleepiness detection from EEG and EOG signals using GAN and LSTM networks}.
\newblock \bibinfo{journal}{\emph{Neurocomputing}}  \bibinfo{volume}{408} (\bibinfo{year}{2020}), \bibinfo{pages}{100--111}.
\newblock


\bibitem[Jo et~al\mbox{.}(2023)]%
        {re91}
\bibfield{author}{\bibinfo{person}{Sangmin Jo}, \bibinfo{person}{Jaehyun Jeon}, \bibinfo{person}{Seungwoo Jeong}, {and} \bibinfo{person}{Heung-Il Suk}.} \bibinfo{year}{2023}\natexlab{}.
\newblock \showarticletitle{Channel-Aware Self-Supervised Learning for EEG-based BCI}. In \bibinfo{booktitle}{\emph{2023 11th International Winter Conference on Brain-Computer Interface (BCI)}}. IEEE, \bibinfo{pages}{1--4}.
\newblock


\bibitem[Kalafatovich et~al\mbox{.}(2020)]%
        {re64}
\bibfield{author}{\bibinfo{person}{Jenifer Kalafatovich}, \bibinfo{person}{Minji Lee}, {and} \bibinfo{person}{Seong-Whan Lee}.} \bibinfo{year}{2020}\natexlab{}.
\newblock \showarticletitle{Decoding visual recognition of objects from eeg signals based on attention-driven convolutional neural network}. In \bibinfo{booktitle}{\emph{2020 IEEE International Conference on Systems, Man, and Cybernetics (SMC)}}. IEEE, \bibinfo{pages}{2985--2990}.
\newblock


\bibitem[Kan et~al\mbox{.}(2023)]%
        {re175}
\bibfield{author}{\bibinfo{person}{Haoning Kan}, \bibinfo{person}{Jiale Yu}, \bibinfo{person}{Jiajin Huang}, \bibinfo{person}{Zihe Liu}, \bibinfo{person}{Heqian Wang}, {and} \bibinfo{person}{Haiyan Zhou}.} \bibinfo{year}{2023}\natexlab{}.
\newblock \showarticletitle{Self-supervised group meiosis contrastive learning for eeg-based emotion recognition}.
\newblock \bibinfo{journal}{\emph{Applied Intelligence}} (\bibinfo{year}{2023}), \bibinfo{pages}{1--19}.
\newblock


\bibitem[Katsigiannis and Ramzan(2017)]%
        {rp27}
\bibfield{author}{\bibinfo{person}{Stamos Katsigiannis} {and} \bibinfo{person}{Naeem Ramzan}.} \bibinfo{year}{2017}\natexlab{}.
\newblock \showarticletitle{DREAMER: A database for emotion recognition through EEG and ECG signals from wireless low-cost off-the-shelf devices}.
\newblock \bibinfo{journal}{\emph{IEEE journal of biomedical and health informatics}} \bibinfo{volume}{22}, \bibinfo{number}{1} (\bibinfo{year}{2017}), \bibinfo{pages}{98--107}.
\newblock


\bibitem[Kemp et~al\mbox{.}(2000)]%
        {rp3}
\bibfield{author}{\bibinfo{person}{Bob Kemp}, \bibinfo{person}{Aeilko~H Zwinderman}, \bibinfo{person}{Bert Tuk}, \bibinfo{person}{Hilbert~AC Kamphuisen}, {and} \bibinfo{person}{Josefien~JL Oberye}.} \bibinfo{year}{2000}\natexlab{}.
\newblock \showarticletitle{Analysis of a sleep-dependent neuronal feedback loop: the slow-wave microcontinuity of the EEG}.
\newblock \bibinfo{journal}{\emph{IEEE Transactions on Biomedical Engineering}} \bibinfo{volume}{47}, \bibinfo{number}{9} (\bibinfo{year}{2000}), \bibinfo{pages}{1185--1194}.
\newblock


\bibitem[Kendall et~al\mbox{.}(2018)]%
        {re177}
\bibfield{author}{\bibinfo{person}{Alex Kendall}, \bibinfo{person}{Yarin Gal}, {and} \bibinfo{person}{Roberto Cipolla}.} \bibinfo{year}{2018}\natexlab{}.
\newblock \showarticletitle{Multi-task learning using uncertainty to weigh losses for scene geometry and semantics}. In \bibinfo{booktitle}{\emph{Proceedings of the IEEE conference on computer vision and pattern recognition}}. \bibinfo{pages}{7482--7491}.
\newblock


\bibitem[Khalighi et~al\mbox{.}(2016)]%
        {rp18}
\bibfield{author}{\bibinfo{person}{Sirvan Khalighi}, \bibinfo{person}{Teresa Sousa}, \bibinfo{person}{Jos{\'e}~Moutinho Santos}, {and} \bibinfo{person}{Urbano Nunes}.} \bibinfo{year}{2016}\natexlab{}.
\newblock \showarticletitle{ISRUC-Sleep: A comprehensive public dataset for sleep researchers}.
\newblock \bibinfo{journal}{\emph{Computer methods and programs in biomedicine}}  \bibinfo{volume}{124} (\bibinfo{year}{2016}), \bibinfo{pages}{180--192}.
\newblock


\bibitem[Khan et~al\mbox{.}(2022)]%
        {rp28}
\bibfield{author}{\bibinfo{person}{Hassan~Aqeel Khan}, \bibinfo{person}{Rahat Ul~Ain}, \bibinfo{person}{Awais~Mehmood Kamboh}, \bibinfo{person}{Hammad~Tanveer Butt}, \bibinfo{person}{Saima Shafait}, \bibinfo{person}{Wasim Alamgir}, \bibinfo{person}{Didier Stricker}, {and} \bibinfo{person}{Faisal Shafait}.} \bibinfo{year}{2022}\natexlab{}.
\newblock \showarticletitle{The NMT scalp EEG dataset: an open-source annotated dataset of healthy and pathological EEG recordings for predictive modeling}.
\newblock \bibinfo{journal}{\emph{Frontiers in neuroscience}}  \bibinfo{volume}{15} (\bibinfo{year}{2022}), \bibinfo{pages}{755817}.
\newblock


\bibitem[Ko and Suk(2022)]%
        {re89}
\bibfield{author}{\bibinfo{person}{Wonjun Ko} {and} \bibinfo{person}{Heung-Il Suk}.} \bibinfo{year}{2022}\natexlab{}.
\newblock \showarticletitle{Eeg-oriented self-supervised learning and cluster-aware adaptation}. In \bibinfo{booktitle}{\emph{Proceedings of the 31st ACM International Conference on Information \& Knowledge Management}}. \bibinfo{pages}{4143--4147}.
\newblock


\bibitem[Koelstra et~al\mbox{.}(2011)]%
        {rp12}
\bibfield{author}{\bibinfo{person}{Sander Koelstra}, \bibinfo{person}{Christian Muhl}, \bibinfo{person}{Mohammad Soleymani}, \bibinfo{person}{Jong-Seok Lee}, \bibinfo{person}{Ashkan Yazdani}, \bibinfo{person}{Touradj Ebrahimi}, \bibinfo{person}{Thierry Pun}, \bibinfo{person}{Anton Nijholt}, {and} \bibinfo{person}{Ioannis Patras}.} \bibinfo{year}{2011}\natexlab{}.
\newblock \showarticletitle{Deap: A database for emotion analysis; using physiological signals}.
\newblock \bibinfo{journal}{\emph{IEEE transactions on affective computing}} \bibinfo{volume}{3}, \bibinfo{number}{1} (\bibinfo{year}{2011}), \bibinfo{pages}{18--31}.
\newblock


\bibitem[Kostas et~al\mbox{.}(2021)]%
        {re104}
\bibfield{author}{\bibinfo{person}{Demetres Kostas}, \bibinfo{person}{Stephane Aroca-Ouellette}, {and} \bibinfo{person}{Frank Rudzicz}.} \bibinfo{year}{2021}\natexlab{}.
\newblock \showarticletitle{BENDR: using transformers and a contrastive self-supervised learning task to learn from massive amounts of EEG data}.
\newblock \bibinfo{journal}{\emph{Frontiers in Human Neuroscience}}  \bibinfo{volume}{15} (\bibinfo{year}{2021}), \bibinfo{pages}{653659}.
\newblock


\bibitem[Kraskov et~al\mbox{.}(2004)]%
        {re80}
\bibfield{author}{\bibinfo{person}{Alexander Kraskov}, \bibinfo{person}{Harald St{\"o}gbauer}, {and} \bibinfo{person}{Peter Grassberger}.} \bibinfo{year}{2004}\natexlab{}.
\newblock \showarticletitle{Estimating mutual information}.
\newblock \bibinfo{journal}{\emph{Physical review E}} \bibinfo{volume}{69}, \bibinfo{number}{6} (\bibinfo{year}{2004}), \bibinfo{pages}{066138}.
\newblock


\bibitem[Kumar et~al\mbox{.}(2022)]%
        {re151}
\bibfield{author}{\bibinfo{person}{Vamsi Kumar}, \bibinfo{person}{Likith Reddy}, \bibinfo{person}{Shivam Kumar~Sharma}, \bibinfo{person}{Kamalaker Dadi}, \bibinfo{person}{Chiranjeevi Yarra}, \bibinfo{person}{Raju~S Bapi}, {and} \bibinfo{person}{Srijithesh Rajendran}.} \bibinfo{year}{2022}\natexlab{}.
\newblock \showarticletitle{mulEEG: a multi-view representation learning on EEG signals}. In \bibinfo{booktitle}{\emph{International Conference on Medical Image Computing and Computer-Assisted Intervention}}. Springer, \bibinfo{pages}{398--407}.
\newblock


\bibitem[Lee et~al\mbox{.}(2022)]%
        {re129}
\bibfield{author}{\bibinfo{person}{Harim Lee}, \bibinfo{person}{Eunseon Seong}, {and} \bibinfo{person}{Dong-Kyu Chae}.} \bibinfo{year}{2022}\natexlab{}.
\newblock \showarticletitle{Self-supervised learning with attention-based latent signal augmentation for sleep staging with limited labeled data}. In \bibinfo{booktitle}{\emph{Proceedings of the Thirty-First International Joint Conference on Artificial Intelligence, IJCAI-22, LD Raedt, Ed. International Joint Conferences on Artificial Intelligence Organization}}, Vol.~\bibinfo{volume}{7}. \bibinfo{pages}{3868--3876}.
\newblock


\bibitem[Lee et~al\mbox{.}(2019)]%
        {rp17}
\bibfield{author}{\bibinfo{person}{Min-Ho Lee}, \bibinfo{person}{O-Yeon Kwon}, \bibinfo{person}{Yong-Jeong Kim}, \bibinfo{person}{Hong-Kyung Kim}, \bibinfo{person}{Young-Eun Lee}, \bibinfo{person}{John Williamson}, \bibinfo{person}{Siamac Fazli}, {and} \bibinfo{person}{Seong-Whan Lee}.} \bibinfo{year}{2019}\natexlab{}.
\newblock \showarticletitle{EEG dataset and OpenBMI toolbox for three BCI paradigms: An investigation into BCI illiteracy}.
\newblock \bibinfo{journal}{\emph{GigaScience}} \bibinfo{volume}{8}, \bibinfo{number}{5} (\bibinfo{year}{2019}), \bibinfo{pages}{giz002}.
\newblock


\bibitem[Lesaja et~al\mbox{.}(2022)]%
        {re131}
\bibfield{author}{\bibinfo{person}{Srdjan Lesaja}, \bibinfo{person}{Morgan Stuart}, \bibinfo{person}{Jerry~J Shih}, \bibinfo{person}{Pedram~Z Soroush}, \bibinfo{person}{Tanja Schultz}, \bibinfo{person}{Milos Manic}, {and} \bibinfo{person}{Dean~J Krusienski}.} \bibinfo{year}{2022}\natexlab{}.
\newblock \showarticletitle{Self-Supervised Learning of Neural Speech Representations From Unlabeled Intracranial Signals}.
\newblock \bibinfo{journal}{\emph{IEEE Access}}  \bibinfo{volume}{10} (\bibinfo{year}{2022}), \bibinfo{pages}{133526--133538}.
\newblock


\bibitem[Li et~al\mbox{.}(2022c)]%
        {re130}
\bibfield{author}{\bibinfo{person}{Rui Li}, \bibinfo{person}{Yiting Wang}, \bibinfo{person}{Wei-Long Zheng}, {and} \bibinfo{person}{Bao-Liang Lu}.} \bibinfo{year}{2022}\natexlab{c}.
\newblock \showarticletitle{A Multi-view Spectral-Spatial-Temporal Masked Autoencoder for Decoding Emotions with Self-supervised Learning}. In \bibinfo{booktitle}{\emph{Proceedings of the 30th ACM International Conference on Multimedia}}. \bibinfo{pages}{6--14}.
\newblock


\bibitem[Li and Metsis(2022)]%
        {re167}
\bibfield{author}{\bibinfo{person}{Xiaomin Li} {and} \bibinfo{person}{Vangelis Metsis}.} \bibinfo{year}{2022}\natexlab{}.
\newblock \showarticletitle{Spp-eegnet: An input-agnostic self-supervised eeg representation model for inter-dataset transfer learning}. In \bibinfo{booktitle}{\emph{International Conference on Computing and Information Technology}}. Springer, \bibinfo{pages}{173--182}.
\newblock


\bibitem[Li et~al\mbox{.}(2022a)]%
        {re85}
\bibfield{author}{\bibinfo{person}{Yang Li}, \bibinfo{person}{Ji Chen}, \bibinfo{person}{Fu Li}, \bibinfo{person}{Boxun Fu}, \bibinfo{person}{Hao Wu}, \bibinfo{person}{Youshuo Ji}, \bibinfo{person}{Yijin Zhou}, \bibinfo{person}{Yi Niu}, \bibinfo{person}{Guangming Shi}, {and} \bibinfo{person}{Wenming Zheng}.} \bibinfo{year}{2022}\natexlab{a}.
\newblock \showarticletitle{GMSS: Graph-based multi-task self-supervised learning for EEG emotion recognition}.
\newblock \bibinfo{journal}{\emph{IEEE Transactions on Affective Computing}} (\bibinfo{year}{2022}).
\newblock


\bibitem[Li et~al\mbox{.}(2022b)]%
        {re97}
\bibfield{author}{\bibinfo{person}{Yamei Li}, \bibinfo{person}{Shengqiong Luo}, \bibinfo{person}{Haibo Zhang}, \bibinfo{person}{Yinkai Zhang}, \bibinfo{person}{Yuan Zhang}, {and} \bibinfo{person}{Benny Lo}.} \bibinfo{year}{2022}\natexlab{b}.
\newblock \showarticletitle{MtCLSS: Multi-Task Contrastive Learning for Semi-Supervised Pediatric Sleep Staging}.
\newblock \bibinfo{journal}{\emph{IEEE Journal of Biomedical and Health Informatics}} (\bibinfo{year}{2022}).
\newblock


\bibitem[Li et~al\mbox{.}(2019)]%
        {re46}
\bibfield{author}{\bibinfo{person}{Yitong Li}, \bibinfo{person}{Michael Murias}, \bibinfo{person}{Samantha Major}, \bibinfo{person}{Geraldine Dawson}, {and} \bibinfo{person}{David Carlson}.} \bibinfo{year}{2019}\natexlab{}.
\newblock \showarticletitle{On target shift in adversarial domain adaptation}. In \bibinfo{booktitle}{\emph{The 22nd International Conference on Artificial Intelligence and Statistics}}. PMLR, \bibinfo{pages}{616--625}.
\newblock


\bibitem[Liu et~al\mbox{.}(2020)]%
        {re135}
\bibfield{author}{\bibinfo{person}{Junxiu Liu}, \bibinfo{person}{Guopei Wu}, \bibinfo{person}{Yuling Luo}, \bibinfo{person}{Senhui Qiu}, \bibinfo{person}{Su Yang}, \bibinfo{person}{Wei Li}, {and} \bibinfo{person}{Yifei Bi}.} \bibinfo{year}{2020}\natexlab{}.
\newblock \showarticletitle{EEG-based emotion classification using a deep neural network and sparse autoencoder}.
\newblock \bibinfo{journal}{\emph{Frontiers in Systems Neuroscience}}  \bibinfo{volume}{14} (\bibinfo{year}{2020}), \bibinfo{pages}{43}.
\newblock


\bibitem[Liu et~al\mbox{.}(2022)]%
        {re74}
\bibfield{author}{\bibinfo{person}{Yixin Liu}, \bibinfo{person}{Ming Jin}, \bibinfo{person}{Shirui Pan}, \bibinfo{person}{Chuan Zhou}, \bibinfo{person}{Yu Zheng}, \bibinfo{person}{Feng Xia}, {and} \bibinfo{person}{S~Yu Philip}.} \bibinfo{year}{2022}\natexlab{}.
\newblock \showarticletitle{Graph self-supervised learning: A survey}.
\newblock \bibinfo{journal}{\emph{IEEE Transactions on Knowledge and Data Engineering}} \bibinfo{volume}{35}, \bibinfo{number}{6} (\bibinfo{year}{2022}), \bibinfo{pages}{5879--5900}.
\newblock


\bibitem[L{\'o}pez et~al\mbox{.}(2017)]%
        {rp2}
\bibfield{author}{\bibinfo{person}{Silvia L{\'o}pez}, \bibinfo{person}{I Obeid}, {and} \bibinfo{person}{J Picone}.} \bibinfo{year}{2017}\natexlab{}.
\newblock \emph{\bibinfo{title}{Automated interpretation of abnormal adult electroencephalograms}}.
\newblock \bibinfo{thesistype}{Ph.\,D. Dissertation}.
\newblock


\bibitem[Miranda-Correa et~al\mbox{.}(2018)]%
        {rp26}
\bibfield{author}{\bibinfo{person}{Juan~Abdon Miranda-Correa}, \bibinfo{person}{Mojtaba~Khomami Abadi}, \bibinfo{person}{Nicu Sebe}, {and} \bibinfo{person}{Ioannis Patras}.} \bibinfo{year}{2018}\natexlab{}.
\newblock \showarticletitle{Amigos: A dataset for affect, personality and mood research on individuals and groups}.
\newblock \bibinfo{journal}{\emph{IEEE Transactions on Affective Computing}} \bibinfo{volume}{12}, \bibinfo{number}{2} (\bibinfo{year}{2018}), \bibinfo{pages}{479--493}.
\newblock


\bibitem[Mirzaei and Ghasemi(2021)]%
        {re137}
\bibfield{author}{\bibinfo{person}{Sayeh Mirzaei} {and} \bibinfo{person}{Parisa Ghasemi}.} \bibinfo{year}{2021}\natexlab{}.
\newblock \showarticletitle{EEG motor imagery classification using dynamic connectivity patterns and convolutional autoencoder}.
\newblock \bibinfo{journal}{\emph{Biomedical Signal Processing and Control}}  \bibinfo{volume}{68} (\bibinfo{year}{2021}), \bibinfo{pages}{102584}.
\newblock


\bibitem[Mohsenvand et~al\mbox{.}(2020)]%
        {re146}
\bibfield{author}{\bibinfo{person}{Mostafa~Neo Mohsenvand}, \bibinfo{person}{Mohammad~Rasool Izadi}, {and} \bibinfo{person}{Pattie Maes}.} \bibinfo{year}{2020}\natexlab{}.
\newblock \showarticletitle{Contrastive representation learning for electroencephalogram classification}. In \bibinfo{booktitle}{\emph{Machine Learning for Health}}. PMLR, \bibinfo{pages}{238--253}.
\newblock


\bibitem[Montero~Quispe et~al\mbox{.}(2022)]%
        {rr1}
\bibfield{author}{\bibinfo{person}{Kevin~G Montero~Quispe}, \bibinfo{person}{Daniel~MS Utyiama}, \bibinfo{person}{Eulanda~M Dos~Santos}, \bibinfo{person}{Hor{\'a}cio~ABF Oliveira}, {and} \bibinfo{person}{Eduardo~JP Souto}.} \bibinfo{year}{2022}\natexlab{}.
\newblock \showarticletitle{Applying self-supervised representation learning for emotion recognition using physiological signals}.
\newblock \bibinfo{journal}{\emph{Sensors}} \bibinfo{volume}{22}, \bibinfo{number}{23} (\bibinfo{year}{2022}), \bibinfo{pages}{9102}.
\newblock


\bibitem[Noroozi and Favaro(2016)]%
        {re53}
\bibfield{author}{\bibinfo{person}{Mehdi Noroozi} {and} \bibinfo{person}{Paolo Favaro}.} \bibinfo{year}{2016}\natexlab{}.
\newblock \showarticletitle{Unsupervised learning of visual representations by solving jigsaw puzzles}. In \bibinfo{booktitle}{\emph{European conference on computer vision}}. Springer, \bibinfo{pages}{69--84}.
\newblock


\bibitem[Oh et~al\mbox{.}(2014)]%
        {re69}
\bibfield{author}{\bibinfo{person}{Seung-Hyeon Oh}, \bibinfo{person}{Yu-Ri Lee}, {and} \bibinfo{person}{Hyoung-Nam Kim}.} \bibinfo{year}{2014}\natexlab{}.
\newblock \showarticletitle{A novel EEG feature extraction method using Hjorth parameter}.
\newblock \bibinfo{journal}{\emph{International Journal of Electronics and Electrical Engineering}} \bibinfo{volume}{2}, \bibinfo{number}{2} (\bibinfo{year}{2014}), \bibinfo{pages}{106--110}.
\newblock


\bibitem[Oord et~al\mbox{.}(2018)]%
        {re143}
\bibfield{author}{\bibinfo{person}{Aaron van~den Oord}, \bibinfo{person}{Yazhe Li}, {and} \bibinfo{person}{Oriol Vinyals}.} \bibinfo{year}{2018}\natexlab{}.
\newblock \showarticletitle{Representation learning with contrastive predictive coding}.
\newblock \bibinfo{journal}{\emph{arXiv preprint arXiv:1807.03748}} (\bibinfo{year}{2018}).
\newblock


\bibitem[O'reilly et~al\mbox{.}(2014)]%
        {rp4}
\bibfield{author}{\bibinfo{person}{Christian O'reilly}, \bibinfo{person}{Nadia Gosselin}, \bibinfo{person}{Julie Carrier}, {and} \bibinfo{person}{Tore Nielsen}.} \bibinfo{year}{2014}\natexlab{}.
\newblock \showarticletitle{Montreal Archive of Sleep Studies: an open-access resource for instrument benchmarking and exploratory research}.
\newblock \bibinfo{journal}{\emph{Journal of sleep research}} \bibinfo{volume}{23}, \bibinfo{number}{6} (\bibinfo{year}{2014}), \bibinfo{pages}{628--635}.
\newblock


\bibitem[Ou et~al\mbox{.}(2022)]%
        {re88}
\bibfield{author}{\bibinfo{person}{Yanghan Ou}, \bibinfo{person}{Siqin Sun}, \bibinfo{person}{Haitao Gan}, \bibinfo{person}{Ran Zhou}, {and} \bibinfo{person}{Zhi Yang}.} \bibinfo{year}{2022}\natexlab{}.
\newblock \showarticletitle{An improved self-supervised learning for EEG classification}.
\newblock \bibinfo{journal}{\emph{Math. Biosci. Eng}}  \bibinfo{volume}{19} (\bibinfo{year}{2022}), \bibinfo{pages}{6907--6922}.
\newblock


\bibitem[Palo et~al\mbox{.}(2015)]%
        {re29}
\bibfield{author}{\bibinfo{person}{HK Palo}, \bibinfo{person}{Mihir~Narayana Mohanty}, {and} \bibinfo{person}{Mahesh Chandra}.} \bibinfo{year}{2015}\natexlab{}.
\newblock \showarticletitle{Use of different features for emotion recognition using MLP network}. In \bibinfo{booktitle}{\emph{Computational Vision and Robotics: Proceedings of ICCVR 2014}}. Springer, \bibinfo{pages}{7--15}.
\newblock


\bibitem[Partovi et~al\mbox{.}(2023)]%
        {re95}
\bibfield{author}{\bibinfo{person}{Andi Partovi}, \bibinfo{person}{Anthony~N Burkitt}, {and} \bibinfo{person}{David Grayden}.} \bibinfo{year}{2023}\natexlab{}.
\newblock \showarticletitle{A Self-Supervised Task-Agnostic Embedding for EEG Signals}. In \bibinfo{booktitle}{\emph{2023 11th International IEEE/EMBS Conference on Neural Engineering (NER)}}. IEEE, \bibinfo{pages}{1--4}.
\newblock


\bibitem[Peng et~al\mbox{.}(2022)]%
        {rp29}
\bibfield{author}{\bibinfo{person}{Ruimin Peng}, \bibinfo{person}{Changming Zhao}, \bibinfo{person}{Jun Jiang}, \bibinfo{person}{Guangtao Kuang}, \bibinfo{person}{Yuqi Cui}, \bibinfo{person}{Yifan Xu}, \bibinfo{person}{Hao Du}, \bibinfo{person}{Jianbo Shao}, {and} \bibinfo{person}{Dongrui Wu}.} \bibinfo{year}{2022}\natexlab{}.
\newblock \showarticletitle{TIE-EEGNet: Temporal information enhanced EEGNet for seizure subtype classification}.
\newblock \bibinfo{journal}{\emph{IEEE Transactions on Neural Systems and Rehabilitation Engineering}}  \bibinfo{volume}{30} (\bibinfo{year}{2022}), \bibinfo{pages}{2567--2576}.
\newblock


\bibitem[Peng et~al\mbox{.}(2023)]%
        {re132}
\bibfield{author}{\bibinfo{person}{Ruimin Peng}, \bibinfo{person}{Changming Zhao}, \bibinfo{person}{Yifan Xu}, \bibinfo{person}{Jun Jiang}, \bibinfo{person}{Guangtao Kuang}, \bibinfo{person}{Jianbo Shao}, {and} \bibinfo{person}{Dongrui Wu}.} \bibinfo{year}{2023}\natexlab{}.
\newblock \showarticletitle{WAVELET2VEC: A Filter Bank Masked Autoencoder for EEG-Based Seizure Subtype Classification}. In \bibinfo{booktitle}{\emph{ICASSP 2023-2023 IEEE International Conference on Acoustics, Speech and Signal Processing (ICASSP)}}. IEEE, \bibinfo{pages}{1--5}.
\newblock


\bibitem[Petrantonakis and Hadjileontiadis(2009)]%
        {re70}
\bibfield{author}{\bibinfo{person}{Panagiotis~C Petrantonakis} {and} \bibinfo{person}{Leontios~J Hadjileontiadis}.} \bibinfo{year}{2009}\natexlab{}.
\newblock \showarticletitle{Emotion recognition from EEG using higher order crossings}.
\newblock \bibinfo{journal}{\emph{IEEE Transactions on information Technology in Biomedicine}} \bibinfo{volume}{14}, \bibinfo{number}{2} (\bibinfo{year}{2009}), \bibinfo{pages}{186--197}.
\newblock


\bibitem[PhysioBank(2000)]%
        {rp10}
\bibfield{author}{\bibinfo{person}{PhysioToolkit PhysioBank}.} \bibinfo{year}{2000}\natexlab{}.
\newblock \showarticletitle{Physionet: components of a new research resource for complex physiologic signals}.
\newblock \bibinfo{journal}{\emph{Circulation}} \bibinfo{volume}{101}, \bibinfo{number}{23} (\bibinfo{year}{2000}), \bibinfo{pages}{e215--e220}.
\newblock


\bibitem[Pulver et~al\mbox{.}(2023)]%
        {re128}
\bibfield{author}{\bibinfo{person}{Dustin Pulver}, \bibinfo{person}{Prithila Angkan}, \bibinfo{person}{Paul Hungler}, {and} \bibinfo{person}{Ali Etemad}.} \bibinfo{year}{2023}\natexlab{}.
\newblock \showarticletitle{EEG-based Cognitive Load Classification using Feature Masked Autoencoding and Emotion Transfer Learning}. In \bibinfo{booktitle}{\emph{Proceedings of the 25th International Conference on Multimodal Interaction}}. \bibinfo{pages}{190--197}.
\newblock


\bibitem[Quan et~al\mbox{.}(1997)]%
        {rp8}
\bibfield{author}{\bibinfo{person}{Stuart~F Quan}, \bibinfo{person}{Barbara~V Howard}, \bibinfo{person}{Conrad Iber}, \bibinfo{person}{James~P Kiley}, \bibinfo{person}{F~Javier Nieto}, \bibinfo{person}{George~T O'Connor}, \bibinfo{person}{David~M Rapoport}, \bibinfo{person}{Susan Redline}, \bibinfo{person}{John Robbins}, \bibinfo{person}{Jonathan~M Samet}, {et~al\mbox{.}}} \bibinfo{year}{1997}\natexlab{}.
\newblock \showarticletitle{The sleep heart health study: design, rationale, and methods}.
\newblock \bibinfo{journal}{\emph{Sleep}} \bibinfo{volume}{20}, \bibinfo{number}{12} (\bibinfo{year}{1997}), \bibinfo{pages}{1077--1085}.
\newblock


\bibitem[Radford et~al\mbox{.}(2021)]%
        {re158}
\bibfield{author}{\bibinfo{person}{Alec Radford}, \bibinfo{person}{Jong~Wook Kim}, \bibinfo{person}{Chris Hallacy}, \bibinfo{person}{Aditya Ramesh}, \bibinfo{person}{Gabriel Goh}, \bibinfo{person}{Sandhini Agarwal}, \bibinfo{person}{Girish Sastry}, \bibinfo{person}{Amanda Askell}, \bibinfo{person}{Pamela Mishkin}, \bibinfo{person}{Jack Clark}, {et~al\mbox{.}}} \bibinfo{year}{2021}\natexlab{}.
\newblock \showarticletitle{Learning transferable visual models from natural language supervision}. In \bibinfo{booktitle}{\emph{International conference on machine learning}}. PMLR, \bibinfo{pages}{8748--8763}.
\newblock


\bibitem[Rafiei et~al\mbox{.}(2022)]%
        {re40}
\bibfield{author}{\bibinfo{person}{Mohammad~H Rafiei}, \bibinfo{person}{Lynne~V Gauthier}, \bibinfo{person}{Hojjat Adeli}, {and} \bibinfo{person}{Daniel Takabi}.} \bibinfo{year}{2022}\natexlab{}.
\newblock \showarticletitle{Self-supervised learning for electroencephalography}.
\newblock \bibinfo{journal}{\emph{IEEE Transactions on Neural Networks and Learning Systems}} (\bibinfo{year}{2022}).
\newblock


\bibitem[Roach and Mathalon(2008)]%
        {re108}
\bibfield{author}{\bibinfo{person}{Brian~J Roach} {and} \bibinfo{person}{Daniel~H Mathalon}.} \bibinfo{year}{2008}\natexlab{}.
\newblock \showarticletitle{Event-related EEG time-frequency analysis: an overview of measures and an analysis of early gamma band phase locking in schizophrenia}.
\newblock \bibinfo{journal}{\emph{Schizophrenia bulletin}} \bibinfo{volume}{34}, \bibinfo{number}{5} (\bibinfo{year}{2008}), \bibinfo{pages}{907--926}.
\newblock


\bibitem[Rodenbeck et~al\mbox{.}(2006)]%
        {re17}
\bibfield{author}{\bibinfo{person}{Andrea Rodenbeck}, \bibinfo{person}{Ralf Binder}, \bibinfo{person}{Peter Geisler}, \bibinfo{person}{Heidi Danker-Hopfe}, \bibinfo{person}{Reimer Lund}, \bibinfo{person}{Friedhart Raschke}, \bibinfo{person}{Hans-G{\"u}nther Wee{\ss}}, {and} \bibinfo{person}{Hartmut Schulz}.} \bibinfo{year}{2006}\natexlab{}.
\newblock \showarticletitle{A review of sleep EEG patterns. Part I: A compilation of amended rules for their visual recognition according to Rechtschaffen and Kales}.
\newblock \bibinfo{journal}{\emph{Somnologie}} \bibinfo{volume}{10}, \bibinfo{number}{4} (\bibinfo{year}{2006}), \bibinfo{pages}{159--175}.
\newblock


\bibitem[Sabbagh et~al\mbox{.}(2020)]%
        {re22}
\bibfield{author}{\bibinfo{person}{David Sabbagh}, \bibinfo{person}{Pierre Ablin}, \bibinfo{person}{Ga{\"e}l Varoquaux}, \bibinfo{person}{Alexandre Gramfort}, {and} \bibinfo{person}{Denis~A Engemann}.} \bibinfo{year}{2020}\natexlab{}.
\newblock \showarticletitle{Predictive regression modeling with MEG/EEG: from source power to signals and cognitive states}.
\newblock \bibinfo{journal}{\emph{NeuroImage}}  \bibinfo{volume}{222} (\bibinfo{year}{2020}), \bibinfo{pages}{116893}.
\newblock


\bibitem[Schalk et~al\mbox{.}(2004)]%
        {rp5}
\bibfield{author}{\bibinfo{person}{Gerwin Schalk}, \bibinfo{person}{Dennis~J McFarland}, \bibinfo{person}{Thilo Hinterberger}, \bibinfo{person}{Niels Birbaumer}, {and} \bibinfo{person}{Jonathan~R Wolpaw}.} \bibinfo{year}{2004}\natexlab{}.
\newblock \showarticletitle{BCI2000: a general-purpose brain-computer interface (BCI) system}.
\newblock \bibinfo{journal}{\emph{IEEE Transactions on biomedical engineering}} \bibinfo{volume}{51}, \bibinfo{number}{6} (\bibinfo{year}{2004}), \bibinfo{pages}{1034--1043}.
\newblock


\bibitem[Schneider et~al\mbox{.}(2019)]%
        {re102}
\bibfield{author}{\bibinfo{person}{Steffen Schneider}, \bibinfo{person}{Alexei Baevski}, \bibinfo{person}{Ronan Collobert}, {and} \bibinfo{person}{Michael Auli}.} \bibinfo{year}{2019}\natexlab{}.
\newblock \showarticletitle{wav2vec: Unsupervised pre-training for speech recognition}.
\newblock \bibinfo{journal}{\emph{arXiv preprint arXiv:1904.05862}} (\bibinfo{year}{2019}).
\newblock


\bibitem[Schroff et~al\mbox{.}(2015)]%
        {re142}
\bibfield{author}{\bibinfo{person}{Florian Schroff}, \bibinfo{person}{Dmitry Kalenichenko}, {and} \bibinfo{person}{James Philbin}.} \bibinfo{year}{2015}\natexlab{}.
\newblock \showarticletitle{Facenet: A unified embedding for face recognition and clustering}. In \bibinfo{booktitle}{\emph{Proceedings of the IEEE conference on computer vision and pattern recognition}}. \bibinfo{pages}{815--823}.
\newblock


\bibitem[Shah et~al\mbox{.}(2018)]%
        {rp13}
\bibfield{author}{\bibinfo{person}{Vinit Shah}, \bibinfo{person}{Eva Von~Weltin}, \bibinfo{person}{Silvia Lopez}, \bibinfo{person}{James~Riley McHugh}, \bibinfo{person}{Lillian Veloso}, \bibinfo{person}{Meysam Golmohammadi}, \bibinfo{person}{Iyad Obeid}, {and} \bibinfo{person}{Joseph Picone}.} \bibinfo{year}{2018}\natexlab{}.
\newblock \showarticletitle{The temple university hospital seizure detection corpus}.
\newblock \bibinfo{journal}{\emph{Frontiers in neuroinformatics}}  \bibinfo{volume}{12} (\bibinfo{year}{2018}), \bibinfo{pages}{83}.
\newblock


\bibitem[Shen et~al\mbox{.}(2022)]%
        {re161}
\bibfield{author}{\bibinfo{person}{Xinke Shen}, \bibinfo{person}{Xianggen Liu}, \bibinfo{person}{Xin Hu}, \bibinfo{person}{Dan Zhang}, {and} \bibinfo{person}{Sen Song}.} \bibinfo{year}{2022}\natexlab{}.
\newblock \showarticletitle{Contrastive learning of subject-invariant eeg representations for cross-subject emotion recognition}.
\newblock \bibinfo{journal}{\emph{IEEE Transactions on Affective Computing}} (\bibinfo{year}{2022}).
\newblock


\bibitem[Shoeb(2009)]%
        {rp20}
\bibfield{author}{\bibinfo{person}{Ali~Hossam Shoeb}.} \bibinfo{year}{2009}\natexlab{}.
\newblock \emph{\bibinfo{title}{Application of machine learning to epileptic seizure onset detection and treatment}}.
\newblock \bibinfo{thesistype}{Ph.\,D. Dissertation}. \bibinfo{school}{Massachusetts Institute of Technology}.
\newblock


\bibitem[Shoeibi et~al\mbox{.}(2021)]%
        {re68}
\bibfield{author}{\bibinfo{person}{Afshin Shoeibi}, \bibinfo{person}{Navid Ghassemi}, \bibinfo{person}{Roohallah Alizadehsani}, \bibinfo{person}{Modjtaba Rouhani}, \bibinfo{person}{Hossein Hosseini-Nejad}, \bibinfo{person}{Abbas Khosravi}, \bibinfo{person}{Maryam Panahiazar}, {and} \bibinfo{person}{Saeid Nahavandi}.} \bibinfo{year}{2021}\natexlab{}.
\newblock \showarticletitle{A comprehensive comparison of handcrafted features and convolutional autoencoders for epileptic seizures detection in EEG signals}.
\newblock \bibinfo{journal}{\emph{Expert Systems with Applications}}  \bibinfo{volume}{163} (\bibinfo{year}{2021}), \bibinfo{pages}{113788}.
\newblock


\bibitem[Singh and Malhotra(2022)]%
        {re110}
\bibfield{author}{\bibinfo{person}{Kuldeep Singh} {and} \bibinfo{person}{Jyoteesh Malhotra}.} \bibinfo{year}{2022}\natexlab{}.
\newblock \showarticletitle{Smart neurocare approach for detection of epileptic seizures using deep learning based temporal analysis of EEG patterns}.
\newblock \bibinfo{journal}{\emph{Multimedia Tools and Applications}} \bibinfo{volume}{81}, \bibinfo{number}{20} (\bibinfo{year}{2022}), \bibinfo{pages}{29555--29586}.
\newblock


\bibitem[Siuly et~al\mbox{.}(2016)]%
        {re21}
\bibfield{author}{\bibinfo{person}{Siuly Siuly}, \bibinfo{person}{Yan Li}, {and} \bibinfo{person}{Yanchun Zhang}.} \bibinfo{year}{2016}\natexlab{}.
\newblock \showarticletitle{EEG signal analysis and classification}.
\newblock \bibinfo{journal}{\emph{IEEE Trans Neural Syst Rehabilit Eng}}  \bibinfo{volume}{11} (\bibinfo{year}{2016}), \bibinfo{pages}{141--144}.
\newblock


\bibitem[Soleymani et~al\mbox{.}(2011)]%
        {rp23}
\bibfield{author}{\bibinfo{person}{Mohammad Soleymani}, \bibinfo{person}{Jeroen Lichtenauer}, \bibinfo{person}{Thierry Pun}, {and} \bibinfo{person}{Maja Pantic}.} \bibinfo{year}{2011}\natexlab{}.
\newblock \showarticletitle{A multimodal database for affect recognition and implicit tagging}.
\newblock \bibinfo{journal}{\emph{IEEE transactions on affective computing}} \bibinfo{volume}{3}, \bibinfo{number}{1} (\bibinfo{year}{2011}), \bibinfo{pages}{42--55}.
\newblock


\bibitem[Song et~al\mbox{.}(2019)]%
        {rp16}
\bibfield{author}{\bibinfo{person}{Tengfei Song}, \bibinfo{person}{Wenming Zheng}, \bibinfo{person}{Cheng Lu}, \bibinfo{person}{Yuan Zong}, \bibinfo{person}{Xilei Zhang}, {and} \bibinfo{person}{Zhen Cui}.} \bibinfo{year}{2019}\natexlab{}.
\newblock \showarticletitle{MPED: A multi-modal physiological emotion database for discrete emotion recognition}.
\newblock \bibinfo{journal}{\emph{IEEE Access}}  \bibinfo{volume}{7} (\bibinfo{year}{2019}), \bibinfo{pages}{12177--12191}.
\newblock


\bibitem[Song et~al\mbox{.}(2023)]%
        {re157}
\bibfield{author}{\bibinfo{person}{Yonghao Song}, \bibinfo{person}{Bingchuan Liu}, \bibinfo{person}{Xiang Li}, \bibinfo{person}{Nanlin Shi}, \bibinfo{person}{Yijun Wang}, {and} \bibinfo{person}{Xiaorong Gao}.} \bibinfo{year}{2023}\natexlab{}.
\newblock \showarticletitle{Decoding Natural Images from EEG for Object Recognition}.
\newblock \bibinfo{journal}{\emph{arXiv preprint arXiv:2308.13234}} (\bibinfo{year}{2023}).
\newblock


\bibitem[Tangermann et~al\mbox{.}(2012)]%
        {rp6}
\bibfield{author}{\bibinfo{person}{Michael Tangermann}, \bibinfo{person}{Klaus-Robert M{\"u}ller}, \bibinfo{person}{Ad Aertsen}, \bibinfo{person}{Niels Birbaumer}, \bibinfo{person}{Christoph Braun}, \bibinfo{person}{Clemens Brunner}, \bibinfo{person}{Robert Leeb}, \bibinfo{person}{Carsten Mehring}, \bibinfo{person}{Kai~J Miller}, \bibinfo{person}{Gernot Mueller-Putz}, {et~al\mbox{.}}} \bibinfo{year}{2012}\natexlab{}.
\newblock \showarticletitle{Review of the BCI competition IV}.
\newblock \bibinfo{journal}{\emph{Frontiers in neuroscience}} (\bibinfo{year}{2012}), \bibinfo{pages}{55}.
\newblock


\bibitem[Temko et~al\mbox{.}(2015)]%
        {rp7}
\bibfield{author}{\bibinfo{person}{Andriy Temko}, \bibinfo{person}{Achintya Sarkar}, {and} \bibinfo{person}{Gordon Lightbody}.} \bibinfo{year}{2015}\natexlab{}.
\newblock \showarticletitle{Detection of seizures in intracranial EEG: UPenn and Mayo Clinic's seizure detection challenge}. In \bibinfo{booktitle}{\emph{2015 37th Annual International Conference of the IEEE Engineering in Medicine and Biology Society (EMBC)}}. IEEE, \bibinfo{pages}{6582--6585}.
\newblock


\bibitem[Teplan et~al\mbox{.}(2002)]%
        {re1}
\bibfield{author}{\bibinfo{person}{Michal Teplan} {et~al\mbox{.}}} \bibinfo{year}{2002}\natexlab{}.
\newblock \showarticletitle{Fundamentals of EEG measurement}.
\newblock \bibinfo{journal}{\emph{Measurement science review}} \bibinfo{volume}{2}, \bibinfo{number}{2} (\bibinfo{year}{2002}), \bibinfo{pages}{1--11}.
\newblock


\bibitem[Tishby et~al\mbox{.}(2000)]%
        {re81}
\bibfield{author}{\bibinfo{person}{Naftali Tishby}, \bibinfo{person}{Fernando~C Pereira}, {and} \bibinfo{person}{William Bialek}.} \bibinfo{year}{2000}\natexlab{}.
\newblock \showarticletitle{The information bottleneck method}.
\newblock \bibinfo{journal}{\emph{arXiv preprint physics/0004057}} (\bibinfo{year}{2000}).
\newblock


\bibitem[{\"U}beyli(2009)]%
        {re71}
\bibfield{author}{\bibinfo{person}{Elif~Derya {\"U}beyli}.} \bibinfo{year}{2009}\natexlab{}.
\newblock \showarticletitle{Statistics over features: EEG signals analysis}.
\newblock \bibinfo{journal}{\emph{Computers in Biology and Medicine}} \bibinfo{volume}{39}, \bibinfo{number}{8} (\bibinfo{year}{2009}), \bibinfo{pages}{733--741}.
\newblock


\bibitem[Vaswani et~al\mbox{.}(2017)]%
        {re34}
\bibfield{author}{\bibinfo{person}{Ashish Vaswani}, \bibinfo{person}{Noam Shazeer}, \bibinfo{person}{Niki Parmar}, \bibinfo{person}{Jakob Uszkoreit}, \bibinfo{person}{Llion Jones}, \bibinfo{person}{Aidan~N Gomez}, \bibinfo{person}{{\L}ukasz Kaiser}, {and} \bibinfo{person}{Illia Polosukhin}.} \bibinfo{year}{2017}\natexlab{}.
\newblock \showarticletitle{Attention is all you need}.
\newblock \bibinfo{journal}{\emph{Advances in neural information processing systems}}  \bibinfo{volume}{30} (\bibinfo{year}{2017}).
\newblock


\bibitem[Wagh et~al\mbox{.}(2021)]%
        {re160}
\bibfield{author}{\bibinfo{person}{Neeraj Wagh}, \bibinfo{person}{Jionghao Wei}, \bibinfo{person}{Samarth Rawal}, \bibinfo{person}{Brent Berry}, \bibinfo{person}{Leland Barnard}, \bibinfo{person}{Benjamin Brinkmann}, \bibinfo{person}{Gregory Worrell}, \bibinfo{person}{David Jones}, {and} \bibinfo{person}{Yogatheesan Varatharajah}.} \bibinfo{year}{2021}\natexlab{}.
\newblock \showarticletitle{Domain-guided self-supervision of eeg data improves downstream classification performance and generalizability}. In \bibinfo{booktitle}{\emph{Machine Learning for Health}}. PMLR, \bibinfo{pages}{130--142}.
\newblock


\bibitem[Wang et~al\mbox{.}(2023)]%
        {re92}
\bibfield{author}{\bibinfo{person}{Xingyi Wang}, \bibinfo{person}{Yuliang Ma}, \bibinfo{person}{Jared Cammon}, \bibinfo{person}{Feng Fang}, \bibinfo{person}{Yunyuan Gao}, {and} \bibinfo{person}{Yingchun Zhang}.} \bibinfo{year}{2023}\natexlab{}.
\newblock \showarticletitle{Self-Supervised EEG Emotion Recognition Models Based on CNN}.
\newblock \bibinfo{journal}{\emph{IEEE Transactions on Neural Systems and Rehabilitation Engineering}}  \bibinfo{volume}{31} (\bibinfo{year}{2023}), \bibinfo{pages}{1952--1962}.
\newblock


\bibitem[Wang and Qi(2022)]%
        {re79}
\bibfield{author}{\bibinfo{person}{Xiao Wang} {and} \bibinfo{person}{Guo-Jun Qi}.} \bibinfo{year}{2022}\natexlab{}.
\newblock \showarticletitle{Contrastive learning with stronger augmentations}.
\newblock \bibinfo{journal}{\emph{IEEE transactions on pattern analysis and machine intelligence}} \bibinfo{volume}{45}, \bibinfo{number}{5} (\bibinfo{year}{2022}), \bibinfo{pages}{5549--5560}.
\newblock


\bibitem[Wen and Zhang(2018)]%
        {re136}
\bibfield{author}{\bibinfo{person}{Tingxi Wen} {and} \bibinfo{person}{Zhongnan Zhang}.} \bibinfo{year}{2018}\natexlab{}.
\newblock \showarticletitle{Deep convolution neural network and autoencoders-based unsupervised feature learning of EEG signals}.
\newblock \bibinfo{journal}{\emph{IEEE Access}}  \bibinfo{volume}{6} (\bibinfo{year}{2018}), \bibinfo{pages}{25399--25410}.
\newblock


\bibitem[Weng et~al\mbox{.}(2022)]%
        {re60}
\bibfield{author}{\bibinfo{person}{Weining Weng}, \bibinfo{person}{Yang Gu}, \bibinfo{person}{Yiqiang Chen}, \bibinfo{person}{Guoqiang Wang}, {and} \bibinfo{person}{Nianfeng Shi}.} \bibinfo{year}{2022}\natexlab{}.
\newblock \showarticletitle{An Efficient Spatial-Temporal Representation Method for EEG Emotion Recognition}. In \bibinfo{booktitle}{\emph{2022 IEEE Smartworld, Ubiquitous Intelligence \& Computing, Scalable Computing \& Communications, Digital Twin, Privacy Computing, Metaverse, Autonomous \& Trusted Vehicles (SmartWorld/UIC/ScalCom/DigitalTwin/PriComp/Meta)}}. IEEE, \bibinfo{pages}{458--467}.
\newblock


\bibitem[Weng et~al\mbox{.}(2023)]%
        {re150}
\bibfield{author}{\bibinfo{person}{Weining Weng}, \bibinfo{person}{Yang Gu}, \bibinfo{person}{Qihui Zhang}, \bibinfo{person}{Yingying Huang}, \bibinfo{person}{Chunyan Miao}, {and} \bibinfo{person}{Yiqiang Chen}.} \bibinfo{year}{2023}\natexlab{}.
\newblock \showarticletitle{A Knowledge-Driven Cross-view Contrastive Learning for EEG Representation}.
\newblock \bibinfo{journal}{\emph{arXiv preprint arXiv:2310.03747}} (\bibinfo{year}{2023}).
\newblock


\bibitem[Wu et~al\mbox{.}(2022)]%
        {re116}
\bibfield{author}{\bibinfo{person}{Di Wu}, \bibinfo{person}{Siyuan Li}, \bibinfo{person}{Jie Yang}, {and} \bibinfo{person}{Mohamad Sawan}.} \bibinfo{year}{2022}\natexlab{}.
\newblock \showarticletitle{neuro2vec: Masked fourier spectrum prediction for neurophysiological representation learning}.
\newblock \bibinfo{journal}{\emph{arXiv preprint arXiv:2204.12440}} (\bibinfo{year}{2022}).
\newblock


\bibitem[Wu et~al\mbox{.}(2020)]%
        {re119}
\bibfield{author}{\bibinfo{person}{Zonghan Wu}, \bibinfo{person}{Shirui Pan}, \bibinfo{person}{Fengwen Chen}, \bibinfo{person}{Guodong Long}, \bibinfo{person}{Chengqi Zhang}, {and} \bibinfo{person}{S~Yu Philip}.} \bibinfo{year}{2020}\natexlab{}.
\newblock \showarticletitle{A comprehensive survey on graph neural networks}.
\newblock \bibinfo{journal}{\emph{IEEE transactions on neural networks and learning systems}} \bibinfo{volume}{32}, \bibinfo{number}{1} (\bibinfo{year}{2020}), \bibinfo{pages}{4--24}.
\newblock


\bibitem[Xi et~al\mbox{.}(2022)]%
        {rr2}
\bibfield{author}{\bibinfo{person}{Liang Xi}, \bibinfo{person}{Zichao Yun}, \bibinfo{person}{Han Liu}, \bibinfo{person}{Ruidong Wang}, \bibinfo{person}{Xunhua Huang}, {and} \bibinfo{person}{Haoyi Fan}.} \bibinfo{year}{2022}\natexlab{}.
\newblock \showarticletitle{Semi-supervised time series classification model with self-supervised learning}.
\newblock \bibinfo{journal}{\emph{Engineering Applications of Artificial Intelligence}}  \bibinfo{volume}{116} (\bibinfo{year}{2022}), \bibinfo{pages}{105331}.
\newblock


\bibitem[Xiao et~al\mbox{.}(2021)]%
        {re163}
\bibfield{author}{\bibinfo{person}{Qinfeng Xiao}, \bibinfo{person}{Jing Wang}, \bibinfo{person}{Jianan Ye}, \bibinfo{person}{Hongjun Zhang}, \bibinfo{person}{Yuyan Bu}, \bibinfo{person}{Yiqiong Zhang}, {and} \bibinfo{person}{Hao Wu}.} \bibinfo{year}{2021}\natexlab{}.
\newblock \showarticletitle{Self-supervised learning for sleep stage classification with predictive and discriminative contrastive coding}. In \bibinfo{booktitle}{\emph{ICASSP 2021-2021 IEEE International Conference on Acoustics, Speech and Signal Processing (ICASSP)}}. IEEE, \bibinfo{pages}{1290--1294}.
\newblock


\bibitem[Xiao et~al\mbox{.}(2024)]%
        {re166}
\bibfield{author}{\bibinfo{person}{Tiantian Xiao}, \bibinfo{person}{Ziwei Wang}, \bibinfo{person}{Yongfeng Zhang}, \bibinfo{person}{Shuai Wang}, \bibinfo{person}{Hailing Feng}, \bibinfo{person}{Yanna Zhao}, {et~al\mbox{.}}} \bibinfo{year}{2024}\natexlab{}.
\newblock \showarticletitle{Self-supervised Learning with Attention Mechanism for EEG-based seizure detection}.
\newblock \bibinfo{journal}{\emph{Biomedical Signal Processing and Control}}  \bibinfo{volume}{87} (\bibinfo{year}{2024}), \bibinfo{pages}{105464}.
\newblock


\bibitem[Xu et~al\mbox{.}(2020)]%
        {re94}
\bibfield{author}{\bibinfo{person}{Junjie Xu}, \bibinfo{person}{Yaojia Zheng}, \bibinfo{person}{Yifan Mao}, \bibinfo{person}{Ruixuan Wang}, {and} \bibinfo{person}{Wei-Shi Zheng}.} \bibinfo{year}{2020}\natexlab{}.
\newblock \showarticletitle{Anomaly detection on electroencephalography with self-supervised learning}. In \bibinfo{booktitle}{\emph{2020 IEEE International Conference on Bioinformatics and Biomedicine (BIBM)}}. IEEE, \bibinfo{pages}{363--368}.
\newblock


\bibitem[Yang et~al\mbox{.}(2023)]%
        {re147}
\bibfield{author}{\bibinfo{person}{Chaoqi Yang}, \bibinfo{person}{Cao Xiao}, \bibinfo{person}{M~Brandon Westover}, \bibinfo{person}{Jimeng Sun}, {et~al\mbox{.}}} \bibinfo{year}{2023}\natexlab{}.
\newblock \showarticletitle{Self-supervised electroencephalogram representation learning for automatic sleep staging: model development and evaluation study}.
\newblock \bibinfo{journal}{\emph{JMIR AI}} \bibinfo{volume}{2}, \bibinfo{number}{1} (\bibinfo{year}{2023}), \bibinfo{pages}{e46769}.
\newblock


\bibitem[Ye et~al\mbox{.}(2021)]%
        {re165}
\bibfield{author}{\bibinfo{person}{Jianan Ye}, \bibinfo{person}{Qinfeng Xiao}, \bibinfo{person}{Jing Wang}, \bibinfo{person}{Hongjun Zhang}, \bibinfo{person}{Jiaoxue Deng}, {and} \bibinfo{person}{Youfang Lin}.} \bibinfo{year}{2021}\natexlab{}.
\newblock \showarticletitle{Cosleep: A multi-view representation learning framework for self-supervised learning of sleep stage classification}.
\newblock \bibinfo{journal}{\emph{IEEE Signal Processing Letters}}  \bibinfo{volume}{29} (\bibinfo{year}{2021}), \bibinfo{pages}{189--193}.
\newblock


\bibitem[Ye et~al\mbox{.}(2023)]%
        {re148}
\bibfield{author}{\bibinfo{person}{Weishan Ye}, \bibinfo{person}{Zhiguo Zhang}, \bibinfo{person}{Min Zhang}, \bibinfo{person}{Fei Teng}, \bibinfo{person}{Li Zhang}, \bibinfo{person}{Linling Li}, \bibinfo{person}{Gan Huang}, \bibinfo{person}{Jianhong Wang}, \bibinfo{person}{Dong Ni}, {and} \bibinfo{person}{Zhen Liang}.} \bibinfo{year}{2023}\natexlab{}.
\newblock \showarticletitle{Semi-Supervised Dual-Stream Self-Attentive Adversarial Graph Contrastive Learning for Cross-Subject EEG-based Emotion Recognition}.
\newblock \bibinfo{journal}{\emph{arXiv preprint arXiv:2308.11635}} (\bibinfo{year}{2023}).
\newblock


\bibitem[You et~al\mbox{.}(2023)]%
        {re171}
\bibfield{author}{\bibinfo{person}{Yuyang You}, \bibinfo{person}{Shuohua Chang}, \bibinfo{person}{Zhihong Yang}, {and} \bibinfo{person}{Qihang Sun}.} \bibinfo{year}{2023}\natexlab{}.
\newblock \showarticletitle{PSNSleep: a self-supervised learning method for sleep staging based on Siamese networks with only positive sample pairs}.
\newblock \bibinfo{journal}{\emph{Frontiers in Neuroscience}}  \bibinfo{volume}{17} (\bibinfo{year}{2023}), \bibinfo{pages}{1167723}.
\newblock


\bibitem[Zbontar et~al\mbox{.}(2021)]%
        {re155}
\bibfield{author}{\bibinfo{person}{Jure Zbontar}, \bibinfo{person}{Li Jing}, \bibinfo{person}{Ishan Misra}, \bibinfo{person}{Yann LeCun}, {and} \bibinfo{person}{St{\'e}phane Deny}.} \bibinfo{year}{2021}\natexlab{}.
\newblock \showarticletitle{Barlow twins: Self-supervised learning via redundancy reduction}. In \bibinfo{booktitle}{\emph{International Conference on Machine Learning}}. PMLR, \bibinfo{pages}{12310--12320}.
\newblock


\bibitem[Zhai et~al\mbox{.}(2018)]%
        {re77}
\bibfield{author}{\bibinfo{person}{Junhai Zhai}, \bibinfo{person}{Sufang Zhang}, \bibinfo{person}{Junfen Chen}, {and} \bibinfo{person}{Qiang He}.} \bibinfo{year}{2018}\natexlab{}.
\newblock \showarticletitle{Autoencoder and its various variants}. In \bibinfo{booktitle}{\emph{2018 IEEE international conference on systems, man, and cybernetics (SMC)}}. IEEE, \bibinfo{pages}{415--419}.
\newblock


\bibitem[Zhang et~al\mbox{.}(2022a)]%
        {re154}
\bibfield{author}{\bibinfo{person}{Hongjun Zhang}, \bibinfo{person}{Jing Wang}, \bibinfo{person}{Jiahong Xiong}, \bibinfo{person}{Yuxuan Ding}, \bibinfo{person}{Zhenliang Gan}, {and} \bibinfo{person}{Youfang Lin}.} \bibinfo{year}{2022}\natexlab{a}.
\newblock \showarticletitle{Expert knowledge inspired contrastive learning for sleep staging}. In \bibinfo{booktitle}{\emph{2022 International Joint Conference on Neural Networks (IJCNN)}}. IEEE, \bibinfo{pages}{1--6}.
\newblock


\bibitem[Zhang et~al\mbox{.}(2008)]%
        {re112}
\bibfield{author}{\bibinfo{person}{Lin Zhang}, \bibinfo{person}{Jonathan Samet}, \bibinfo{person}{Brian Caffo}, \bibinfo{person}{Isaac Bankman}, {and} \bibinfo{person}{Naresh~M Punjabi}.} \bibinfo{year}{2008}\natexlab{}.
\newblock \showarticletitle{Power spectral analysis of EEG activity during sleep in cigarette smokers}.
\newblock \bibinfo{journal}{\emph{Chest}} \bibinfo{volume}{133}, \bibinfo{number}{2} (\bibinfo{year}{2008}), \bibinfo{pages}{427--432}.
\newblock


\bibitem[Zhang et~al\mbox{.}(2016)]%
        {re76}
\bibfield{author}{\bibinfo{person}{Richard Zhang}, \bibinfo{person}{Phillip Isola}, {and} \bibinfo{person}{Alexei~A Efros}.} \bibinfo{year}{2016}\natexlab{}.
\newblock \showarticletitle{Colorful image colorization}. In \bibinfo{booktitle}{\emph{Computer Vision--ECCV 2016: 14th European Conference, Amsterdam, The Netherlands, October 11-14, 2016, Proceedings, Part III 14}}. Springer, \bibinfo{pages}{649--666}.
\newblock


\bibitem[Zhang and Chen(2016)]%
        {re30}
\bibfield{author}{\bibinfo{person}{Tao Zhang} {and} \bibinfo{person}{Wanzhong Chen}.} \bibinfo{year}{2016}\natexlab{}.
\newblock \showarticletitle{LMD based features for the automatic seizure detection of EEG signals using SVM}.
\newblock \bibinfo{journal}{\emph{IEEE Transactions on Neural Systems and Rehabilitation Engineering}} \bibinfo{volume}{25}, \bibinfo{number}{8} (\bibinfo{year}{2016}), \bibinfo{pages}{1100--1108}.
\newblock


\bibitem[Zhang et~al\mbox{.}(2022b)]%
        {re133}
\bibfield{author}{\bibinfo{person}{Wenrui Zhang}, \bibinfo{person}{Ling Yang}, \bibinfo{person}{Shijia Geng}, {and} \bibinfo{person}{Shenda Hong}.} \bibinfo{year}{2022}\natexlab{b}.
\newblock \showarticletitle{Self-Supervised Time Series Representation Learning via Cross Reconstruction Transformer}.
\newblock \bibinfo{journal}{\emph{arXiv preprint arXiv:2205.09928}} (\bibinfo{year}{2022}).
\newblock


\bibitem[Zhang et~al\mbox{.}(2022c)]%
        {re152}
\bibfield{author}{\bibinfo{person}{Xiang Zhang}, \bibinfo{person}{Ziyuan Zhao}, \bibinfo{person}{Theodoros Tsiligkaridis}, {and} \bibinfo{person}{Marinka Zitnik}.} \bibinfo{year}{2022}\natexlab{c}.
\newblock \showarticletitle{Self-supervised contrastive pre-training for time series via time-frequency consistency}.
\newblock \bibinfo{journal}{\emph{Advances in Neural Information Processing Systems}}  \bibinfo{volume}{35} (\bibinfo{year}{2022}), \bibinfo{pages}{3988--4003}.
\newblock


\bibitem[Zhang and Yang(2021)]%
        {re176}
\bibfield{author}{\bibinfo{person}{Yu Zhang} {and} \bibinfo{person}{Qiang Yang}.} \bibinfo{year}{2021}\natexlab{}.
\newblock \showarticletitle{A survey on multi-task learning}.
\newblock \bibinfo{journal}{\emph{IEEE Transactions on Knowledge and Data Engineering}} \bibinfo{volume}{34}, \bibinfo{number}{12} (\bibinfo{year}{2021}), \bibinfo{pages}{5586--5609}.
\newblock


\bibitem[Zhang et~al\mbox{.}(2022d)]%
        {re125}
\bibfield{author}{\bibinfo{person}{Zhi Zhang}, \bibinfo{person}{Sheng-hua Zhong}, {and} \bibinfo{person}{Yan Liu}.} \bibinfo{year}{2022}\natexlab{d}.
\newblock \showarticletitle{GANSER: A self-supervised data augmentation framework for EEG-based emotion recognition}.
\newblock \bibinfo{journal}{\emph{IEEE Transactions on Affective Computing}} (\bibinfo{year}{2022}).
\newblock


\bibitem[Zheng et~al\mbox{.}(2018)]%
        {rp15}
\bibfield{author}{\bibinfo{person}{Wei-Long Zheng}, \bibinfo{person}{Wei Liu}, \bibinfo{person}{Yifei Lu}, \bibinfo{person}{Bao-Liang Lu}, {and} \bibinfo{person}{Andrzej Cichocki}.} \bibinfo{year}{2018}\natexlab{}.
\newblock \showarticletitle{Emotionmeter: A multimodal framework for recognizing human emotions}.
\newblock \bibinfo{journal}{\emph{IEEE transactions on cybernetics}} \bibinfo{volume}{49}, \bibinfo{number}{3} (\bibinfo{year}{2018}), \bibinfo{pages}{1110--1122}.
\newblock


\bibitem[Zheng and Lu(2015)]%
        {re44}
\bibfield{author}{\bibinfo{person}{Wei-Long Zheng} {and} \bibinfo{person}{Bao-Liang Lu}.} \bibinfo{year}{2015}\natexlab{}.
\newblock \showarticletitle{Investigating Critical Frequency Bands and Channels for {EEG}-based Emotion Recognition with Deep Neural Networks}.
\newblock \bibinfo{journal}{\emph{IEEE Transactions on Autonomous Mental Development}} \bibinfo{volume}{7}, \bibinfo{number}{3} (\bibinfo{year}{2015}), \bibinfo{pages}{162--175}.
\newblock
\urldef\tempurl%
\url{https://doi.org/10.1109/TAMD.2015.2431497}
\showDOI{\tempurl}


\bibitem[Zheng et~al\mbox{.}(2022)]%
        {re98}
\bibfield{author}{\bibinfo{person}{Yaojia Zheng}, \bibinfo{person}{Zhouwu Liu}, \bibinfo{person}{Rong Mo}, \bibinfo{person}{Ziyi Chen}, \bibinfo{person}{Wei-shi Zheng}, {and} \bibinfo{person}{Ruixuan Wang}.} \bibinfo{year}{2022}\natexlab{}.
\newblock \showarticletitle{Task-oriented self-supervised learning for anomaly detection in electroencephalography}. In \bibinfo{booktitle}{\emph{International Conference on Medical Image Computing and Computer-Assisted Intervention}}. Springer, \bibinfo{pages}{193--203}.
\newblock


\bibitem[Zhu et~al\mbox{.}(2023)]%
        {re127}
\bibfield{author}{\bibinfo{person}{Qiushi Zhu}, \bibinfo{person}{Xiaoying Zhao}, \bibinfo{person}{Jie Zhang}, \bibinfo{person}{Yu Gu}, \bibinfo{person}{Chao Weng}, {and} \bibinfo{person}{Yuchen Hu}.} \bibinfo{year}{2023}\natexlab{}.
\newblock \showarticletitle{Eeg2vec: Self-Supervised Electroencephalographic Representation Learning}.
\newblock \bibinfo{journal}{\emph{arXiv preprint arXiv:2305.13957}} (\bibinfo{year}{2023}).
\newblock


\end{thebibliography}


\end{document}